\documentclass{article}
\usepackage{amssymb}
\usepackage{jheppub}

\def\MCEq{\texttt{MCEq}}

\usepackage[english]{babel}
\usepackage{booktabs}

\usepackage{amsmath}
\usepackage{graphicx}
\usepackage{float}
\usepackage{cleveref}
\usepackage{xcolor}

\preprint{DESY-25-194}
\title{The role of charm and unflavored mesons in prompt atmospheric lepton fluxes}
\author[a]{Laksha Pradip Das,}
\author[a]{Diksha Garg,}
\author[b]{Maria Vittoria Garzelli,}
\author[a]{Mary~Hall Reno,}
\author[b]{G\"unter Sigl}
\affiliation[a]{Department of Physics and Astronomy, University of Iowa, Iowa City, IA 52242, USA}
\affiliation[b]{II Institut f\"{u}r Theoretische Physik, Universit\"{a}t Hamburg, Luruper Chaussee 149, D-22761 Hamburg, Germany}

\emailAdd{lakshapradip-das@uiowa.edu}
\emailAdd{diksha-garg@uiowa.edu}
\emailAdd{maria.vittoria.garzelli@desy.de}
\emailAdd{mary-hall-reno@uiowa.edu}
\emailAdd{guenter.sigl@desy.de}

\abstract{The all-sky very-high-energy ($10^4-10^6$ GeV) atmospheric muon flux measured by IceCube shows a spectral hardening at the highest energies, indicating the presence of a prompt component. IceCube has also measured the atmospheric muon neutrino flux at high energy. However, since this flux is dominated by astrophysical neutrinos, only an upper bound can be placed on the prompt atmospheric $\nu_\mu+\bar{\nu}_\mu$ contribution. In this work, we provide a new evaluation of the prompt atmospheric muon flux including an intrinsic charm component in the cosmic ray-air interactions. The latter enhances the forward production of $\bar{D}^0$, $D^-$, and $\Lambda_c$, which subsequently decay into final states containing muons and muon neutrinos. We show how the increase in the prompt muon flux due to intrinsic charm is accompanied by a corresponding enhancement in the prompt muon neutrino flux. We implement different intrinsic charm production models in \texttt{MCEq} to calculate the resulting lepton fluxes. We discuss the challenges of achieving predictions that are simultaneously consistent with both IceCube’s high-energy atmospheric muon flux measurements and IceCube upper bounds on the prompt muon neutrino flux, and we quantify the resulting discrepancies. As possible solutions, we explore scaling of the unflavored meson contributions to the prompt atmospheric muon flux to assess how such adjustments can reconcile these differences. The tensions emphasized in our work call for a refinement of the hadronic interaction models, especially the production of unflavored mesons, and for new experimental data sensitive to unflavored meson and heavy flavor production with reliable estimates of the associated uncertainties. We suggest that the energy and zenith angle dependence of muon and neutrino flux ratios from future neutrino telescope measurements may help to disentangle different scenarios. 
}
\begin{document}
\maketitle

\section{Introduction}\label{sec:intro}
When high-energy cosmic rays enter the Earth’s atmosphere, they interact with air nuclei and initiate a cascade of secondary particles, including protons, neutrons, pions, kaons, and charm hadrons. These secondary hadrons undergo further interactions and decays, giving rise to the atmospheric fluxes of muons and neutrinos (see, e.g., ref. \cite{Gaisser:2002jj} and references therein). Since the primary cosmic ray spectrum spans many orders of magnitude in energy, the secondary particles are also produced in a wide energy range \cite{Gaisser:2016uoy,Lipari:1993hd}. Among these, high-energy atmospheric leptons ($E > 10^4$ GeV), in particular muons and neutrinos, are of central importance in astroparticle physics. The latter form a major background for neutrino telescopes searching for astrophysical neutrinos \cite{IceCube:2013low,IceCube:2016umi,IceCube:2020acn,IceCube:2024fxo,IceCube:2025dlr,Abbasi:2025rmj}, making precise knowledge of the atmospheric neutrino fluxes essential. Such measurements, together with the complementary input from atmospheric muon fluxes, not only help testing and constraining hadronic interaction models (see, e.g., \cite{Fedynitch:2018cbl, Fedynitch:2022vty,Yanez:2023lsy}), but also enable robust searches for astrophysical neutrino signals. 

Atmospheric leptons are produced through a chain of interactions and decays: cosmic rays interact with air nuclei with average atomic mass number of $\langle A_{\rm air} \rangle = 14.5$ to produce hadrons. Muons and neutrinos come from leptonic or semi-leptonic decays of these hadrons. To evaluate the atmospheric lepton flux, incident cosmic ray nuclei of mass number $A_{N}$, charge $Z$ and energy $E_0$ are treated in the superposition approximation, i.e., the nucleus is treated as an ensemble of $Z$ protons and $A_N-Z$ neutrons, each carrying energy $E_0/A_{N}$ \cite{Gaisser:2016uoy}. The resulting all-nucleon cosmic ray spectrum at the top of the atmosphere is denoted by $\phi^0_{N}(E)$. Although the cosmic ray composition above $E_0\simeq 10^5-10^6$~GeV 
has been measured only indirectly \cite{Kampert:2012mx}, there is reasonably good agreement among cosmic ray composition models that $\phi^0_N(E)\sim E^{-\gamma}$ with the value of $\gamma$ approximately constant but different in different energy ranges (e.g., broken power law) \cite{APEL2012183,PierreAuger:2014gko,Rawlins:2020hzc,Gaisser:2011klf,Gaisser:2013bla,Hoerandel:2002yg,Dembinski:2017zsh}. 

The magnitudes of the atmospheric lepton fluxes from hadron decays depend on the competition between hadron-air interactions and hadron decays. If the hadron interaction length is much longer than its decay length, the lepton flux at energy $E_\ell$ approximately follows the primary cosmic ray nucleon spectrum, scaling as $\sim E_\ell^{-\gamma}$. Conversely, when hadrons are more likely to interact than decay, the corresponding atmospheric lepton flux is suppressed, scaling as $E_\ell^{-\gamma-1}$ \cite{Lipari:1993hd}. For the energies considered here, $E_\ell\gtrsim 10^4$ GeV, the contributions from $\pi^\pm$ and $K$ decays  to muon and neutrino fluxes scale like $\sim E_\ell^{-\gamma-1}$. The combination of these two contributions is referred to as the conventional atmospheric lepton flux. 
In contrast, hadrons with much shorter lifetimes, such as charm hadrons, decay promptly over a wide energy range. The prompt atmospheric lepton flux from charm hadron decays scales as $ \sim E_\ell^{-\gamma}$, up to lepton energies $E_\ell \sim 2\times 10^7$ GeV \cite{Gondolo:1995fq}. In addition to charm hadron decays, prompt muon production also receives contributions from electromagnetic decays of light unflavored mesons such as $\eta,\ \eta^\prime,\ \rho^0,\ \omega,$ and $\phi$, despite their small branching fractions ($\sim 10^{-5}$–$10^{-4}$) \cite{Illana:2010gh}. Their extremely short lifetimes ensure decay before interaction, enhancing the prompt atmospheric muon flux at high energies by a factor of $\sim 2$  relative to the prompt neutrino flux.

\begin{figure}[t]
    \centering    \includegraphics[width=0.45\linewidth]{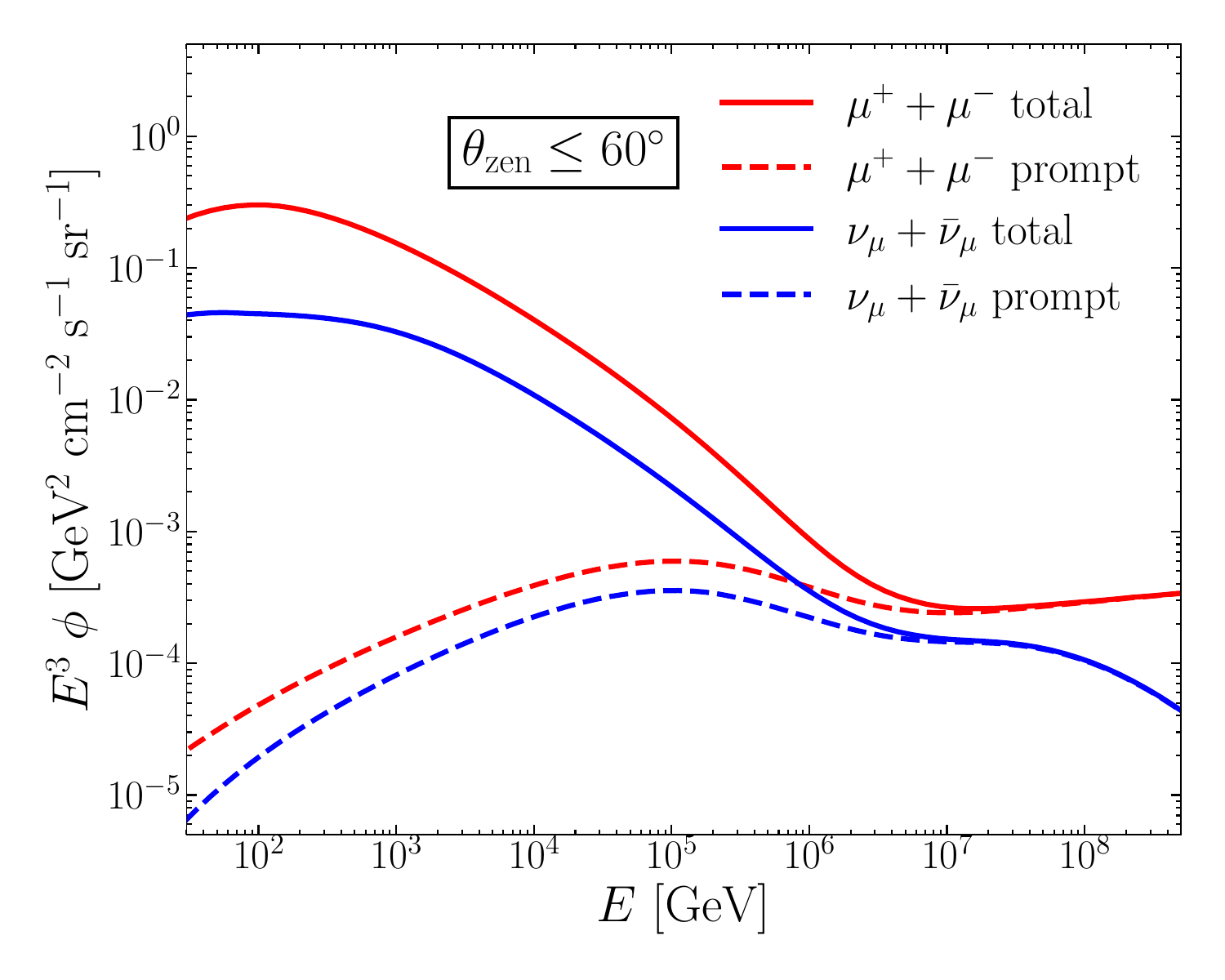}
    \caption{The atmospheric ($\mu^+ + \mu^-$) flux  and atmospheric ($\nu_\mu+\bar\nu_\mu$) flux, both scaled by $E^3$ and angle-averaged over zenith angles less than 60$^\circ$, from \texttt{MCEq} \cite{Fedynitch:2018cbl} using \texttt{H3a} cosmic ray flux model~\cite{Gaisser:2011klf} and \texttt{Sibyll-2.3c} \cite{Riehn:2017mfm,Fedynitch:2018cbl} for interactions.
    The dashed lines show the prompt contributions, and the solid lines show the total contributions (conventional + prompt).}
    \label{fig:average-flux-atm}
\end{figure}

The energy behaviors of the conventional and prompt lepton fluxes combine to give a characteristic energy spectrum that indirectly reveals the hadronic origins of the leptons. 
\Cref{fig:average-flux-atm} illustrates the angle-averaged $\mu^+ + \mu^-$ and $\nu_\mu + \bar\nu_\mu$ atmospheric fluxes, showing the transition from the conventional to the prompt regime. At low energies, the $\mu^++\mu^-$ flux exceeds the $\nu_\mu+\bar\nu_\mu$ flux since in $\pi\to\mu\nu_\mu$ decays the muon carries most of the available energy. At high energies, the additional prompt muon contributions from unflavored meson decays become the dominant difference between the $\mu^++\mu^-$ and $\nu_\mu+\bar\nu_\mu$ fluxes. The charm decays produce nearly equal fluxes of muons and muon neutrinos, with the latter exceeding the former by about $15$–$20\%$ due to decay kinematics~\cite{Illana:2010gh,Volkova:2011zza}, as we also confirm in  Appendix \ref{appendix:a}. 

Prompt neutrinos from charm decays in the atmosphere are often considered an irreducible background to astrophysical neutrinos, since both fluxes are nearly isotropic at Earth. Their spectral behaviors differ, offering possible means of discrimination between the prompt atmospheric background and astrophysical neutrinos. 
In this paper, we focus on the role of enhanced charm production in the atmosphere using the atmospheric muon flux as a complementary probe to atmospheric neutrinos. 
While many previous studies  have focused primarily on neutrino fluxes~\cite{Laha:2016dri,Halzen:2016thi,Giannini:2018utr,Goncalves:2021yvw,Ostapchenko:2022thy,Garzelli:2023jlq}, our analysis expands to include both muons and neutrinos (see also, ref. \cite{Das:2025snq}).
As we describe in the next section, measurements of the atmospheric muon flux, particularly in the energy range where the prompt muon flux begins to dominate, exceed theory predictions, although with large experimental error bars \cite{Soldin:2023lbr, Soldin:2018vak,IceCube:2015wro}. 
To solve this discrepancy, we first consider potential contributions from intrinsic charm. The intrinsic charm contributions are also constrained by prompt neutrino measurements. Therefore, with existing upper bounds on the prompt atmospheric muon neutrino flux from IceCube \cite{IceCube:2016umi,IceCube:2015wro,Abbasi:2025rmj}, we also assess the extent to which the contribution of unflavored mesons to the prompt atmospheric muon flux may be enhanced.  

The paper is organized as follows: in Section \ref{sec:atmfluxes}, we discuss the cascade equations that describe the development of the atmospheric lepton fluxes as cosmic rays and the hadrons they produce propagate through the atmosphere. We also show results from the cascade equations for muons and muon neutrinos, along with IceCube data and upper bounds on the prompt neutrino flux, respectively. In Section \ref{sec:intrinsiccharm}, we introduce the framework we use to incorporate intrinsic charm effects. Section \ref{sec:results} presents our results including framework-dependent constraints on intrinsic charm, Monte Carlo generator dependence and cosmic ray flux model dependence, followed by a discussion in Section \ref{sec:discussion}.  Within this framework, we conclude in Section \ref{sec:conclusion} that enhanced charm production cannot account for both the measured atmospheric muon flux \cite{Soldin:2023lbr, Soldin:2018vak,IceCube:2015wro} and the IceCube upper bound \cite{IceCube:2016umi,IceCube:2015wro,Abbasi:2025rmj} on the prompt $\nu_\mu+\bar\nu_\mu$ flux.

For completeness, Appendix \ref{appendix:a}  includes details of the charm particle decays to leptons, needed to evaluate the lepton fluxes. A brief discussion of an approximate semi-analytic description of the prompt atmospheric muon flux based on the $Z-$moment method is provided in Appendix \ref{appendix:z}. A comparison of our default prompt atmospheric  $\nu_\mu+\bar\nu_\mu$ flux with several recent perturbative QCD evaluations of the same quantity in the energy range relevant to the IceCube upper bounds \cite{Abbasi:2025rmj} is shown in Appendix \ref{appendix:b}.

\section{Atmospheric lepton fluxes}\label{sec:atmfluxes}
\subsection{Cascade equations}
\label{subsec:cascade}

Quantitative evaluations of the atmospheric flux begin with the incident all-nucleon cosmic ray spectrum at the top of the atmosphere, $\phi_N^0(E)$. The flux  per solid angle $\phi_N^0$ is isotropic  at high energies. 
The cascade equations that describe the development of particle fluxes with subsequent interactions and decays \cite{Lipari:1993hd,Gondolo:1995fq}. 
The cascade equations are evaluated as a function of column depth $X$ which depends on the atmospheric density ($\rho$), path length ($l$) traversed by the particle in the atmosphere, altitude ($h$) and zenith angle ($\theta$),
 \begin{equation}\label{eq:col_dpeth}
    X(l,\theta) = \int_l^\infty d l' \ \rho(h(l',\theta))\,.
\end{equation}
The cascade equation for nucleons is
 \begin{eqnarray}
 \frac{d\phi_N(E,X)}{dX}&=&-\frac{\phi_N(E,X)}{\lambda_N(E)} + S(N\to N; E, X)\, ,
 \end{eqnarray}
where the source term for nucleon-nucleon regeneration is determined by
 \begin{equation}
 \label{eq:source_term} S(k\to j; E, X) = \int_E^{\infty}dE ' \frac{\phi_k(E',X)}{\lambda_k(E')}\frac{dn(k\to j;E',E)}{dE}\, ,
 \end{equation}
 for $k=j=N$. The energy distribution of outgoing particle $j$ with incoming particle $k$ is given by
 \begin{equation} 
 \frac{dn(k\to j;E',E)}{dE}=\frac{1}{\sigma_{kA}(E')}\frac{d\sigma(k\to j;E',E)}{dE}\, .
 \end{equation}
where $E^\prime$ and $E$ are energies of incoming particle $k$ and outgoing particle $j$, respectively.
 The interaction length depends on the cross section, with $\lambda_k= 1/(N_A \sigma _{kA})$. For other hadrons, additional loss in flux occurs from decays, so
 \begin{equation}
\frac{d\phi_j(E,X)}{dX}=-\frac{\phi_j(E,X)}{\lambda_j(E)} - \frac{\phi_j(E,X)}{\lambda_j^{\rm dec}(E)}
+ \sum _k S(k\to j; E, X)\, ,
\label{eq:hadron-cascade}
\end{equation}
where the hadron decay length is 
$\lambda^{\rm dec}_j (E)= \rho(h)\gamma_j (E) c\tau_j$ and $\gamma_j (E)= E/m_j$.

Muons and neutrinos are produced by hadron decays in the atmosphere (neglecting the photon conversion to $\mu^++\mu^-$ processes, discussed below). Their cascade equations are
\begin{eqnarray}
\label{eq:neutrino-cascade}
 \frac{d\phi_\nu(E,X)}{dX}&=&
\sum _j S(j\to \nu; E, X)\, \\
 \frac{d\phi_\mu(E,X)}{dX}&=& - \frac{\partial }{\partial E}\Bigl( \beta(E)\phi_\mu(E,X)\Bigr)+
\sum _j S(j\to \mu; E, X)\, . \label{eq:muon-cascade}
\end{eqnarray}
The muon cascade equation has a term $\beta(E)\equiv dE/dX$  that accounts for electromagnetic energy loss due to ionization and radiative processes. The source terms for decays to muons and neutrinos involve the decay distributions rather than the differential cross sections, and the decay lengths rather than the interaction lengths. 

The total column depth traversed by particles depends strongly on the zenith angle of their trajectory. For vertical incidence ($\theta = 0^\circ$), the atmospheric column depth is $X_{\rm total} \simeq 1036\ \text{g/cm}^2$, while for horizontal incidence ($\theta = 90^\circ$), it increases to $X_{\rm total} \simeq 3.65 \times 10^4\ \text{g/cm}^2$. Consequently, the relative probabilities for hadronic interactions and decays, as well as the associated energy losses of secondary particles, can vary significantly with zenith angle. This in turn makes the atmospheric lepton flux dependent on the zenith angle. 

The cascade equations can be solved approximately in the hadron-decay-dominated and hadron-interaction-dominated regimes \cite{Lipari:1993hd,Gondolo:1995fq}. This method, called the $Z$-moment method, approximates 
eq.~(\ref{eq:source_term}) for production and decay processes by putting $\phi_k(E',X)/\phi_k(E,X)\simeq\phi_k(E',0)/\phi_k(E,0)$ so that
 \begin{equation}
 \label{eq:source_term1} S(k\to j; E, X) \simeq \frac{\phi_k(E,X)}{\lambda_k(E)}
 \int _E^\infty dE'
 \frac{\phi_k(E',0)}{\phi_k(E,0)}\frac{\lambda_k(E)}{\lambda_k(E')}\frac{dn(k\to j;E',E)}{dE}= \frac{\phi_k(E,X)}{\lambda_k(E)} Z_{kj}(E)\, .
 \end{equation}
The atmospheric lepton flux at sea level is written as combinations of $Z$-moments, interaction lengths and lifetimes. 
The $Z-$moment solution for prompt lepton fluxes is briefly described in Appendix \ref{appendix:z}.

An alternative approach for evaluating the atmospheric lepton fluxes is to solve the cascade equations directly. The Matrix Cascade Equations (\texttt{MCEq}) code~\cite{fedynitch2015calculationconventionalpromptlepton,Fedynitch:2018cbl} solves the cascade equations using a matrix approach. While the \texttt{MCEq} results are more precise, given hadronic production inputs, the $Z$-moment method is straightforward to implement. The two approaches differ by $5-20\%$ depending on energy and angle \cite{Gaisser:2019xlw}. In this paper, we use the \texttt{MCEq} code with some modifications outlined below to evaluate all atmospheric lepton fluxes. Another approach is to use Monte Carlo air shower simulations. Recent air shower flux calculations~\cite{LudwigNeste:2025mcf} show good agreement with \texttt{MCEq} results.

\subsection{Muon and muon neutrino fluxes from \texttt{MCEq}}

\begin{figure}
    \centering
\includegraphics[width=1\linewidth]{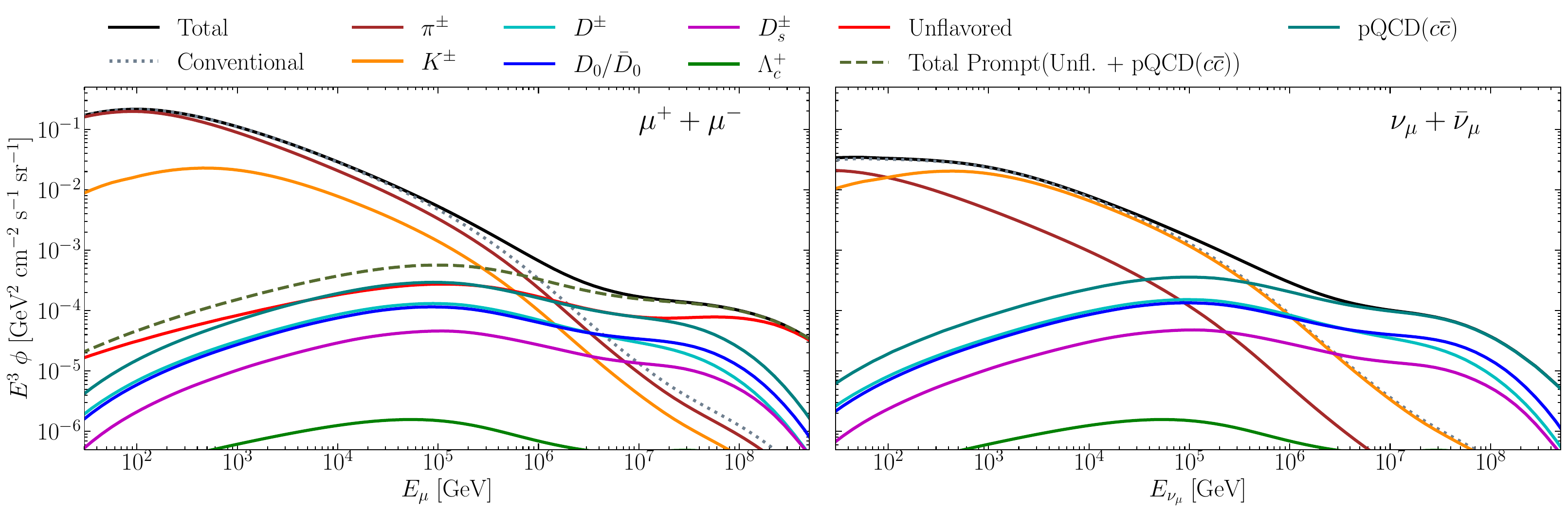}
    \caption{Results for $E^3\phi(E)$ for $\mu^++\mu^-$ and $\nu_\mu+\bar\nu_\mu$ for zenith angle $\theta=0^\circ$ (vertical), showing the individual contributions of light, heavy and unflavored mesons, as determined by \texttt{MCEq} using the \texttt{Sibyll-2.3c} hadronic interaction model and the \texttt{H3a} cosmic ray flux model.}~
    \label{fig:fluxes-std}
\end{figure}

\texttt{MCEq} is equipped with several hadronic interaction models from event generators (e.g., \texttt{Sibyll-2.3c}, \texttt{DPMJETIII306}, \texttt{QGSJETII04}, \texttt{EPOSLHC}) and cosmic ray nucleon flux models.
We use the model \texttt{Sibyll-2.3c}~\cite{Riehn:2017mfm,Fedynitch:2018cbl} for hadronic interactions, and the \texttt{H3a} cosmic ray flux and composition model~\cite{Gaisser:2011klf}. We discuss the impact of different hadronic models and all-nucleon cosmic ray flux inputs in Section \ref{subsec:uncertainties}.

Charm production and its contribution to the prompt atmospheric neutrino flux have been extensively studied \cite{Gondolo:1995fq,Pasquali:1998xf,Enberg:2008te,Bhattacharya:2015jpa,Bhattacharya:2016jce,Gauld:2015yia,Gauld:2015kvh,Garzelli:2015psa,Garzelli:2016xmx,Ostapchenko:2022thy,Bai:2022xad,Benzke:2017yjn,Zenaiev:2019ktw}. Next-to-leading order (NLO) perturbative QCD evaluations of open charm production (sometimes called ``extrinsic charm'') have yielded a range of predictions that depend on parton distribution function choices, the charm quark mass and renormalization/factorization scales. As elaborated in Appendix \ref{appendix:b}, in the energy range relevant to the IceCube upper bounds on the prompt atmospheric $\nu_\mu+\bar\nu_\mu$ flux, the \texttt{Sibyll-2.3c} neutrino flux prediction is above the central values of other predictions, but within the large uncertainty bands of the perturbative QCD calculations. 
Given this consistency with NLO QCD-based approaches to charm production and the fact that the  recent IceCube upper bound on the prompt atmospheric neutrino flux relies on the shape of the \texttt{Sibyll-2.3c} flux, we adopt the \texttt{Sibyll-2.3c} prompt atmospheric lepton flux results from charm as our baselines. \texttt{Sibyll-2.3c} has the added feature that it includes prompt contributions from light unflavored meson decays to $\mu^+\mu^-$, which can not be obtained through pQCD, so it provides a unified approach to the prompt atmospheric muon flux.

The implementation of charm production in \texttt{Sibyll-2.3c} follows a phenomenological model based on soft and hard minijets \cite{Fedynitch:2018cbl}. In addition to consistency with NLO QCD-based prompt atmospheric neutrino flux evaluations as demonstrated in Appendix \ref{appendix:b},
for LHCb kinematic distributions for charmed-meson  production,  the overall combination of perturbative and non-perturbative contributions in \texttt{Sibyll-2.3c} is consistent with the next-to-leading order QCD predictions of refs. \cite{Gauld:2015yia,Garzelli:2016xmx,PROSA:2015yid} 
at LHC collider energies (see, e.g., figure 32 of ref. \cite{Fedynitch:2018cbl}).
Further cross-checks have been carried out  over the wide center-of-mass energy range interesting for the production of prompt lepton fluxes, showing agreement within the large error bands of the pQCD calculations. As detailed in ref. \cite{Fedynitch:2018cbl}, \texttt{Sibyll-2.3c} results are consistent with the inclusive $c\bar{c}$ cross section as a function of center-of-mass energy, with the transverse momentum distributions of $D$ mesons at ALICE and at LHCb, as well as with lower energy measurements of charm in $pp$ and $p\pi$ experiments. While some components of the implementation of charm production in \texttt{Sibyll-2.3c} are described as ``intrinsic'' charm, the charm hadrons do not have the same energy distribution, shape and normalization, as in the model for intrinsic charm that we introduce in Section \ref{sec:intrinsiccharm}. Our implementation of intrinsic charm does not refer to an intrinsic charm parton distribution function, as discussed below.
Even though \texttt{Sibyll-2.3c}’s implementation of
charm production has a weighted mix of perturbative and non-perturbative components, we will nevertheless refer to \texttt{Sibyll-2.3c}’s contribution to the prompt atmospheric lepton fluxes as coming from perturbative QCD (pQCD($c\bar{c}$)) to distinguish it from the intrinsic charm contribution. 

\Cref{fig:fluxes-std} shows the \texttt{MCEq} results for the separate contributions of pions, kaons and (prompt) charm hadrons to the  muon (left) and muon neutrino (right) atmospheric fluxes for zenith angle $\theta_{\rm zen}=0^\circ$ (vertical). As noted above, the muon flux also includes contributions from light unflavored mesons to its prompt component. The standard release of \texttt{MCEq} does not include $\Lambda_c$ production and decay. 
We include perturbative $\Lambda_c$ production by extracting the $D^0$ production spectrum from \texttt{MCEq} and applying a constant suppression factor of 0.10 that reflects the relative charm quark fragmentation fractions to $\Lambda_c$ compared to $D^0$ in proton-proton collisions~\cite{Lisovyi:2015uqa}. The final results of our paper are not very sensitive to this suppression factor because charm contributions to the atmospheric $\mu^++\mu^-$, $\nu_\mu+\bar\nu_\mu$ and $\nu_e+\bar\nu_e$ fluxes are dominated by the semi-leptonic decays of $D^\pm,\ D^0$ and $\bar D^0$ mesons. For semi-leptonic $\Lambda_c$ decays, we follow refs.~\cite{Bugaev:1998bi} and~\cite{Buras:1976dg} for the energy distributions of muons and neutrinos, which we have also described in Appendix \ref{appendix:a}. While the details of the semi-leptonic $\Lambda_c$ decay distribution are not essential for the pQCD($c\bar{c})$ contributions to the prompt lepton fluxes, a $\Lambda_c$ is always produced in our implementation in intrinsic-charm induced events, as discussed in \cref{sec:intrinsiccharm}. 

\Cref{fig:fluxes-std} exhibits an high-energy behavior that reflects the dominance of the prompt contributions to the atmospheric lepton fluxes, visible in their harder spectra compared to the conventional fluxes.
The plots show fluxes for vertical trajectories entering from the zenith, $\theta_{\rm zen}=0^\circ$. 
The zenith angle dependence is primarily in the conventional flux, where pions and kaons are more likely to interact than decay. The conventional flux increases with zenith angle according to $1/\cos\theta_{\rm zen}$ for $\theta_{\rm zen}\leq 70^\circ$ \cite{Klimushin:2000cy}. 
At very high energies, the prompt contributions will also vary with zenith angle, however, at the energies we consider for the rest of the paper, 
$E\lesssim 10^6$~GeV, the prompt atmospheric lepton fluxes are essentially isotropic.

The strong suppression of the prompt flux relative to the conventional flux in the lower energy range shown in \cref{fig:fluxes-std}
originates from the production and decay properties of the parent particles. For charm baryons, their relatively large mass makes their production less abundant than the production of lighter mesons such as pions and kaons, introducing a significant production suppression factor. In the case of unflavored mesons, their branching fractions to di-muon pairs is small, so even with a relatively large rate of unflavored meson production, their contributions to the muon flux is comparatively small.

We can also obtain di-muon pairs through the process $\gamma \to \mu^+\mu^-$. There are two possible sources for this channel. The first is from diffuse cosmic rays interacting in the atmosphere and producing high energy photons, which can subsequently convert into muon pairs. However, this contribution is expected to be very small even at the highest $\gamma$-ray energies~\cite{Illana:2010gh}. The second source is the diffuse $\gamma$-ray flux from the Galactic plane. According to ref.~\cite{LHAASO:2023gne}, at around 100 TeV the diffuse Galactic $\gamma$-ray flux is approximately $10^{-5}$ of the diffuse cosmic ray flux. Thus, diffuse Galactic $\gamma$-ray flux will also provide negligible contribution to the overall atmospheric muon flux.

\begin{figure}[ht]
    \centering
    \includegraphics[width=0.495\linewidth]{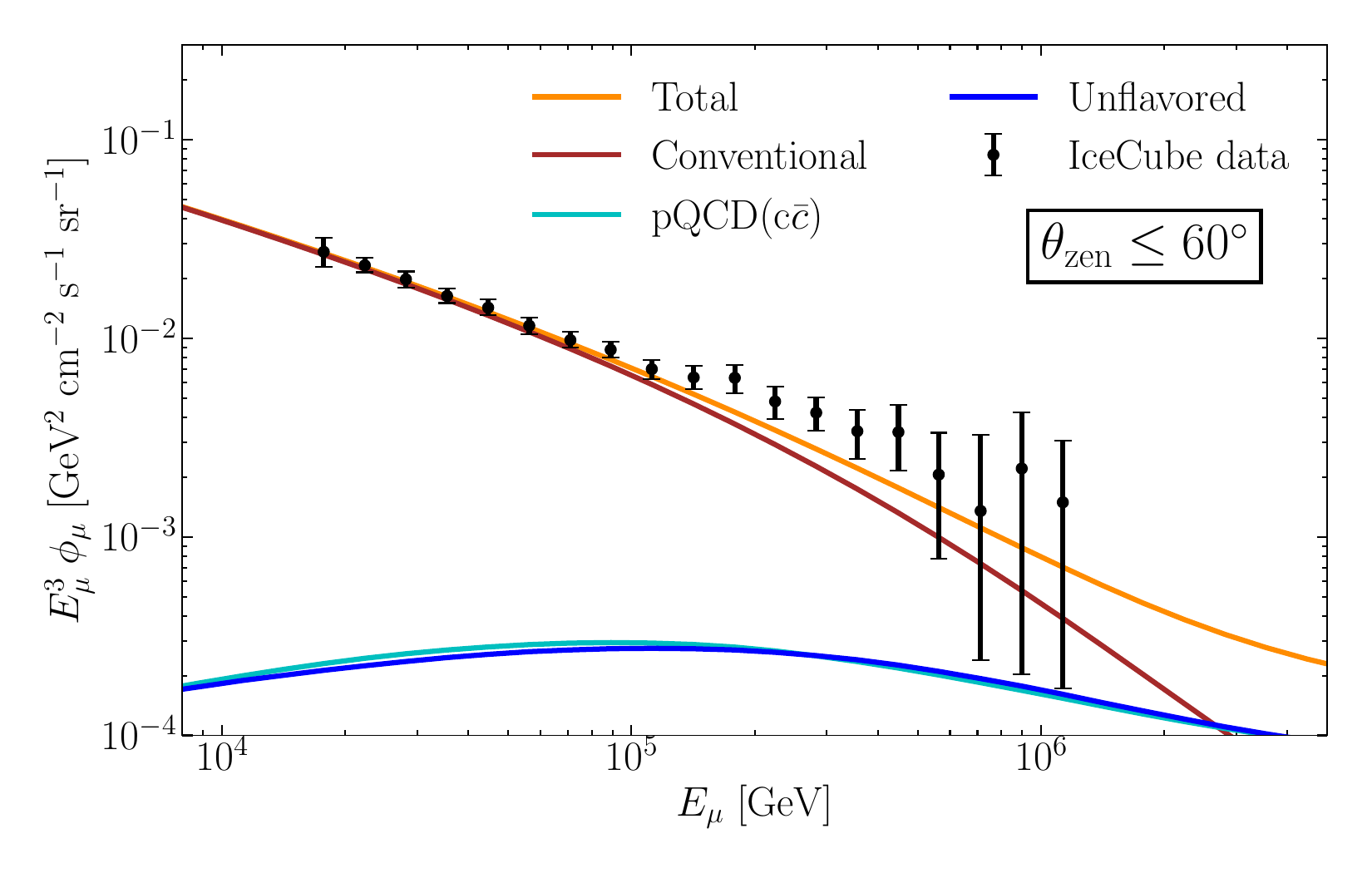}
        \includegraphics[width=0.495\linewidth]{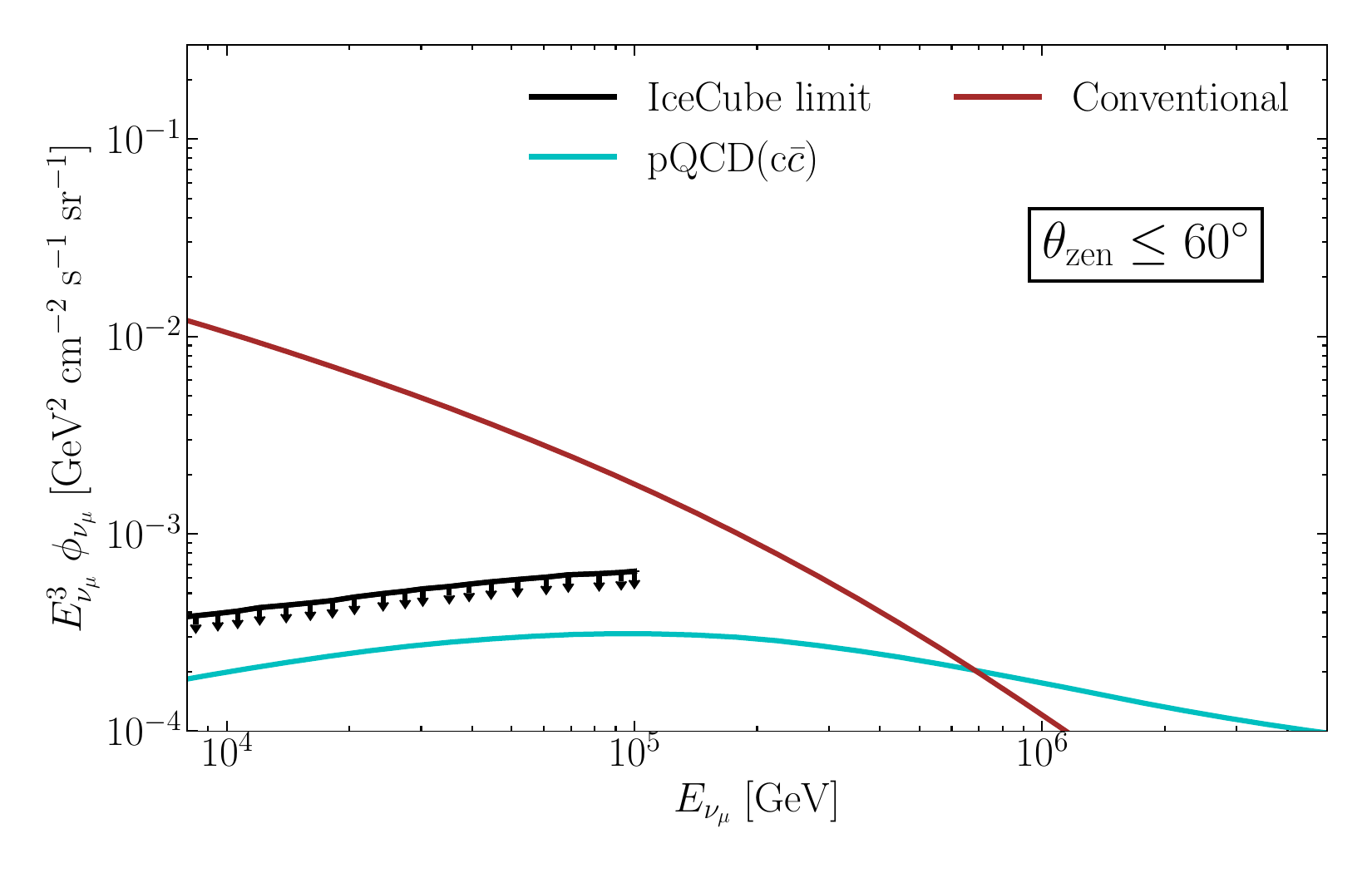}
    \caption{Left: \texttt{MCEq} predictions for angle-averaged (up to zenith angle of $60^\circ$) conventional and prompt ($\mu^{+}+\mu^{-}$) flux, compared to the experimentally measured spectrum of high-energy muons using data for $\theta_{\rm zen}\leq 60^\circ$ from refs. \cite{Soldin:2023lbr,Soldin:2018vak,IceCube:2015wro}. The unflavored and pQCD($c \bar{c}$) curves are nearly equal in the muon energy range shown in the figure. Right: Comparison of \texttt{MCEq} pQCD($c\bar{c}$) prompt ($\nu_\mu + \bar{\nu}_{\mu}$) flux with the IceCube upper limit on the prompt flux from Ref.~\cite{Abbasi:2025rmj}. The angle-averaged (up to zenith angle of $60^\circ$) conventional neutrino flux from \texttt{MCEq} is also shown. The prompt neutrino flux is isotropic in the energy range considered here.
    The cosmic ray all-nucleon flux model \texttt{H3a} and the hadronic interaction model 
    \texttt{Sibyll-2.3c} are used as input of 
    \texttt{MCEq}.
    \label{fig:average-flux-muon-neutrino}
    }
\end{figure}

\subsection{Comparisons with IceCube data and upper limits}

\Cref{fig:average-flux-muon-neutrino} shows the angle-averaged atmospheric conventional and prompt $\mu^++\mu^-$ flux (left panel) and  $\nu_\mu+\bar\nu_\mu$ flux (right panel). For the conventional component, we average over zenith angles $\theta_{\rm zen}=0^\circ-60^\circ$ to avoid nearly horizontal directions where experimental mis-reconstruction of muon events may be significant~\cite{IceCube:2015wro}.
In the left panel, the conventional flux, the pQCD$(c\bar{c})$ contribution, and the unflavored meson contribution are shown separately, along with the total flux (orange curve). The \texttt{Sibyll-2.3c} hadronic interaction model predicts approximately equal contributions to the prompt atmospheric $\mu^++\mu^-$ flux from heavy flavor and light unflavored mesons which makes them difficult to distinguish in the figure. 
Their near equality is accidental for the \texttt{H3a} cosmic ray flux with \texttt{Sibyll-2.3c}. However, the analysis by Illana et al. in ref. \cite{Illana:2010gh} already suggested their contributions could be comparable. Appendix \ref{appendix:z} discusses their nearly identical energy dependence.
The IceCube measurement of the angle-averaged muon flux for $\theta_{\rm zen}<60^\circ$ is shown by the data points with bin-by-bin errorbars~\cite{IceCube:2015wro}. Above $\sim 10^5$ GeV, the \texttt{MCEq} prediction underestimates the observed flux by a factor of $\sim 1.5-2$ (see the red points in~\cref{fig:best_fit_muons_R}), although the statistical uncertainties are larger as the energy increases.
The right panel of \cref{fig:average-flux-muon-neutrino} shows IceCube's 2025 upper bound on the prompt neutrino flux \cite{Abbasi:2025rmj} assuming isotropy~\cite{Enberg:2008te}, a reliable approximation in the energy range shown in the figure. The IceCube upper limit on the prompt $\nu_\mu+\bar\nu_\mu$ flux still allows room for additional prompt contributions. In what follows, we explore possible additional sources of prompt leptons, i.e., both heavy flavored hadrons generated by mechanisms involving intrinsic charm and light unflavored mesons, and study their interplay in the production of prompt atmospheric muons and muon neutrinos.

\section{Intrinsic charm}\label{sec:intrinsiccharm}
Constraints on intrinsic charm contributions to the atmospheric neutrino flux have been discussed in the literature (see, e.g., refs. \cite{Ostapchenko:2022thy,Garzelli:2023jlq,Laha:2016dri,Halzen:2016thi}). 
The potential discrepancy between the atmospheric muon measurements and the predictions from \texttt{MCEq} suggest missing contributions to the atmospheric muon flux. One approach is to add intrinsic charm contributions to the prompt atmospheric muon flux, adjusting the normalization of the intrinsic charm contributions to account for the deficit in the theory predictions for muons, then checking for consistency with the experimental upper bound on the flux of atmospheric muon neutrinos. 
The goal here is to provide a framework to evaluate how enhanced charm production impacts both muon and muon neutrino fluxes, not to provide a rigorous constraint on intrinsic charm
from a global fit of experimental data.

In fact, there are several approaches for incorporating intrinsic charm in particle interactions. The pioneering work of Brodsky, Hoyer, Peterson and Sakai (BHPS) \cite{Brodsky:1980pb,Brodsky:1981se} introduced a 5-quark Fock state for the proton, $|uudc\bar{c}\rangle$, in which the coalescence of three quarks into a $\Lambda_c$ yields an average momentum fraction  of the $\Lambda_c$ with respect to the 5-quark proton Fock state (5C) of $\langle x_\Lambda\rangle_{5C}=4/7$, while for the $D$ meson, $\langle x_D\rangle_{5C}=3/7$. A model with 5-quark and even 7-quark states that includes coalescence and independent fragmentation appears in ref. \cite{Gutierrez:1998bc}. We restrict our attention to the simplified model with 5-quark coalescence, in which the $D$ and $\Lambda_c$ carry a large fraction of the proton momentum, but they are produced and emitted only rarely,
as this component of the proton wave function is suppressed by a very small coefficient  of order $10^{-3}$, as discussed below.
Thus, we estimate that the charm momentum fraction of the proton is less than $\sim 0.1\%$.

The charm hadron momentum distributions can be further interpreted in terms of their constituent charm quark momentum distributions, the $c(x)$ and $\bar{c}(x)$ quark emerging from gluon to charm-anticharm fluctuations in the proton. From the 5-quark coalescence picture of BHPS, the average values of charm quark momentum fraction in the 5-quark state are $\langle x_c\rangle_{5C} =\langle x_{\bar{c}}\rangle_{5C} = 2/7$ in this simplified model. Guided by the functional form of BPHS, by meson-baryon cloud models such as that of Hobbs, Londergan and Melnitchouk (HLM) \cite{Hobbs:2013bia}, by Regge theory \cite{Kaidalov:1985jg, Kaidalov:2003wp} and by other approaches \cite{Ball:2016neh,Ball:2022qks}, extensive literature on intrinsic/fitted charm PDFs has emerged \cite{Jimenez-Delgado:2014zga,Ball:2016neh,  Hou:2017khm,Ball:2022qks,Guzzi:2022rca,Steffens:1999hx,Pumplin:2005yf,Dulat:2013hea,Brodsky:2015fna}, with parameterizations inspired by the aforementioned models at a scale $\mu\sim m_c$ \cite{Ball:2015tna,Ball:2022qks,Guzzi:2022rca}.
Measurements at LHCb of charm in association with $Z-$boson production, although not conclusive, suggest that there could be a valence-like charm component in proton PDFs \cite{LHCb:2021stx}.
PDF analyses, for example in refs. \cite{Jimenez-Delgado:2014zga,Ball:2016neh,Hou:2017khm,Ball:2022qks,Guzzi:2022rca}, find that the fraction of proton momentum carried by intrinsic charm
is $\lesssim 1\%$. However, there is still quite open debate about whether intrinsic/fitted charm in PDFs provides a conceptually sound description of the intrinsic charm phenomenon and is required or not to explain the available data (see, e.g., the discussion in ref. \cite{Guzzi:2022rca}).
The very small intrinsic charm PDF component of the momentum of the proton can be thought of arising from the small normalization factor of the contribution of the 5-quark Fock state to the proton wave function. 

An important feature of intrinsic charm in the context of atmospheric lepton fluxes is its contribution to very forward charm hadron production. The  charm hadron momentum distributions from, e.g., refs. \cite{Brodsky:1980pb,Brodsky:1981se,Hobbs:2013bia}, account for the coalescence of the $c$ and $\bar{c}$ with valence quarks. In this 5-quark coalescence picture, a relatively large fraction of the incident baryon momentum is transferred to charm hadrons from valence quarks, in particular 
to $\bar D^0$ and $ \Lambda_c $ in proton-air ($pA$) scattering and to $ D^-$ and $ \Lambda_c $ in neutron-air ($nA$) scattering:
\begin{eqnarray}
    p+A &\to \bar D^0 + \Lambda_c + X\\
     n+A &\to  D^- + \Lambda_c + X\,.
\end{eqnarray}
From the cascade equation, eq. (\ref{eq:hadron-cascade}), one can see that
the steeply falling cosmic ray flux allows for the small fraction of the total charm production cross section which comes from intrinsic charm to have a relatively large contribution to the lepton flux.

Following ref. \cite{Laha:2016dri,Ostapchenko:2022thy,Garzelli:2023jlq,Gutierrez:1998bc,Vogt:2023plx}, we write the differential cross section for proton-air production of charm hadron $h_c$ via non-perturbative intrinsic charm as
\begin{equation}
\label{eq:dsdx-ic}
  \frac{d\sigma_{p-{\rm air}}^{h_c{\rm (intr)}} (E_p,x_{h_c})}{dx_{h_c}} = w_{\rm intr}^c \sigma _{p-{\rm air}}(E_p) f_{h_c}^{\rm (intr)} (x_{h_c})\,,
\end{equation}
where $x_{h_c}=E_{h_c}/E_p$, and similarly for neutron constituents of cosmic rays. The parameter $w_{\rm intr}^c$, whose value can be extracted by fits to experimental data, determines the overall level of the intrinsic charm contribution relative to the inelastic proton-air cross section. The quantity $f_{h_c}^{\rm (intr)}$ is the fragmentation function for the cosmic ray nucleon to produce a charmed baryon or charmed meson $h_c$.  We assume that $w_{\rm intr}^c$ is independent of energy and identical for protons and neutrons. At low energies, the prompt lepton flux is much smaller than the conventional lepton flux. At higher energies, one expects the intrinsic charm fraction of the $p-$air cross section to be approximately energy independent.

\begin{figure}
    \centering
    \includegraphics[width=0.45\linewidth]{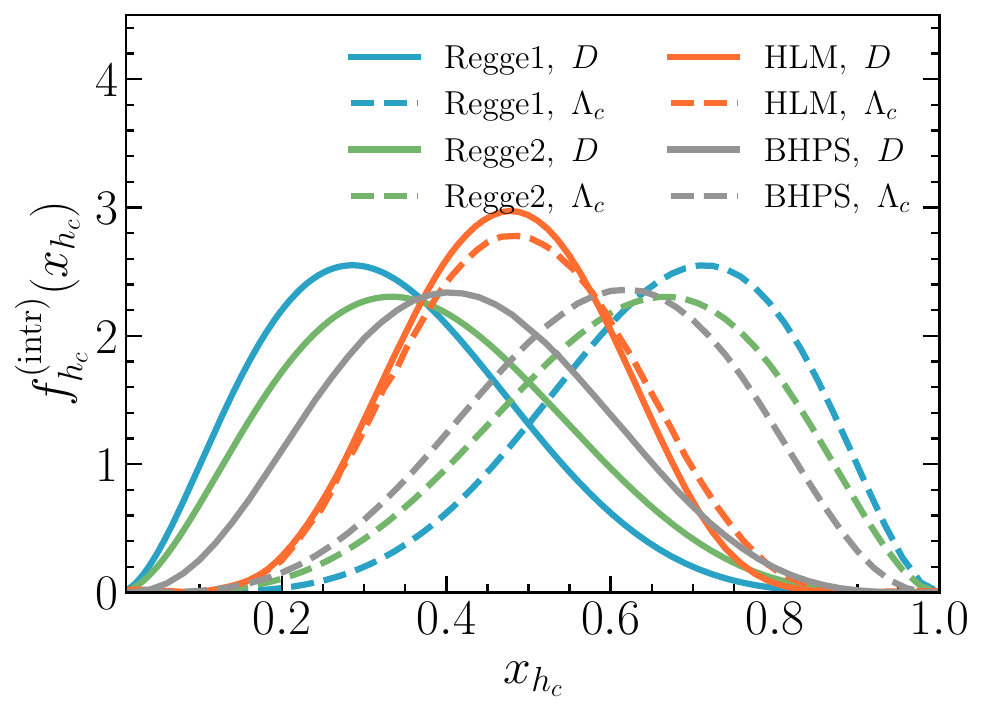}
    \caption{$f_{h_c}^{(\rm intr)}$ of $D$ mesons and $\Lambda_c$ as a function of $x_{h_c}$ (see eq.~\ref{eq:dsdx-ic}) 
    using  the Regge ansatz \cite{Kaidalov:1985jg,Kaidalov:2003wp} with either $a_N=-0.5$ (Regge1) or $a_N=0$ (Regge2), using the Hobbs, Londergan and Melnitchouk (HLM) approach \cite{Hobbs:2013bia} (see text for details) and from Brodsky et al. (BHPS) \cite{Brodsky:1980pb}.}
    \label{fig:fragmentation}
\end{figure}

As noted above, there are several approaches to model the intrinsic charm fragmentation to charm hadrons, e.g., BHPS \cite{Brodsky:1980pb,Brodsky:1981se}, Regge ansatz \cite{Kaidalov:1985jg,Kaidalov:2003wp}
and meson-baryon models, in particular HLM~\cite{Hobbs:2013bia}.
We use the Regge ansatz as our starting point, and also show results for HLM. 
The 5-quark coalescence fragmentation functions of BHPS are similar to the Regge functions \cite{Ostapchenko:2025jlz}, and they sit between the Regge and HLM functions, 
as shown in \cref{fig:fragmentation}.

The Regge fragmentation function (normalized to unity) is \cite{Kaidalov:1985jg,Kaidalov:2003wp} (see also ref. \cite{Ostapchenko:2022thy,Ostapchenko:2025jlz})
\begin{eqnarray}
f_{h_c}^{(\rm intr )} (x)&=& N_{\rm intr}\, x^{-a_\psi}(1-x)^{-a_
\psi + 2(1-a_N)}
    \label{eq:Regge}\\
  N_{\rm intr}   &=& \frac{\Gamma (-2a_\psi+4-2a_N)}{\Gamma(-a_\psi+1)\Gamma(-a_\psi+3-2a_N)}\,.
  \nonumber
\end{eqnarray}
The choices for constants used in eq.~(\ref{eq:Regge}) are $a_\psi = -2.0$, $a_N=-0.5$ (Regge1) or 
$a_N=0$ (Regge2). We use $a_N=-0.5$ (Regge1) as the default for the results shown here, unless otherwise specified.
Furthermore, we assume for the Regge ansatz that
\begin{eqnarray}
    f_{D^-}^{\rm (intr)}(x_D) &=& f_{\bar D^0}^{\rm (intr)}(x_D)\\
    f_{\Lambda_c}^{\rm (intr)}(x_\Lambda)
    &=& f_{\bar D^0}^{\rm (intr)}(1-x_\Lambda)\,.
\end{eqnarray}

Hobbs, Londergan and Melnitchouk \cite{Hobbs:2013bia} determined splitting functions in a meson-baryon model, for which the $\bar D ^{*0}\Lambda_c$ dominates for $pA$ interactions. In  \cref{fig:fragmentation} we show the $\Lambda_c$ distribution using $x_{\Lambda_c}=1-x_{\bar D ^{*0}}$. We account for a diminished $\bar D^0$ momentum fraction from the decay $\bar D ^{*0}\to D^0 \pi^0$ and $\bar D ^{*0}\to D^0 \gamma$ following refs. \cite{Cacciari:2003zu,Cacciari:2005uk}. 
For $nA$ interactions, $D ^{*-}\Lambda_c$ fluctuations dominate. The decays of $\bar D ^{*-}\to D^- \pi^0$, $\bar D ^{*-}\to D^- \gamma$ and $\bar D ^{*-}\to \bar D^0 \pi^-$ have very similar kinematics and yield  $D$ meson $x$-distributions that follows the HLM solid line in fig. \ref{fig:fragmentation}. For reference, fig. \ref{fig:fragmentation} also shows the BHPS distributions for $D$ mesons and $\Lambda_c$. 
For all three approaches, the charmed baryon carries a larger energy fraction than the charmed meson since two valence quarks ($ud$) contribute, although the energy balance is much closer in the HLM approach than with the Regge ansatz. We use the Regge1, Regge2 and HLM distributions in our analysis below to cover the range of models.

In our analysis below, we find numerical values of $w^c_{\rm intr}$ that depend on our choice of \texttt{Sibyll-2.3c} to represent the perturbative QCD contribution to charm production and on eq. (\ref{eq:dsdx-ic}) to represent the intrinsic charm contribution. These numerical values for $w^c_{\rm intr}$ are not strongly dependent on which  $f_{h_c}^{(\rm intr)}(x_{h_c})$ we use, as we show in the next section.

\section{Results}\label{sec:results}
\subsection{Angle-averaged atmospheric lepton fluxes}
\begin{figure}[]
    \centering
    \includegraphics[width=0.495\linewidth]{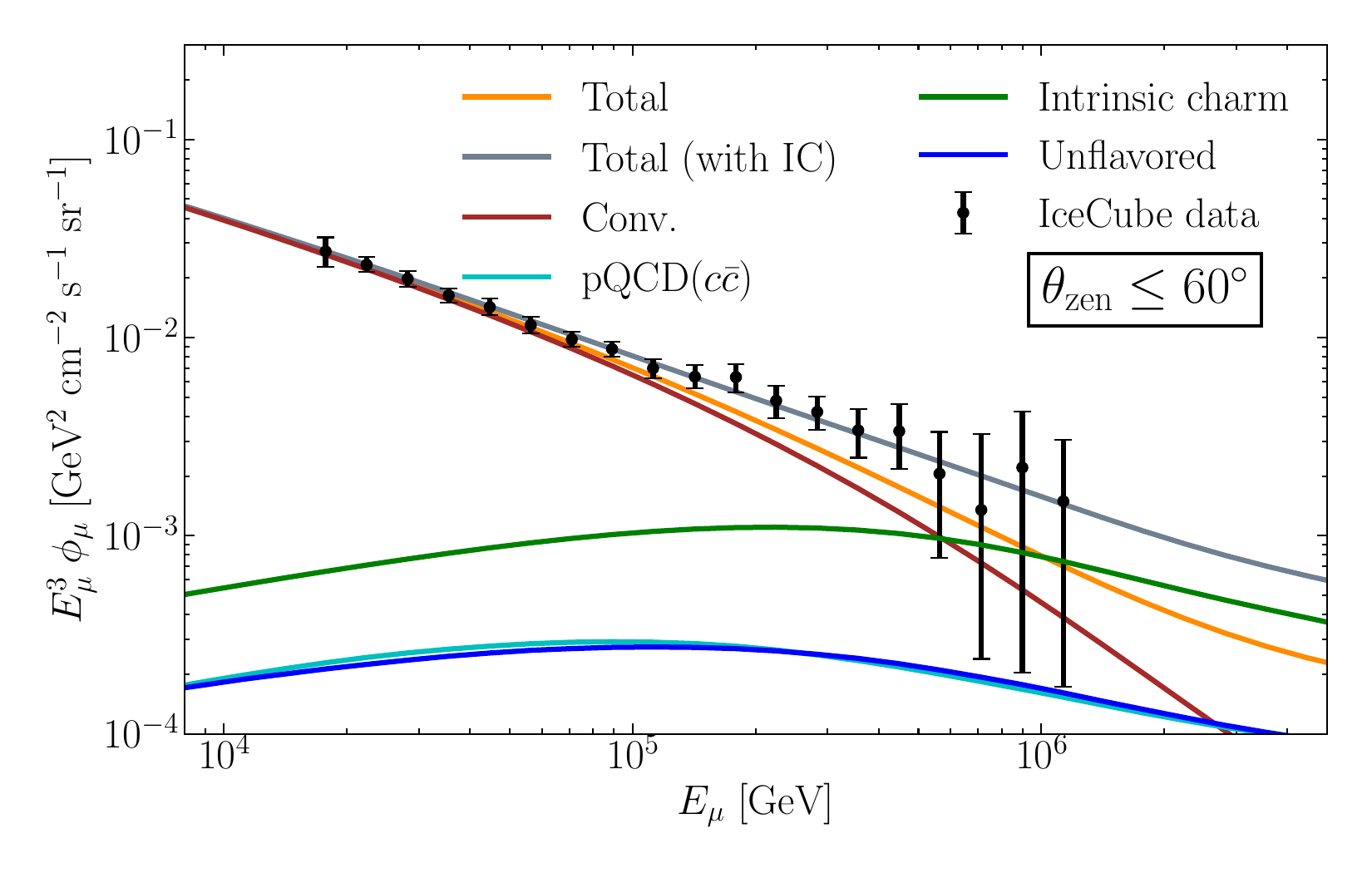}
        \includegraphics[width= 0.495\linewidth]{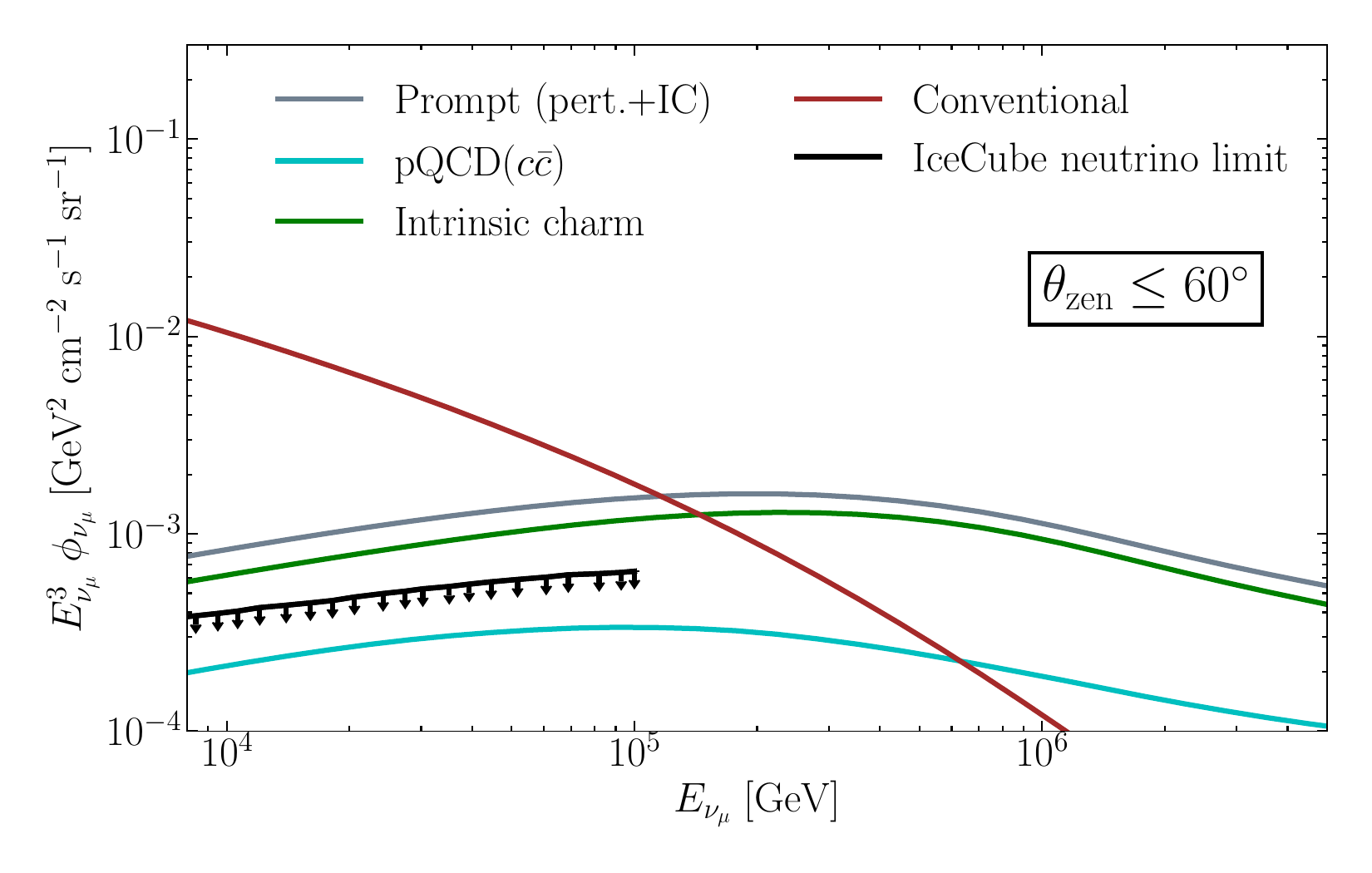}
    \caption{As in \cref{fig:average-flux-muon-neutrino}, the angle-averaged atmospheric fluxes with $\theta_{\rm zen}\leq 60^\circ$. Left: Including intrinsic charm (IC) contribution with the weight $w_c^{\rm intr}=3.93\times10^{-3}$, the best-fit for the IceCube muon data \cite{IceCube:2015wro}. The gray curve shows the sum of conventional and prompt $\mu^++\mu^-$ fluxes including intrinsic charm. Right: The prompt atmospheric muon neutrino flux including intrinsic charm contribution with $w_c^{\rm intr}=3.93\times10^{-3}$ and the upper limit on the prompt flux from IceCube \cite{Abbasi:2025rmj}. 
    The sum of perturbative and intrinsic charm contributions to the prompt $\nu_\mu+\bar\nu_\mu$ flux is shown with the gray curve.}
    \label{fig:best_fit_muons}
\end{figure}

\begin{figure}[]
    \centering     \includegraphics[width=0.495\linewidth]{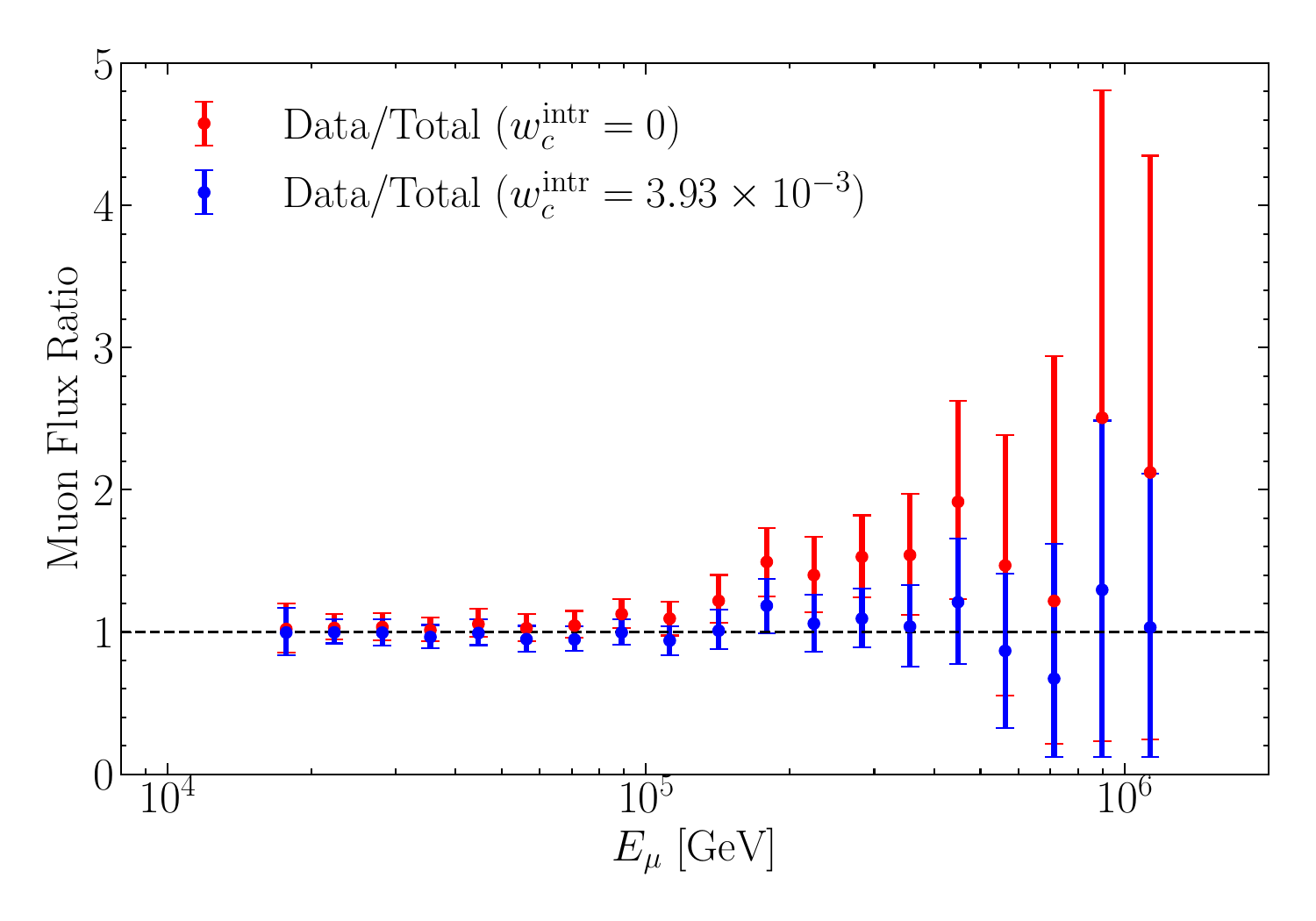} 
    \caption{Ratio of IceCube muon data with experimental error bars \cite{IceCube:2015wro} to the muon flux from \texttt{MCEq} both with (blue) and without (red) including intrinsic charm contribution with $w_{c}^{\rm intr} = 3.93\times10^{-3}$.}
    \label{fig:best_fit_muons_R}
\end{figure}

As we saw in \cref{fig:average-flux-muon-neutrino}, the \texttt{MCEq} prediction for the atmospheric muon flux lies below the IceCube data. In this section, we begin by introducing intrinsic charm (IC) contributions to both muon and muon neutrino fluxes.
The left panel of \cref{fig:best_fit_muons} shows the best-fit intrinsic charm contribution to the angle-averaged ($\theta_{\rm zen}\leq 60^\circ$) atmospheric muon flux, as well as the conventional, prompt light unflavored, prompt pQCD($c\bar{c})$ and the total muon flux contributions calculated using \texttt{MCEq}, along with the IceCube muon data \cite{IceCube:2015wro}. The best-fit value of the $w_{c}^{\rm intr}$ parameter amounts to $w_c^{\rm intr}=3.93\times10^{-3}$ for the default Regge1 intrinsic charm fragmentation model. This value for $w_{c}^{\rm intr}$ implies that $\sim 0.4\%$ of the nucleon-air cross section is due to intrinsic charm. Despite the small value of $w_{c}^{\rm intr}$, the energy distribution of the charm hadrons following \cref{eq:dsdx-ic} is large because the cosmic ray flux falls rapidly with increasing cosmic ray energy.
\Cref{fig:best_fit_muons_R} shows the ratio of the IceCube data (including experimental error bars) to the \texttt{MCEq} prediction both without and with the addition of an intrinsic charm contribution corresponding to  $w_c^{\rm intr}=3.93\times10^{-3}$.

The $1\sigma$ interval for $w_{c}^{\rm intr}$ based on a $\chi^2$ fit is $(2.95-4.91)\times 10^{-3}$. Table \ref{tab:w_intr_table} shows the best fit values and $1\sigma$ intervals for the Regge1, Regge2 and HLM models. All considered models lead to similar ranges of a few $\times 10^{-3}$ for $w_{c}^{\rm intr}$.

\begin{figure}[]
    \centering
    \includegraphics[width=0.495\linewidth]{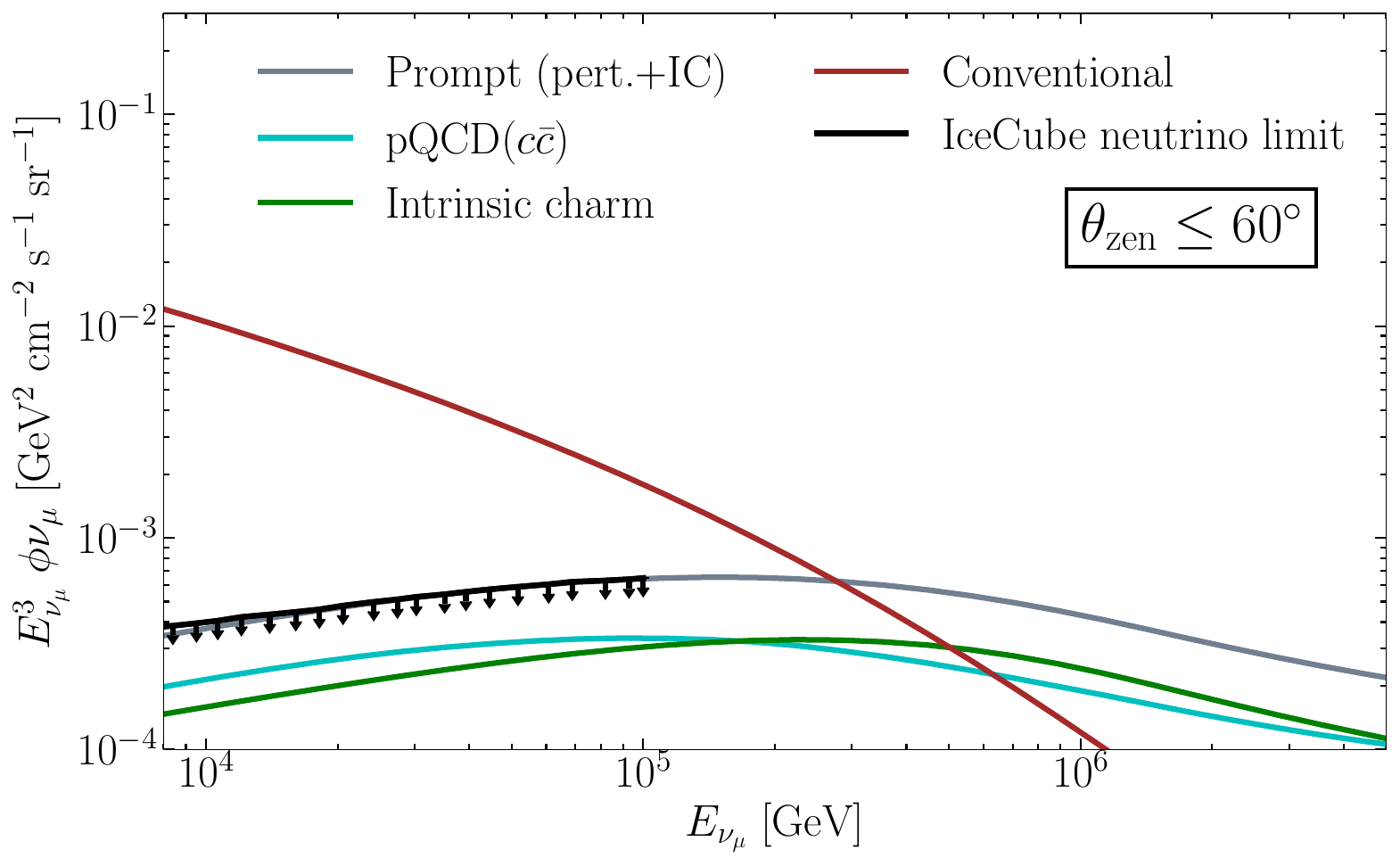}
      \includegraphics[width=0.495\linewidth]{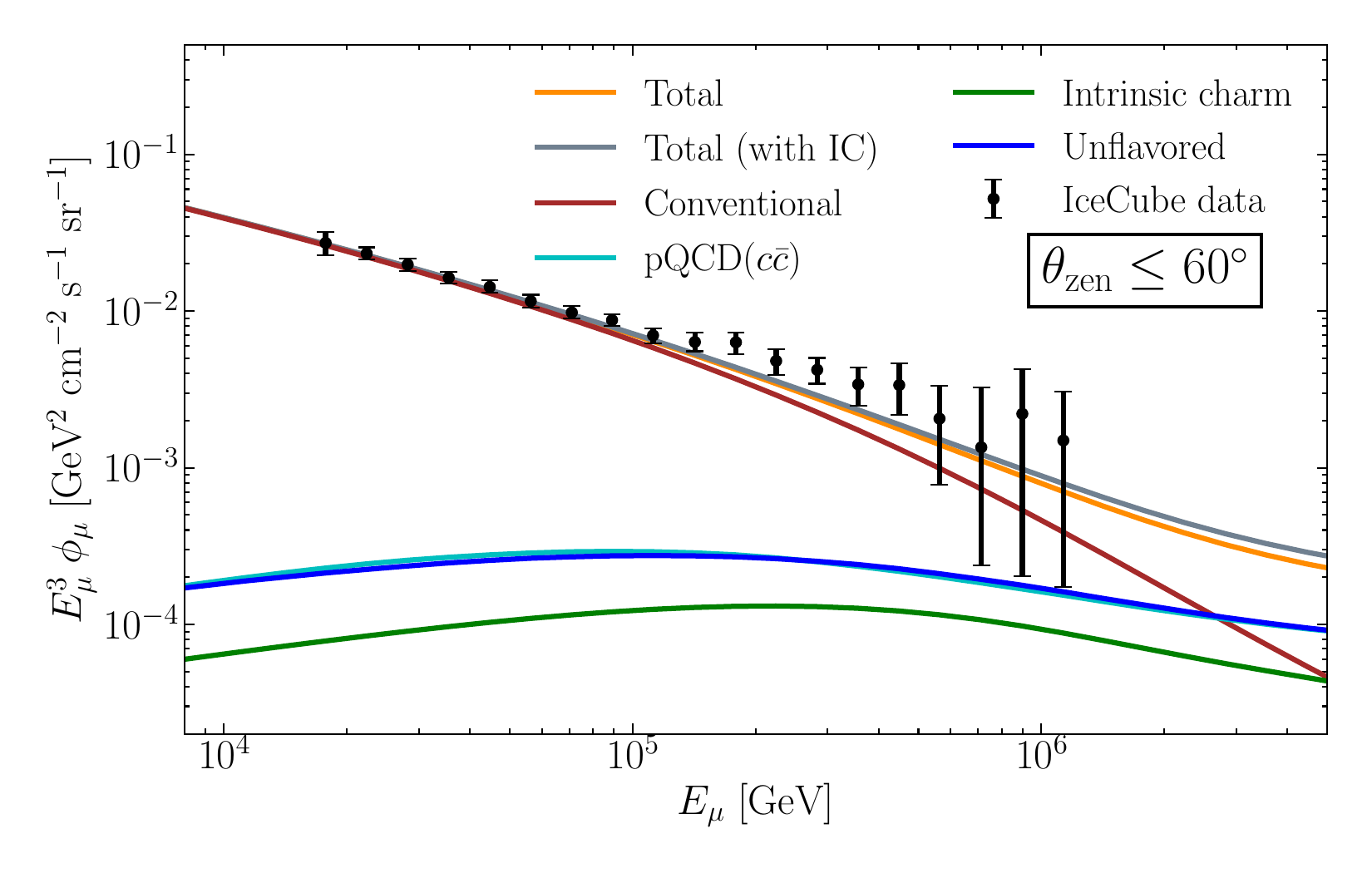}
    \caption{As in \cref{fig:average-flux-muon-neutrino}, angle-averaged with $\theta_{\rm zen}\leq 60^\circ$. Left: Comparison of prompt neutrino spectrum ($\nu_\mu + \bar{\nu}_{\mu}$) including intrinsic charm contribution with $w_c^{\rm intr}=1.01\times10^{-3}$ with the upper limit from IceCube \cite{Abbasi:2025rmj}. Right: Comparison of angle-averaged muon flux ($\mu^+ + \mu^-$) with IceCube data \cite{IceCube:2015wro} after including intrinsic charm contribution with $w_c^{\rm intr}=1.01\times10^{-3}$.
    \label{fig:Neutrino_flux_withIC}}
\end{figure}

On the other hand, the right panel of \cref{fig:best_fit_muons} shows that the prompt perturbative plus intrinsic charm contributions to the prompt atmospheric $\nu_\mu+\bar\nu_\mu$ flux significantly overshoot the IceCube upper bound \cite{Abbasi:2025rmj}. To satisfy the prompt atmospheric $\nu_\mu+\bar\nu_\mu$ flux constraint from ref. \cite{Abbasi:2025rmj}, $w_{ c}^{\rm intr}$ must be reduced, which also implies that intrinsic charm alone cannot account for the apparent deficit in the flux of atmospheric muons at high energies. The IceCube upper bound \cite{Abbasi:2025rmj}  shown in the right panel of fig. \ref{fig:best_fit_muons} relies on the shape of the prompt flux evaluated using charm production from \texttt{Sibyll-2.3c}. The best-fit intrinsic charm contribution that saturates the IceCube upper bound is {$w_c^{\rm intr}=1.01\times10^{-3}$}.  
\Cref{fig:Neutrino_flux_withIC}  shows the intrinsic charm contributions to the atmospheric muon neutrino (left) and muon (right) fluxes for $w_c^{\rm intr}=1.01\times10^{-3}$. The
Regge2 and HLM fragmentation models have similar best-fit values to the Regge1 one. The value of $w_c^{\rm intr}$ is $\sim 1/4$ of the $\mu^++\mu^-$ best-fit value, so the atmospheric muon flux data remain above the sum of the 
flux contributions, now with a reduced intrinsic charm component, as shown in the right panel of \cref{fig:Neutrino_flux_withIC}.

\begin{figure}[]
    \centering
    \includegraphics[width=0.495\linewidth]{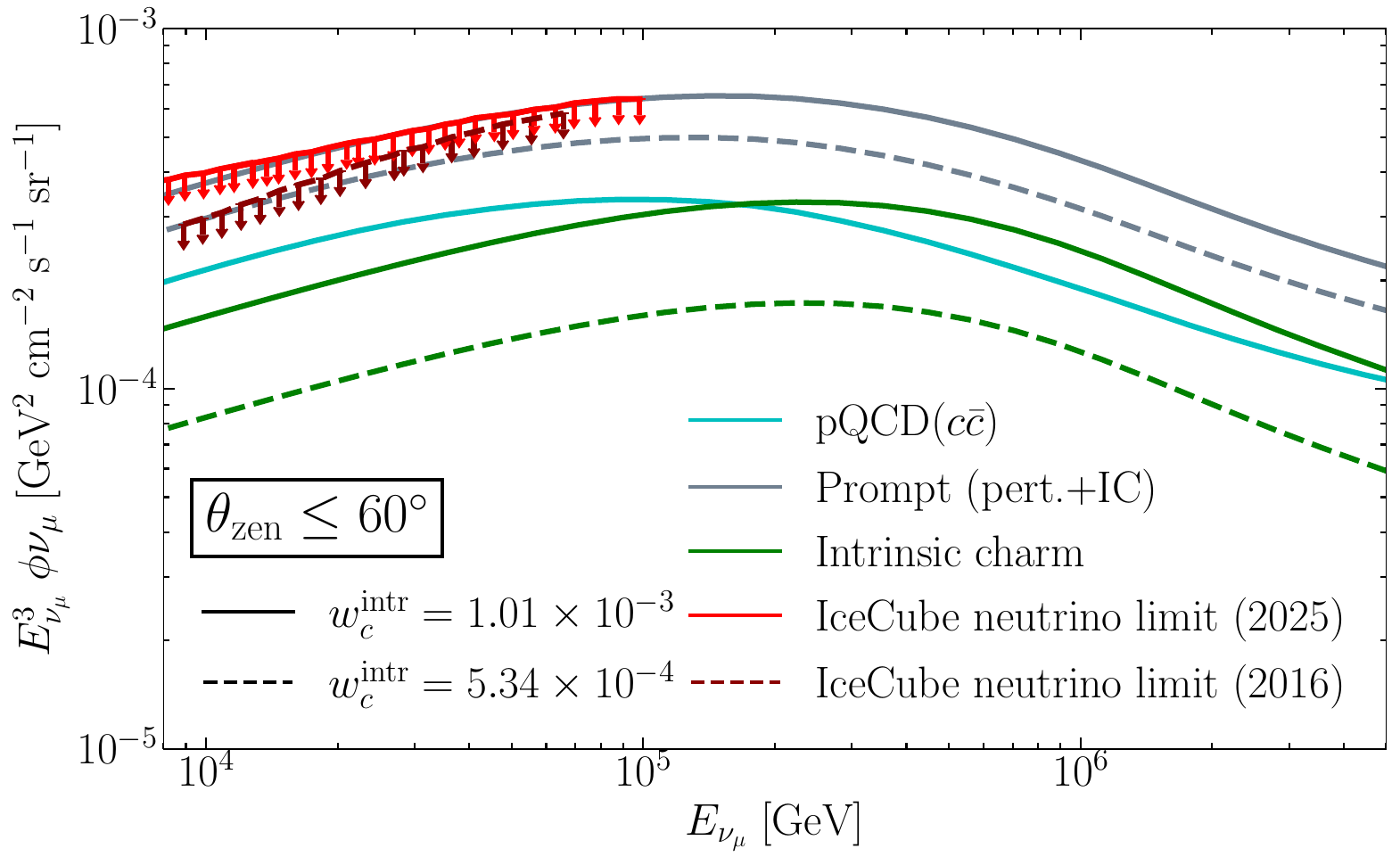}
\caption{Comparisons of prompt neutrino spectrum ($\nu_\mu + \bar{\nu}_{\mu}$) including intrinsic charm contribution with $w_c^{\rm intr}=1.01\times10^{-3}$ to the 2025 upper limit from IceCube \cite{Abbasi:2025rmj}, and with $w_c^{\rm intr}=5.34\times10^{-4}$ to the 2016 upper bound from IceCube \cite{IceCube:2016umi} that relies on the ERS prompt flux spectrum \cite{Enberg:2008te}.
}
\label{fig:2limits}
\end{figure}

As noted above, the IceCube limit on the prompt atmospheric neutrino flux in ref. \cite{Abbasi:2025rmj} is a scaling of the prompt flux evaluated with \texttt{Sibyll-2.3c}. 
On the other hand, the 2016 IceCube upper bound \cite{IceCube:2016umi} relied on the shape of the prompt flux from Enberg et al. (ERS)  \cite{Enberg:2008te} as a function of neutrino energy. The slightly different shapes of the two prompt neutrino fluxes can be seen with the aforementioned upper bounds in \cref{fig:2limits}, where the solid red bound is from ref. \cite{Abbasi:2025rmj} while the dashed red bound is from ref. \cite{IceCube:2016umi}. For the 2016 upper bound, we normalize $w_c^{\rm intr}$ to the value of the IceCube upper bound at $E_\nu=10^4$ GeV, where this bound has a minimum. The scaled intrinsic charm and  pQCD($c\bar{c}$) plus intrinsic charm contributions to the prompt neutrino flux are shown in \cref{fig:2limits} with dashed lines corresponding to $w_c^{\rm intr}=5.34\times 10^{-4}$ resulting from saturating this IceCube bound \cite{IceCube:2016umi}. In the case of the 2025 bound, we normalize $w_c^{\rm intr}$  in such a way that the total prompt neutrino contribution (pQCD + intrinsic charm) saturates the bound at $E_\nu = 10^5$ GeV, obtaining $w_c^{\rm intr} = 1.01\times 10^{-3}$ and the intrinsic charm component shown by solid line in the same figure.

\begin{figure}[]
    \centering
\includegraphics[width=0.495\linewidth]{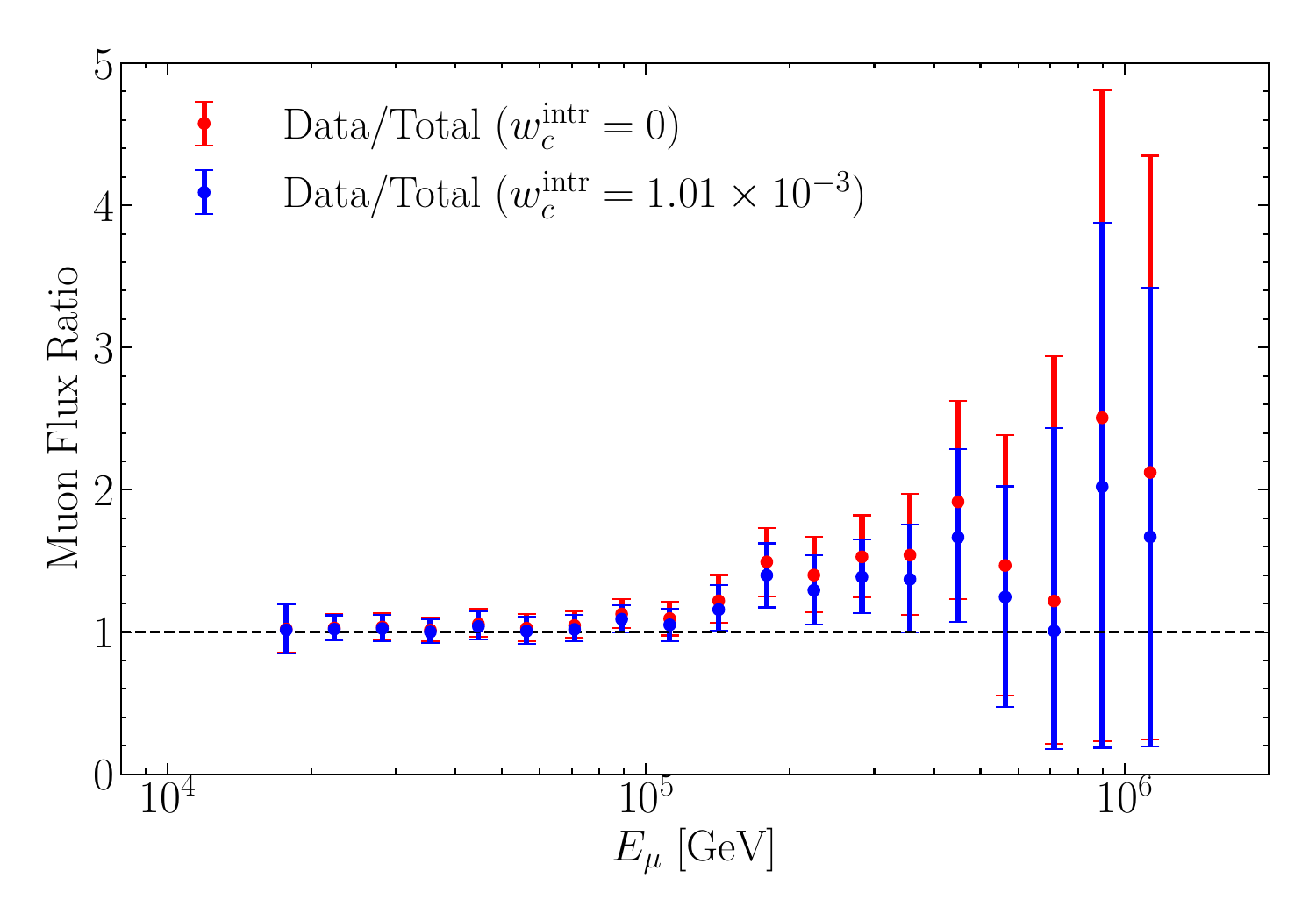}   \caption{Ratio of IceCube muon data with experimental error bars \cite{IceCube:2015wro} to the muon flux from \texttt{MCEq} both with (blue) and without (red) including intrinsic charm contribution with $w_c^{\rm intr} = 1.01\times10^{-3}$. }
    \label{fig:Neutrino_flux_withIC_R}
\end{figure}

\Cref{fig:Neutrino_flux_withIC_R} shows the ratio of IceCube muon data with error bars to the muon flux with and without intrinsic charm, as in \cref{fig:best_fit_muons_R}, but here with $w_c^{\rm intr}=1.01\times10^{-3}$. Such a small value of $w_c^{\rm intr}$ has little impact on the muon flux ratio.

\begin{table}[]
\centering
\vskip 0.1in
\begin{tabular}{@{} l c c @{}}
 \toprule
 \textbf{Model} & \textbf{Best fit} $\mathbf{{\it w_c}^{\rm intr}}(\times 10^{-3})$ & $\mathbf{1\,\sigma}$ \textbf{interval} $(\times 10^{-3})$ \\ 
 \midrule
 \midrule
Regge1$(a_\psi = -2, a_N = -0.5)$ & 3.93 & $2.95 - 4.91$ \\
Regge2$(a_\psi = -2, a_N = 0)$ & 3.55 & $2.66 - 4.43$ \\ 
HLM & 3.03 & $2.28 - 3.79$\\
 \bottomrule
\end{tabular}
\caption{Values of $w_{ c}^{\rm intr}$ for intrinsic-charm model-dependent fit to the angle-averaged muon energy spectrum ($\theta_{\rm zen}\leq$ $60^\circ$). Best fit values of $w_{c}^{\rm intr}$ along with the $1\,\sigma$ interval are shown for the two Regge ansatz models \cite{Kaidalov:1985jg,Kaidalov:2003wp} and for the Hobbs, Londergan and Melnitchouk (HLM) model \cite{Hobbs:2013bia}.}
\label{tab:w_intr_table}
\end{table}

\begin{figure}[]
    \centering    \includegraphics[width=0.495\linewidth]{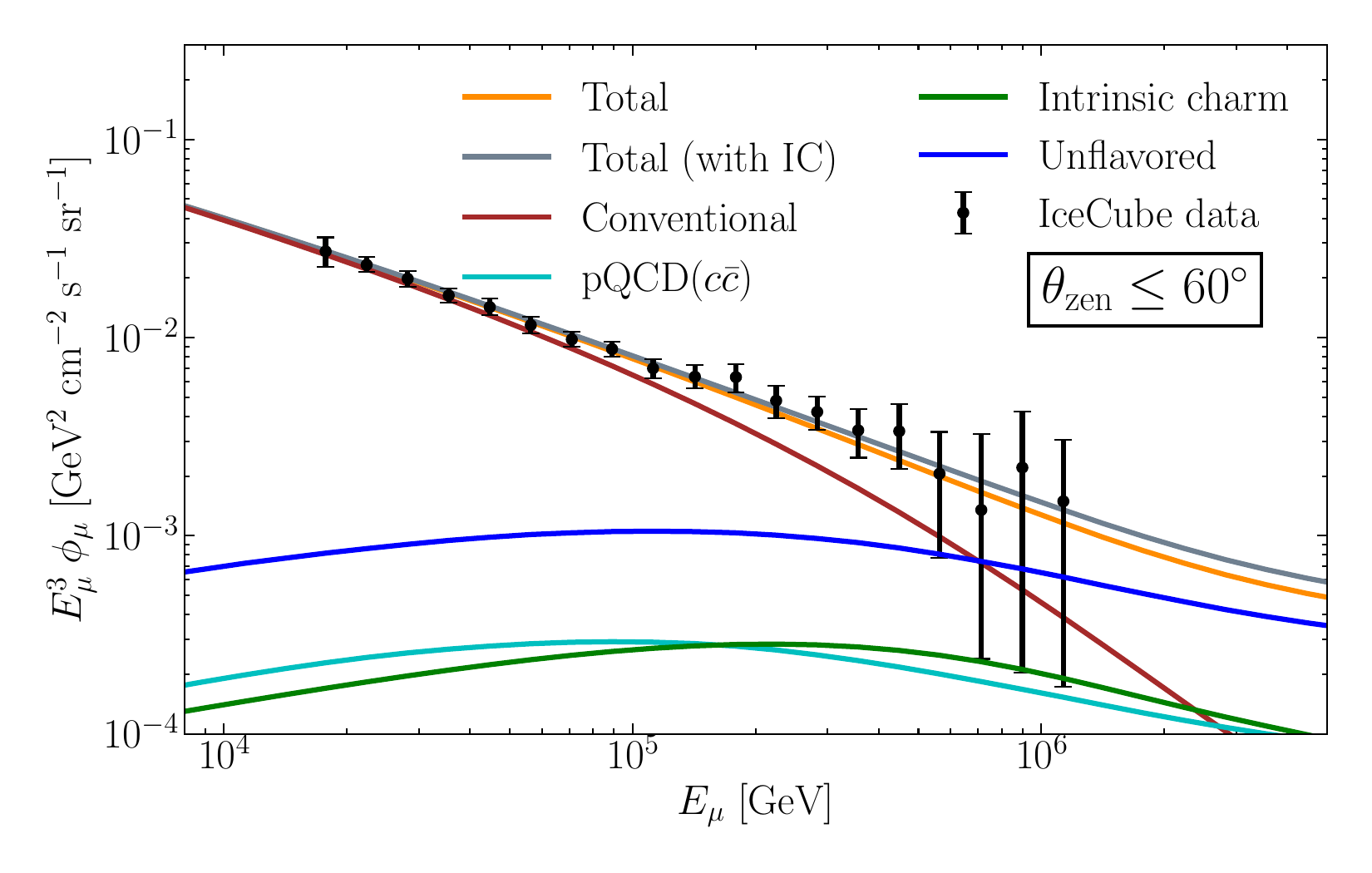}
    \includegraphics[width=0.495\linewidth]{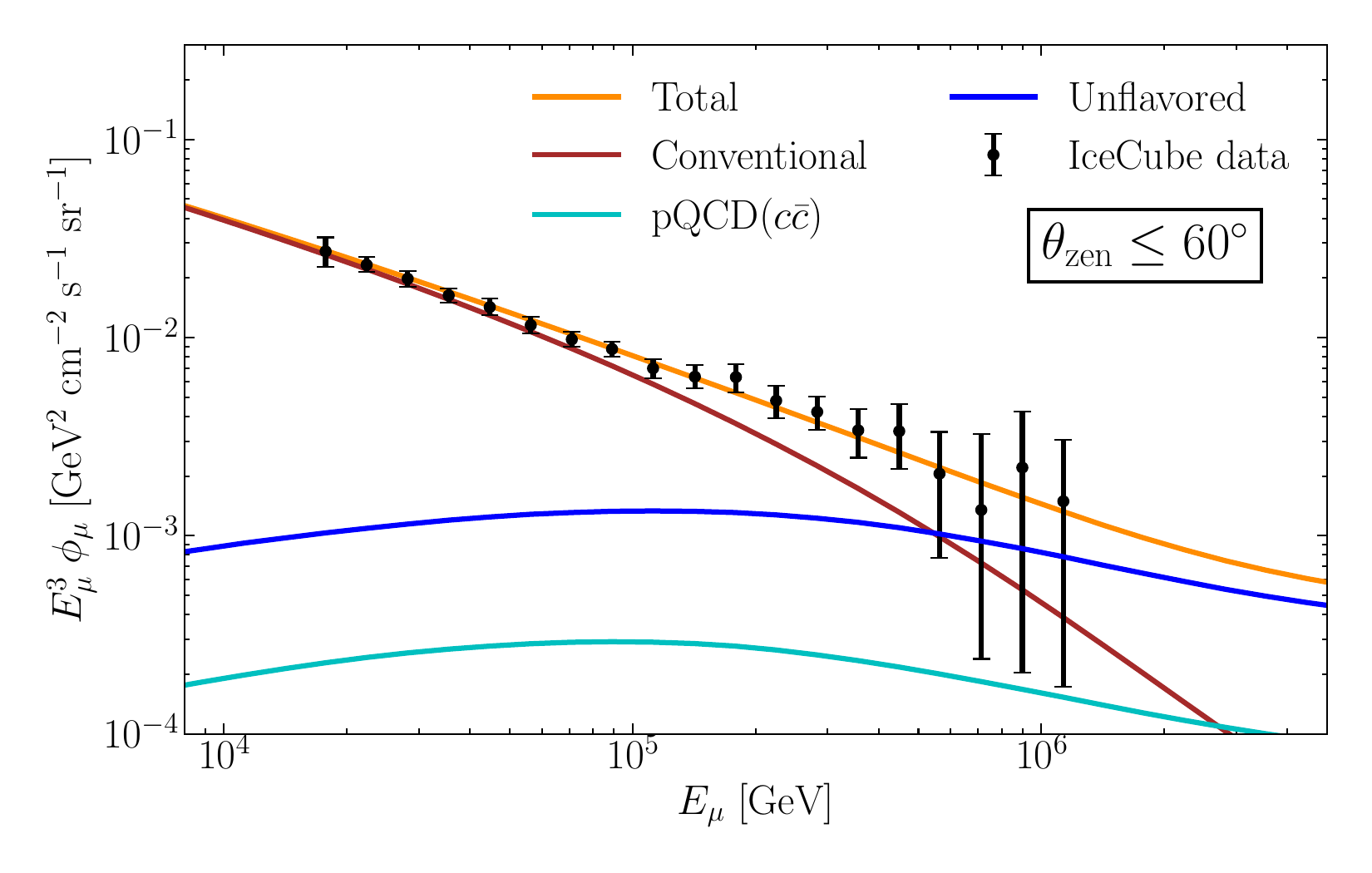}
    \caption{Left: The angle-averaged ($\theta_{\rm zen}\leq 60^\circ$) muon flux ($\mu^++\mu^-$) including an intrinsic charm contribution with $w_c^{\rm intr} = 1.01\times10^{-3}$ and scaling the unflavored flux by the best-fit parameter of $\alpha_{\rm unfl} = 3.84$.
    Right: Without any intrinsic charm contribution, the angle-averaged ($\theta_{\rm zen}\leq 60^\circ$) muon flux with scaling the unflavored flux by the best-fit parameter of $\alpha_{\rm unfl}  = 4.86$.}
    \label{fig:Muon_flux_unfl_scaled}
\end{figure}

As a next step, we consider the possibility of a normalization uncertainty in the unflavored light meson contribution to the prompt muon flux. This could arise from a higher multiplicity and/or a harder energy spectrum of light meson production in cosmic ray-air collisions relative to the \texttt{Sibyll-2.3c} model.
We define a scale factor $\alpha_{\rm unfl}$ that multiplies the \texttt{Sibyll-2.3c} angle-averaged unflavored light meson contribution to the atmospheric muon flux:
\begin{equation}
    \phi_\mu ^{\rm unfl}= \alpha_{\rm unfl}\, \phi_\mu^{\rm MCEq,unfl}\,.
\end{equation}
With $w_c^{\rm intr} =1.01\times10^{-3}$, the best-fit unflavored flux scale factor is $\alpha_{\rm unfl} =3.84$, with a $\chi^2$-fit 1$\sigma$ range of $2.85-4.83$ for the Regge1 model. The Regge2 and HLM values for $\alpha_{\rm unfl}$ are nearly the same, as shown in Table \ref{unfl_table}.
The angle-averaged atmospheric muon flux with the central value of $\alpha_{\rm unfl}$ and   $w_c^{\rm intr}=1.01\times10^{-3
}$ is shown in the left panel of  \cref{fig:Muon_flux_unfl_scaled}. When $w_c^{\rm intr}=0$, the best-fit value of $\alpha_{\rm unfl} =4.86$ yields the angle-averaged muon flux shown in the right panel of \cref{fig:Muon_flux_unfl_scaled}. 

We note that the shapes and normalizations of the prompt contributions to the atmospheric muon flux from light unflavored hadrons and from charm hadrons from \texttt{Sibyll-2.3c} in \MCEq \ are nearly identical. The flux shapes are determined by the hadronic interactions and the prompt decays. The nearly equal normalization is accidental: many more light unflavored mesons are produced than charm hadrons, however, the light-meson branching fractions to $\mu^++\mu^-$ are very small.

\begin{table}[]
\centering

\vskip 0.1in
\begin{tabular}{ @{} l c l c @{}}
 \toprule
 \textbf{Model} & $\mathbf{{\it w_c}^{\rm intr}\times 10^{-3}}$ & $\mathbf{\alpha_{\rm unfl}}$ & $\mathbf{1\sigma}$ \textbf{interval} \\
  \midrule
  \midrule
 Regge1$(a_\psi = -2, a_N = -0.5)$ & 1.01 & 3.84 & 2.85 - 4.83 \\
  Regge2$(a_\psi = -2, a_N = 0)$& 0.92 & 3.83 & 2.84 - 4.82 \\ 
 HLM & 0.78 & 3.84 & 2.85 - 4.83\\
 \bottomrule
\end{tabular}
\caption{Best-fit values and 1$\sigma$ intervals for the scaling factor of the unflavored flux
of $\mu^+ + \mu^-$, $\alpha_{\rm unfl}$, assuming a value of  $w_c^{\rm intr}$ needed to meet the 2025 IceCube prompt neutrino flux upper bound \cite{Abbasi:2025rmj}. For reference, our prescription for satisfying the 2016 IceCube neutrino upper bound on the $\nu_\mu+\bar{\nu}_\mu$ flux \cite{IceCube:2016umi} yields $w_c^{\rm intr}=5.34\times 10^{-4}$ and $\alpha_{\rm unfl} = 4.32$ for the Regge1 model. For no intrinsic charm, $\alpha_{\rm unfl} = 4.86$ with a $1\sigma$ interval of $3.87 - 5.85$.}
\label{unfl_table}
\end{table}

\subsection{Flux ratios}

In addition to the angle-averaged atmospheric muon and muon neutrino fluxes, we also show anti\-particle-particle ratios, the muon to muon neutrino flux, and a ratio to illustrate the angular dependence of the atmospheric muon flux. 

\begin{figure}[h]
    \centering
    \includegraphics[width=0.495\linewidth]{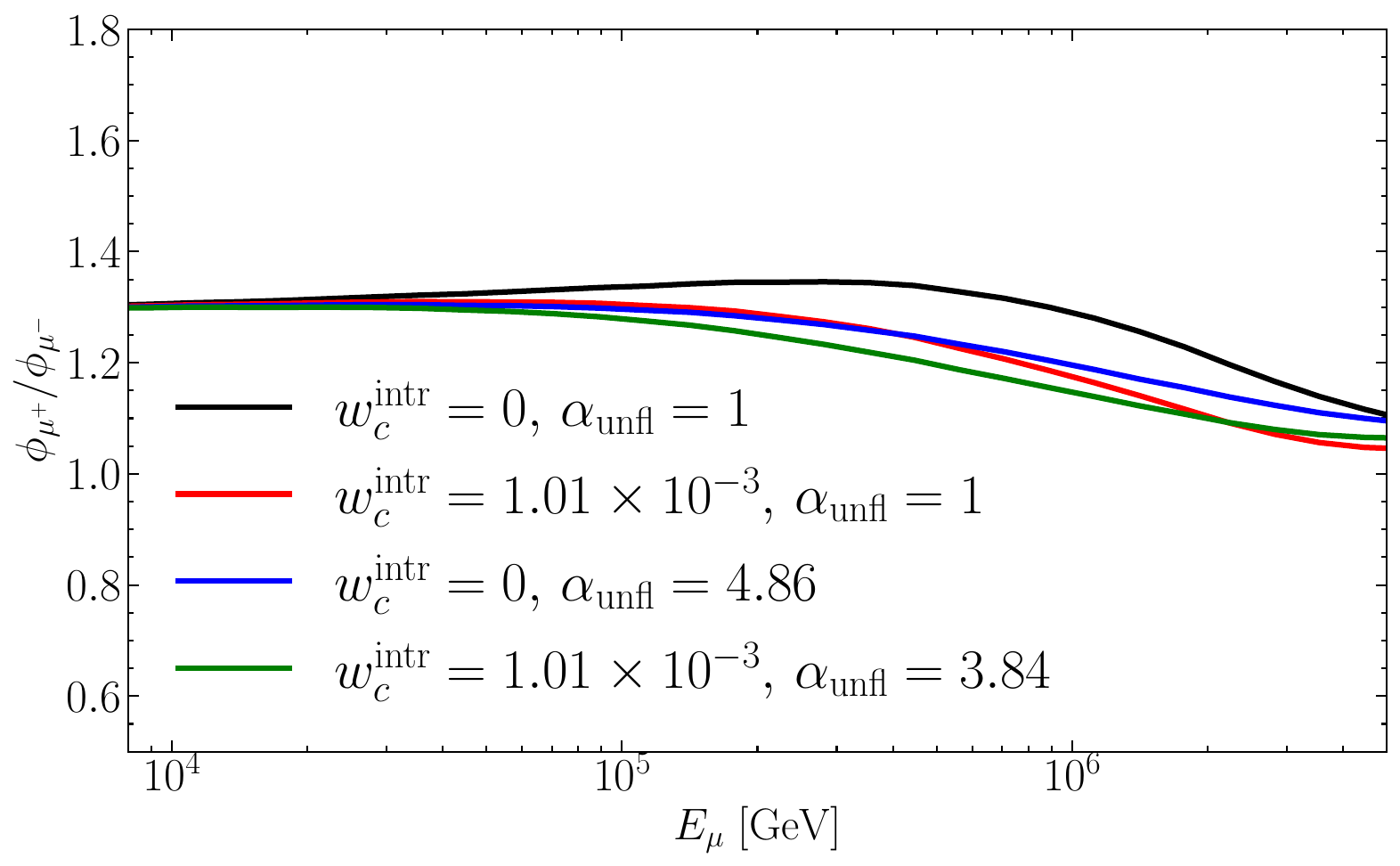}
    \includegraphics[width=0.495\linewidth]{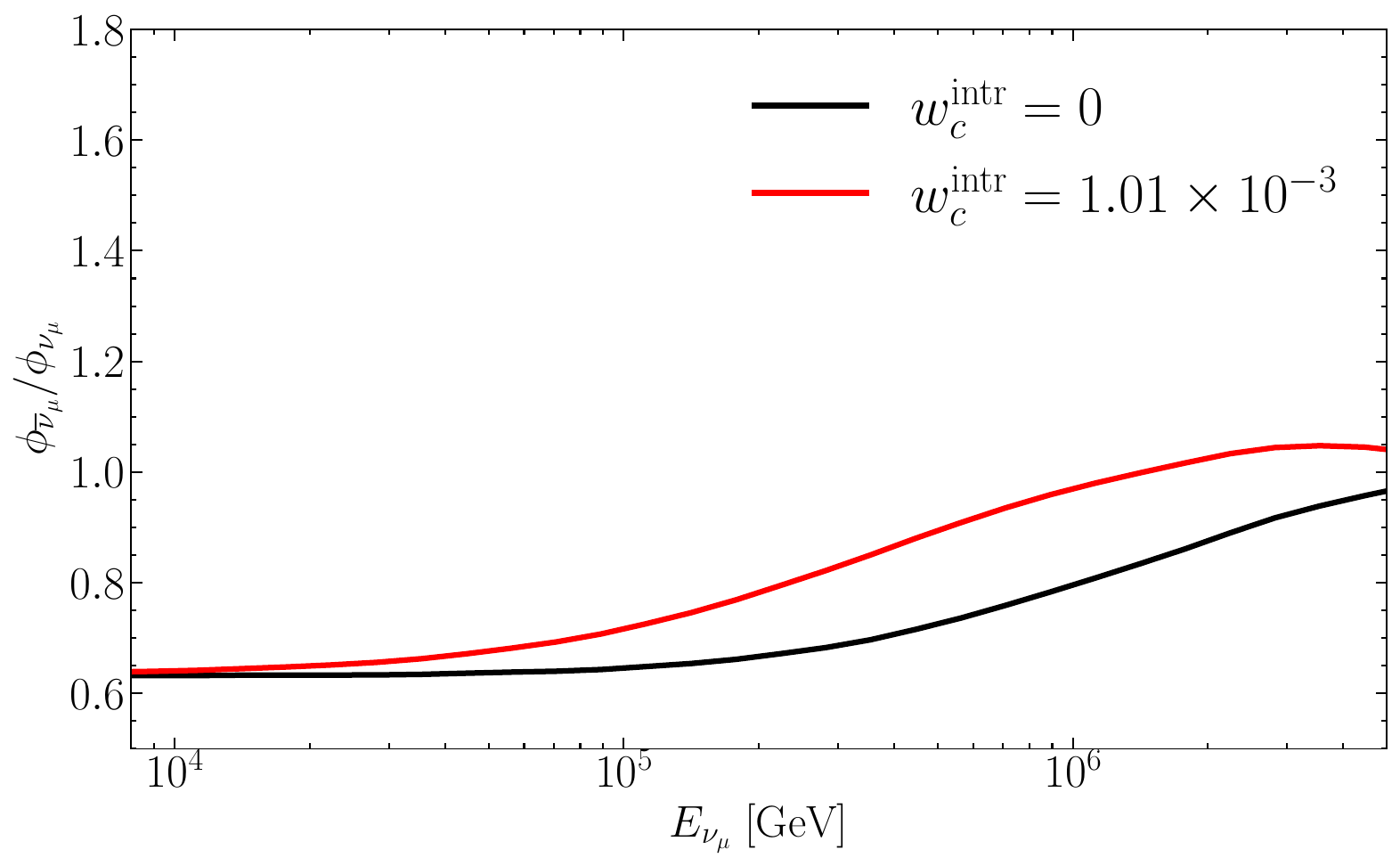}
    \caption{Left: Muon charge ratio without any intrinsic charm, with intrinsic charm and rescaling the unflavored contribution, with all ratios with fluxes angle averaged for $\theta_{\rm zen}\leq 60^\circ$. Right: Anti-neutrino to neutrino flux ratio without and with intrinsic charm. The neutrino and anti-neutrino fluxes do not depend on   the unflavored contribution. Ratios are of fluxes angle averaged for $\theta_{\rm zen}\leq 60^\circ$ and Regge1 intrinsic charm model.}
    \label{fig:charge_ratio}
\end{figure}

The left panel of fig. \ref{fig:charge_ratio} shows the ratios of the angle-averaged fluxes $\phi_{\mu^+}/\phi_{\mu^-}$ for four choices of $(w_c^{\rm intr},\alpha_{\rm unfl})$. With each of these four combinations, with or without intrinsic charm, the flux ratios are nearly identical up to $E_\mu\sim 3\times 10^4$ GeV, where the flux stems predominantly from charged pion and kaon decays. 
There are more $\mu^+$ than $\mu^-$ at fixed energy because of the excess of positive pions with respect to negative ones, e.g., more $\pi^+$ than $\pi^-$, that is traced to the net positive charges of incident cosmic rays.
The ratio changes by $\sim 20\%$ from $[\phi_{\mu^+}/\phi_{\mu^-}]\simeq 1.3$ to $[\phi_{\mu^+}/\phi_{\mu^-}]\simeq 1.1$  as $E_\mu$ increases by a factor of $\sim 100$. The choices of $(w_c^{\rm intr},\alpha_{\rm unfl})$ yield curves between $E_\mu\sim 3\times 10^4$ GeV and $5\times 10^6$ GeV that differ by only $\sim 10\%$. In any case, the muon charges are not distinguished in neutrino telescopes since there are no magnetic fields in detectors. 

The ratios of the angle-averaged fluxes $[\phi_{\bar\nu_\mu}/\phi_{\nu_\mu}]$ are shown in the right panel of fig. \ref{fig:charge_ratio}. The value of this ratio at $E_\nu=10^3$ GeV, amounting to $\sim 0.6$, is less than unity because of the same charge excess of $\pi^+$ relative to $\pi^-$ that yields $[\phi_{\mu^+}/\phi_{\mu^-}]>1$ at the same energy (see discussion in the previous paragraph). The ratio $[\phi_{\bar\nu_\mu}/\phi_{\nu_\mu}]$ increases to $\sim 1$ at the energy where the prompt neutrino flux dominates. The light unflavored mesons do not contribute to the neutrino flux so the neutrino flux does not  depend on $\alpha_{\rm unfl}$. The ratio curves differ by $\sim 10\%$ for $E_\nu=10^6$ GeV. This will have little impact on the ($\nu_\mu+\bar\nu_\mu$) interaction rates, especially since the neutrino and antineutrino cross sections differ by less than 5\%  at this energy \cite{Weigel:2024gzh}.

\begin{figure}[]
    \centering
    \includegraphics[width=0.495\linewidth]{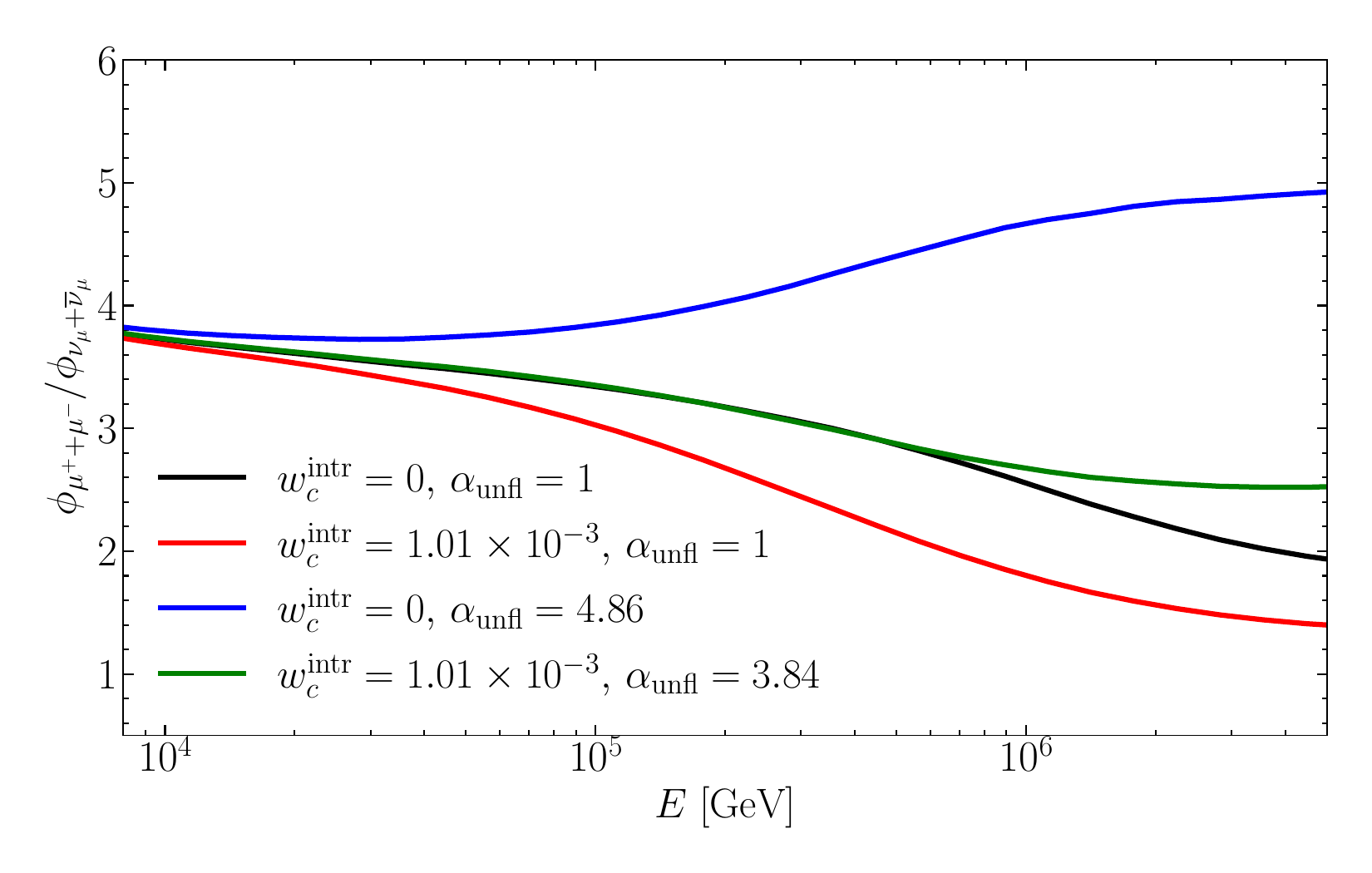}
    \caption{Ratio of muon flux to muon neutrino flux with and without intrinsic charm angle averaged for $\theta_{\rm zen}\leq 60^\circ$. 
  The red (black) curve has $\alpha_{\rm unfl}=1$, and $w_c^{\rm intr}=1.01\times 10^{-3}$ ($w_c^{\rm intr}=0$). The green (blue) curve has $w_c^{\rm intr}=1.01\times 10^{-3},\ \alpha_{\rm unfl}=3.84$ ($w_c^{\rm intr}=0,\ \alpha_{\rm unfl}=4.86$).}
    \label{fig:munu_ratio}
\end{figure}

\Cref{fig:munu_ratio}
shows the ratio of $\phi_{\mu^++\mu^-}/\phi_{\nu_\mu +\bar{\nu}_\mu}$ as a function of energy for different choices for $w_c^{\rm intr}$ and $\alpha_{\rm unfl}$. The two lower curves use as input $\alpha_{\rm unfl}=1$ and either $w_c^{\rm intr}=0$ or $w_c^{\rm intr}=~1.01\times 10^{-3}$.   
The black curve for $w_c^{\rm intr}=0$ tends to a ratio of 2 in the figure since the unflavored and perturbative charm contributions to the muon flux are approximately equal, and the perturbative charm contributions to the neutrino and muon fluxes are nearly equal.
For $E\gtrsim 10^6$ GeV, including intrinsic charm decreases the ratio by $\sim 18\%$, since intrinsic charm is relatively more important for the neutrino flux than for the muon flux. 
At the energies where ratios with $\alpha_{\rm unfl}=1$ shown in \cref{fig:munu_ratio} differ, the astrophysical neutrino flux already dominates the atmospheric neutrino flux. Extracting the ratio in this energy regime will be difficult.
The upper two curves in \cref{fig:munu_ratio} include the best-fit values of $\alpha_{\rm unfl}$ for $w_c^{\rm intr}=0$ and $1.01\times 10^{-3}$. 
The largest values of the ratio $\phi_{\mu^++\mu^-}/\phi_{\nu_\mu +\bar{\nu}_\mu}$ are obtained when the unflavored meson contribution to the prompt $\mu^++\mu^-$ flux is increased with no increase in the prompt $\nu_\mu+\bar\nu_\mu$ flux, namely when $w_c^{\rm intr}=0$ and $\alpha_{\rm unfl}= 4.86$.

\begin{figure}[h]
    \centering
    \includegraphics[width=0.495\linewidth]{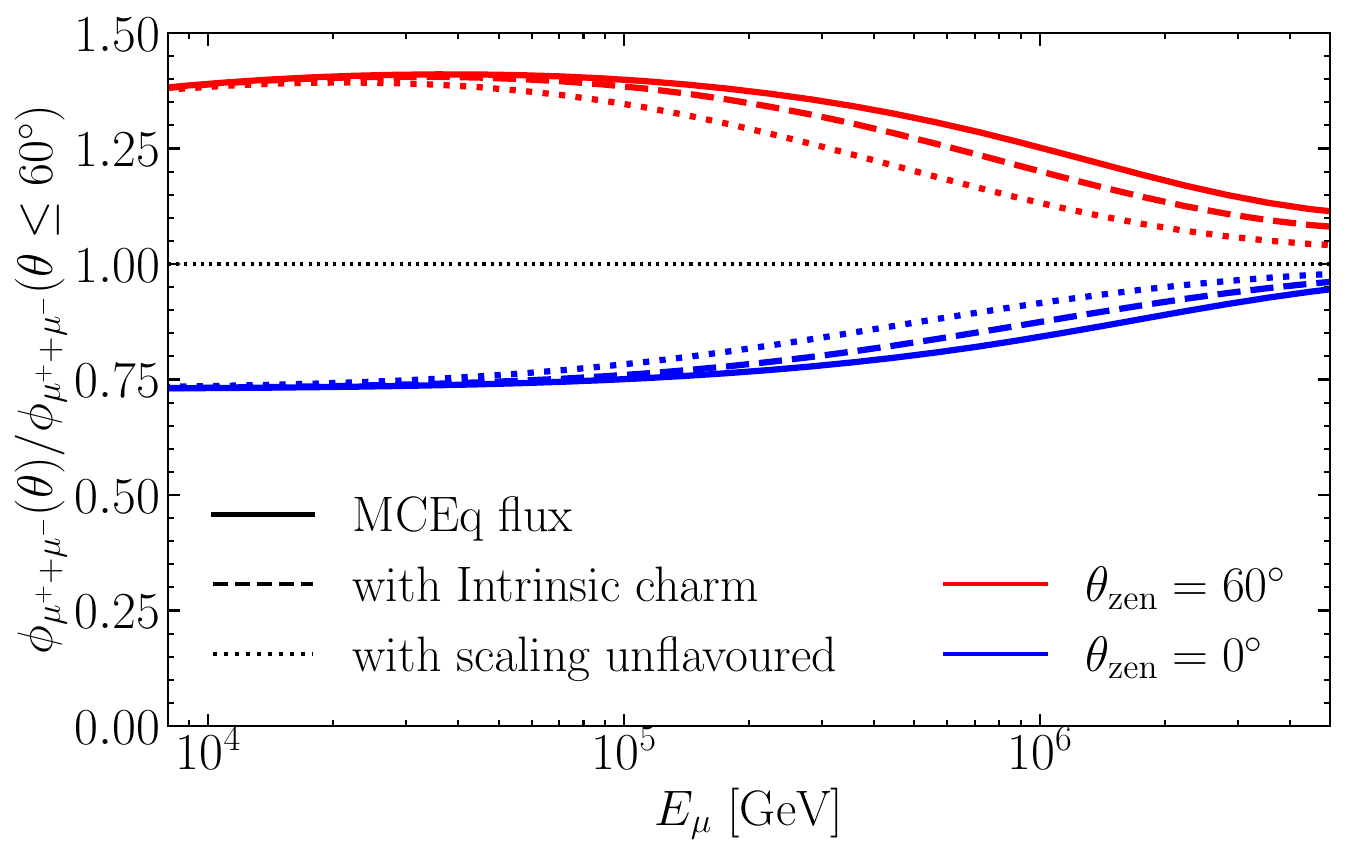}
    \caption{Ratio of muon flux per solid angle $\phi_{\mu^++\mu^-}(\theta) $ at zenith angles of $0^\circ$ and $60^\circ$ to the angle-averaged $\theta_{\rm zen}\leq 60^\circ$ muon flux per solid angle $\phi_{\mu^++\mu^-}$ without any intrinsic charm (labeled \MCEq\ flux), with $w_c^{intr} = 1.01\times10^{-3}$ and $\alpha_{\rm unfl}=1$ (with intrinsic charm), and with $w_c^{intr} = 1.01\times10^{-3}$ and $\alpha_{\rm 
    unfl} = 3.84$ (with scaling of unflavored meson contributions).}
    \label{fig:unfolding}
\end{figure}

Finally, in \cref{fig:unfolding} we show the ratio of the muon flux per solid angle at zenith angles $\theta_{\rm zen} =0^\circ$ and $60^\circ$ relative to the angle-averaged muon flux. For the flux dominated by pion decays, the flux scales as $\sim 1/\cos\theta$ for the zenith angle range considered here \cite{Klimushin:2000cy}. With this functional form, the flux ratios are $\phi_{\mu^++\mu^-}(0^\circ)/\phi_{\mu^++\mu^-}(\leq60^\circ)=0.72$ and $\phi_{\mu^++\mu^-}(60^\circ)/\phi_{\mu^++\mu^-}(\leq60^\circ)=1.44$, consistent with the curves in \cref{fig:unfolding} for $E_\mu\sim 10^4$ GeV. The dominance of prompt contributions, whether from perturbative charm, intrinsic charm or light unflavored mesons, leads to an isotropic distribution and a unit ratio at high energy, independent of angle. \Cref{fig:unfolding} shows the trend towards prompt dominance as the ratio curves move closer  to one.
The transition to prompt-dominated atmospheric muon flux 
as a function of energy 
is more rapid with $\alpha_{\rm unfl}=3.84$ than with just intrinsic charm or with the default \texttt{MCEq}. 

\subsection{Uncertainties}
\label{subsec:uncertainties}

There are a number of uncertainties in the \MCEq\ evaluation of the atmospheric leptons fluxes. We discuss them assuming $w_c^{\rm intr}=0$. In our \texttt{MCEq} computations, we do not include any seasonal variations in the atmospheric muon and muon neutrino fluxes. The IceCube collaboration has measured a $3.9 - 4.6$ \% seasonal modulation in the lepton flux for energies of 125 GeV $-$ 10 TeV~\cite{IceCube:2025wtk}, a small variation compared to other uncertainties and mostly relevant to the conventional atmospheric lepton contributions.

\begin{figure}[]
    \centering
    \includegraphics[width=0.495\linewidth]{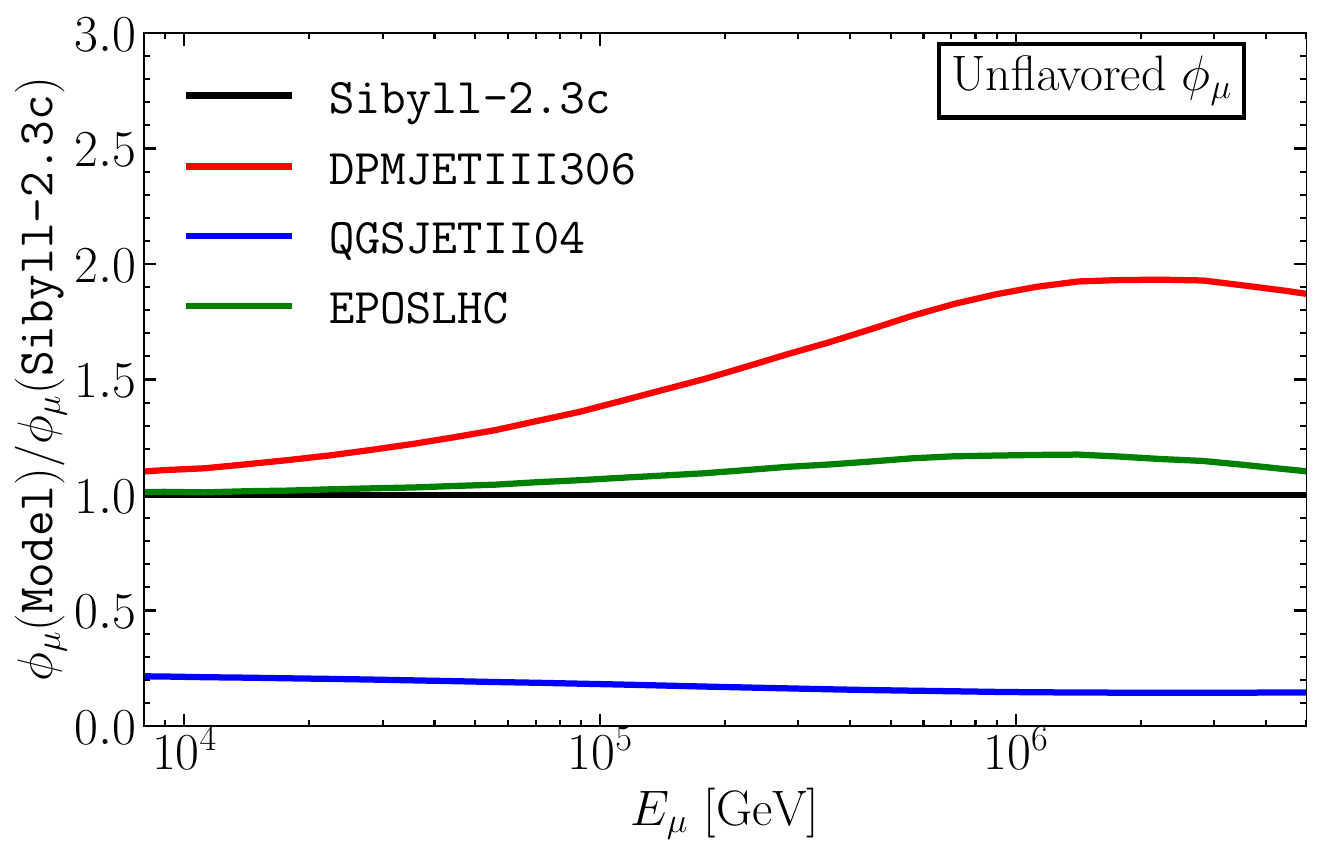}    
    \includegraphics[width=0.495\linewidth]{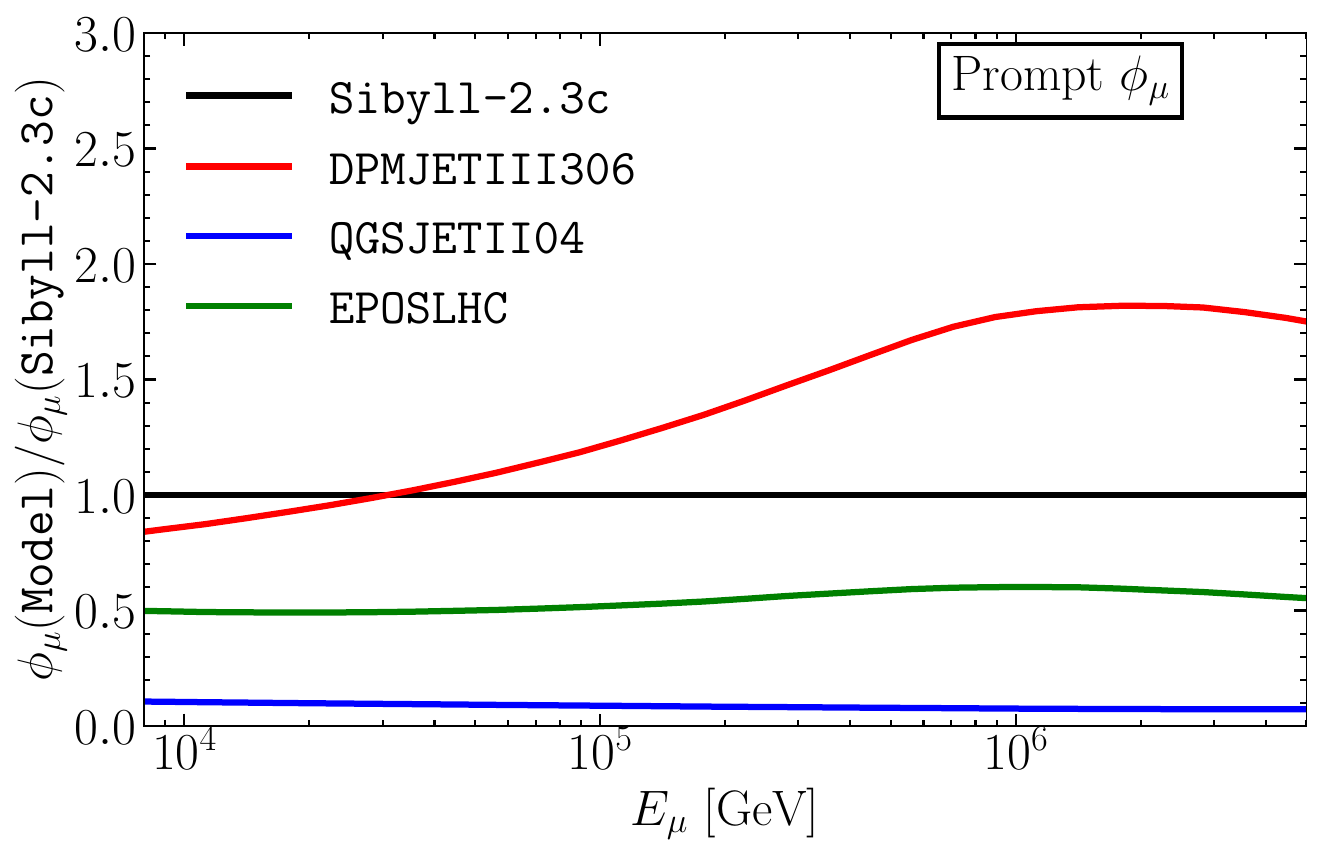}
    \includegraphics[width=0.495\linewidth]{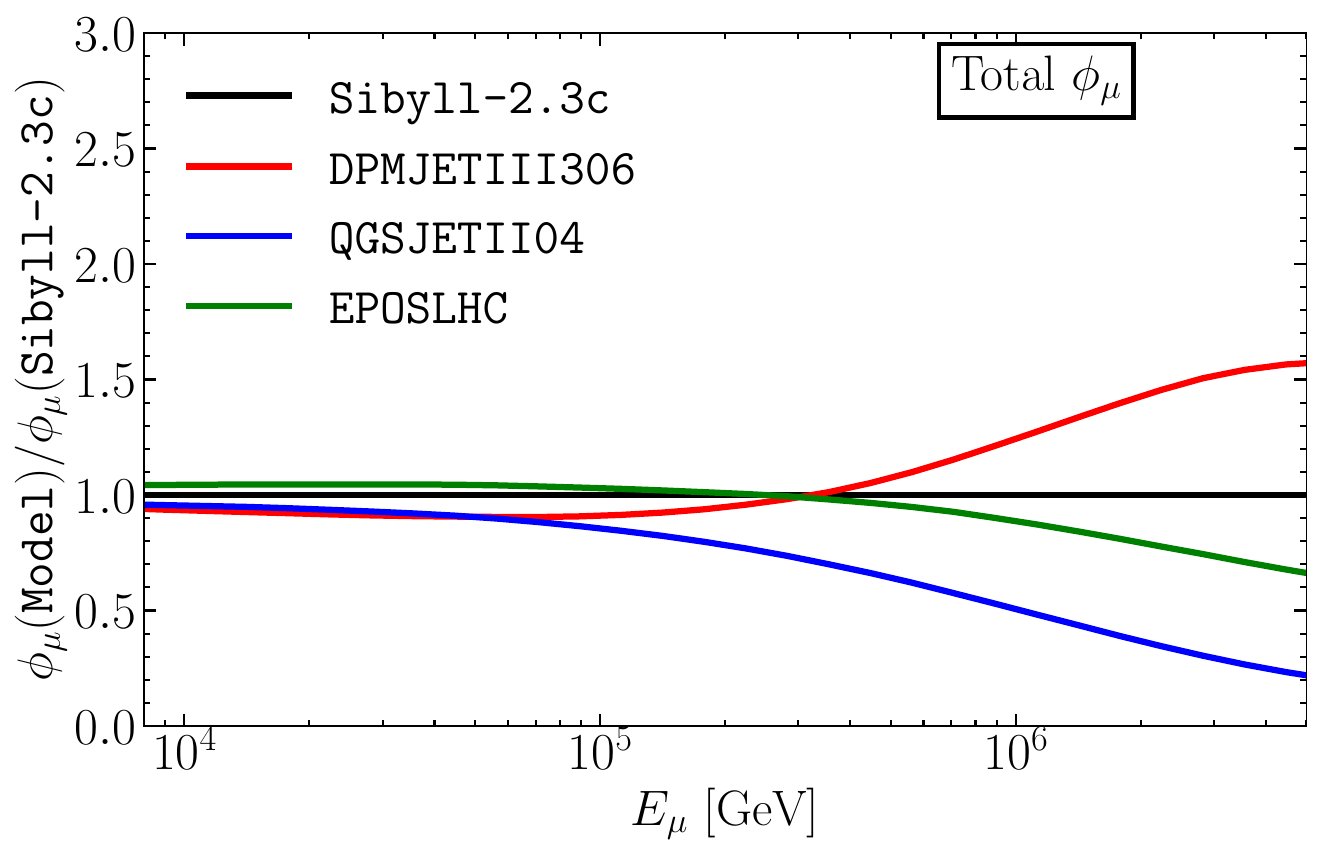}
    \caption{Ratios of the angle-averaged ($\theta_{\rm zen} \leq 60^\circ$) muon fluxes from different interaction models to the muon flux from \texttt{Sibyll-2.3c}, with $w_c^{\rm intr}=0$. Left: Unflavored $\mu^+ + \mu^-$ flux. Right: Prompt $\mu^+ + \mu^-$ flux (unflavored plus charm). Bottom: Total $\mu^+ + \mu^-$ flux (conventional, unflavored and charm). 
     Note that \texttt{QGSJETII04} and \texttt{EPOS-LHC} do not include charm production.}
    \label{fig:mu_uncertainity_int_models}
\end{figure}

A larger uncertainty comes from the hadronic interaction modeling. Within the \MCEq\ framework, a range of Monte Carlo event generators for hadronic interactions are available: besides \texttt{Sibyll-2.3c}, \texttt{EPOS-LHC} \cite{Pierog:2013ria}, \texttt{QGSJETII04} \cite{Ostapchenko:2010vb} and \texttt{DPMJETIII306} \cite{Roesler:2000he} can also be used for predictions. The three panels of fig. \ref{fig:mu_uncertainity_int_models} show ratios of prompt muon fluxes using these three models  to our default result using \texttt{Sibyll-2.3c} for the angle-average ($\theta_{\rm zen}\leq 60^\circ$) for $\mu^++\mu^-$. 
The upper left panel shows the ratios of the prompt fluxes from light unflavored mesons. \texttt{DPMJETIII306} has a larger proportion of prompt muons from light unflavored mesons than \texttt{Sibyll-2.3c} does for $E_\mu\gtrsim 4\times 10^5$ GeV, and almost a factor of two larger for $E_\mu=10^6$ GeV. The \texttt{EPOS-LHC} angle-averaged muon flux from light unflavored mesons nearly equals that from  \texttt{Sibyll-2.3c}. The \texttt{QGSJETII04} results for muons from light unflavored mesons are $\sim 20\%$ of those from \texttt{Sibyll-2.3c}.

With \texttt{Sibyll-2.3c}, we showed that the prompt atmospheric flux of muons from light unflavored mesons is nearly equal to the prompt atmospheric muon flux from charm.
The upper right panel of fig. \ref{fig:mu_uncertainity_int_models} shows the $\mu^++\mu^-$ flux ratios for all prompt atmospheric muons (unflavored plus charm). Neither \texttt{QGSJETII04} nor \texttt{EPOS-LHC} include charm production, so the ratios are consequently reduced. The prompt atmospheric muon flux from \texttt{DPMJETIII306} does include charm contributions.
Including both unflavored mesons and charm, the ratio is $\sim 1.8$ for $E_\mu=10^6$ GeV and $\sim 0.85$ for $E_\mu=10^4$ GeV.
The lower panel of fig. \ref{fig:mu_uncertainity_int_models}  shows the ratios of the total $\mu^++\mu^-$ flux. In the lower energy range of the figure, the ratios are within $\sim 10\%$, while at high energies where the prompt flux dominates (or has no charm contribution as in the cases of \texttt{QGSJETII04} and \texttt{EPOS-LHC}), the curves deviate. Comparing the muon fluxes of \texttt{DPMJETIII306} and \texttt{Sibyll-2.3c}, the total flux ratio is $\sim 0.9$ for $E_\mu=10^5$ GeV, and $\sim 1.3$ for $E_\mu=10^6$ GeV. A best-fit rescaling of the \texttt{DPMJETIII306} unflavored meson contribution gives $\alpha_{\rm unfl}=4.89$ (without the inclusion of an intrinsic charm component). Even though the \texttt{DPMJETIII306} muon flux for $E_\mu=10^6$ GeV is larger, the fit is dominated by the muon data with smaller errors, $E_\mu\lesssim 3\times 10^5$ GeV, where the \texttt{DPMJETIII306} flux is slightly smaller than the \texttt{Sibyll-2.3c} flux, so the value of $\alpha_{\rm unfl}$ is slightly larger for the \texttt{DPMJETIII306} muon flux than for the \texttt{Sibyll-2.3c} flux.

\begin{figure}[]
    \centering
    \includegraphics[width=0.495\linewidth]{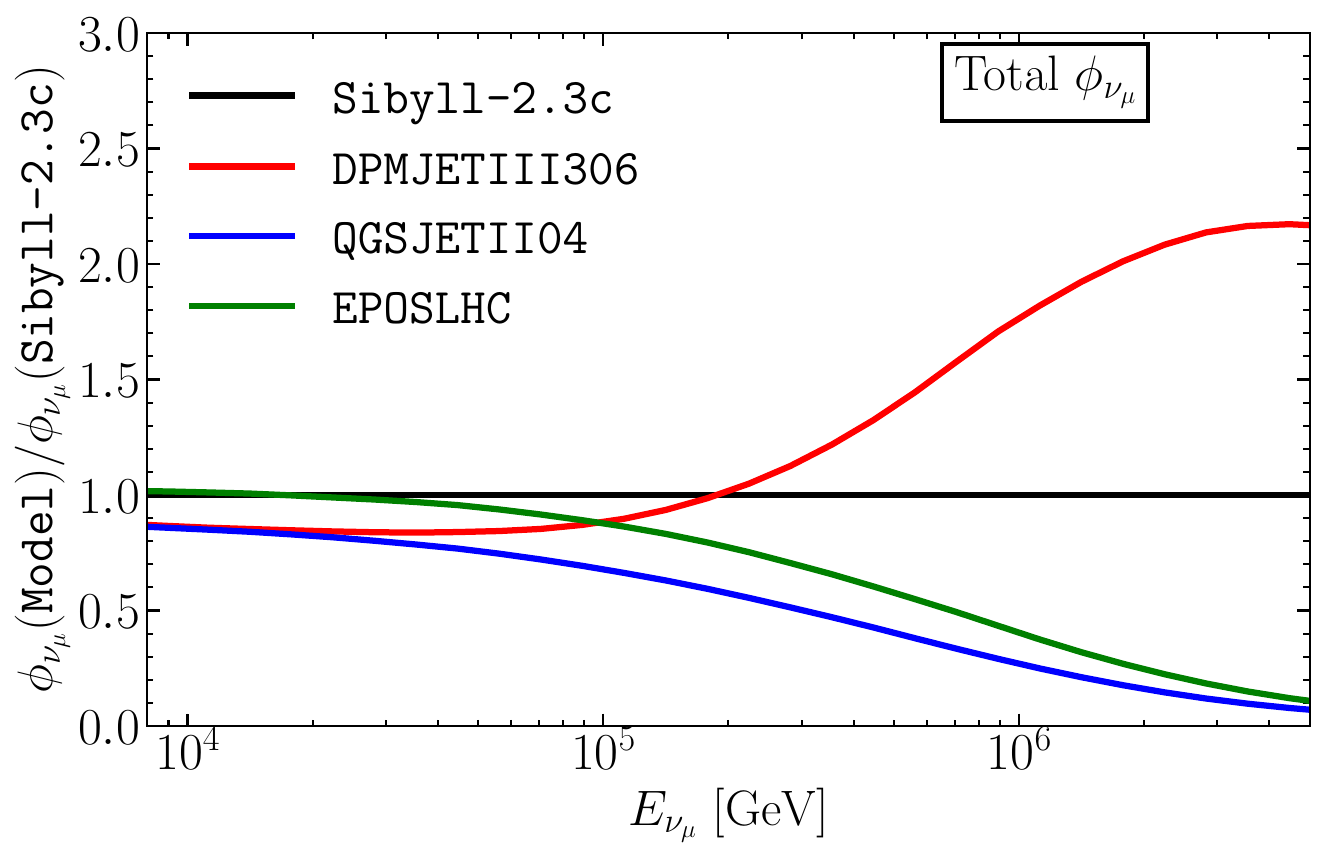}
    \includegraphics[width=0.495\linewidth]{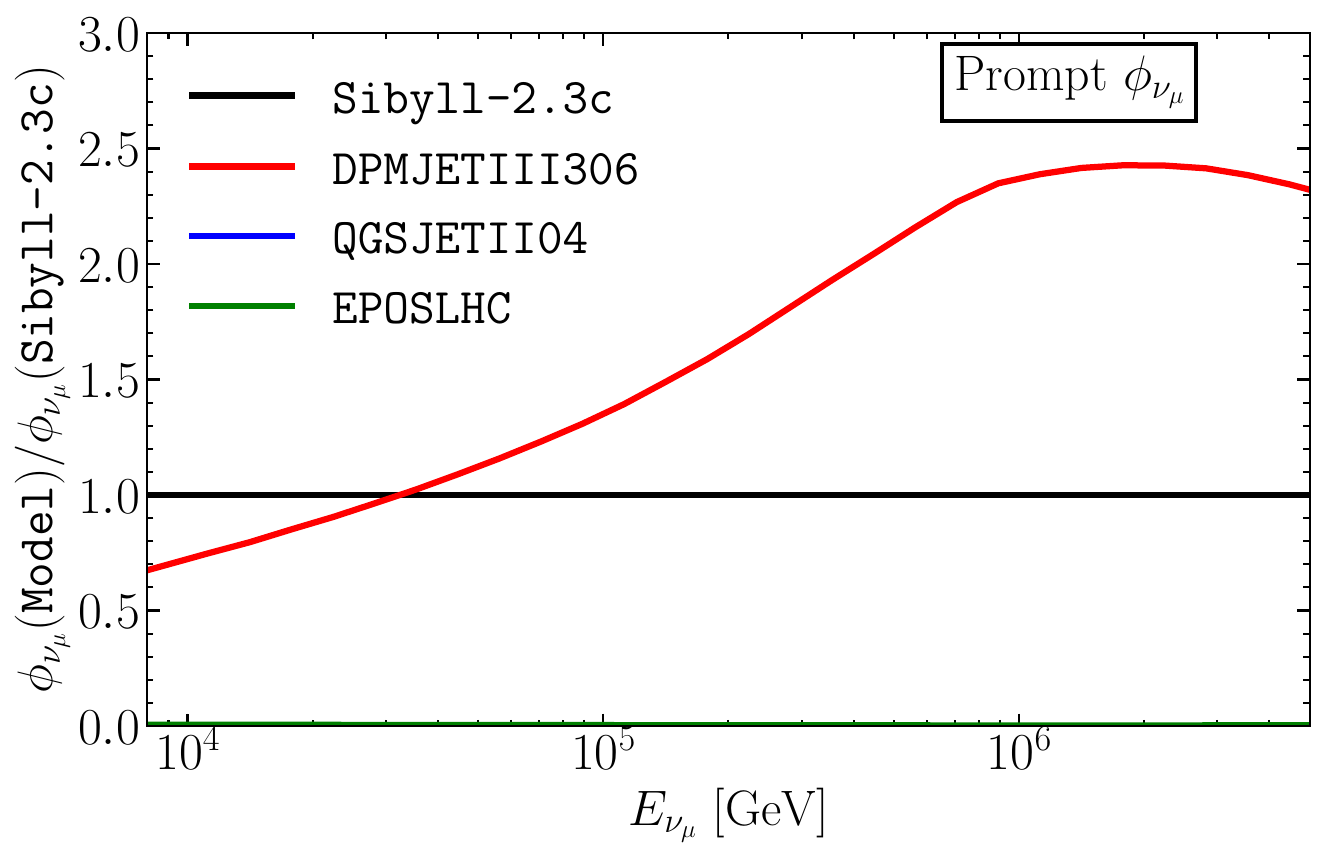}
    \caption{The ratios of the angle-averaged $\nu_\mu+\bar\nu_\mu$ fluxes  ($\theta_{\rm zen} \leq 60^\circ$) from different interaction models to the $\nu_\mu+\bar\nu_\mu$ flux from $\texttt{Sibyll-2.3c}$, all with $w_c^{\rm intr}=0$. Left: Total $\nu_\mu + \overline{\nu}_\mu$ flux. Right: Prompt $\nu_\mu + \overline{\nu}_\mu$
    flux. Note that \texttt{QGSJETII04} and \texttt{EPOS-LHC} do not include charm production.}
    \label{fig:nu_uncertainity_int_models}
\end{figure}

We have already noted that \texttt{Sibyll-2.3c} is a Monte Carlo generator for hadronic interactions that yields results consistent with NLO perturbative QCD evaluations of the prompt atmospheric $\nu_\mu+\bar\nu_\mu$ flux from charm, discussed in Appendix \ref{appendix:b}. Figure \ref{fig:flux-compare} in the appendix shows the perturbative QCD uncertainty together with the \texttt{Sibyll-2.3c} prompt neutrino flux in the energy range of the IceCube upper bound. The uncertainties in the perturbative QCD contributions from charm are large, but in the relevant energy range, the uncertainty bands are below the IceCube upper bounds on the prompt muon neutrino flux. Within the \MCEq\ framework, we consider Monte Carlo generator variations.
\Cref{fig:nu_uncertainity_int_models} shows the ratios of the angle-averaged atmospheric total flux (left) and prompt flux (right) for $\nu_\mu+\bar\nu_\mu$ for different interaction models. There are no unflavored contributions to the prompt $\nu_\mu+\bar\nu_\mu$ flux. For $E_{\nu_\mu}=10^4$ GeV, the interaction model predictions agree to within approximately 10\%, while again, \texttt{DPMJETIII306} has relatively more prompt contributions to the $\nu_\mu+\bar\nu_\mu$ flux than  \texttt{Sibyll-2.3c}.

\begin{figure}[h]
\centering
\includegraphics[width=0.495\linewidth]{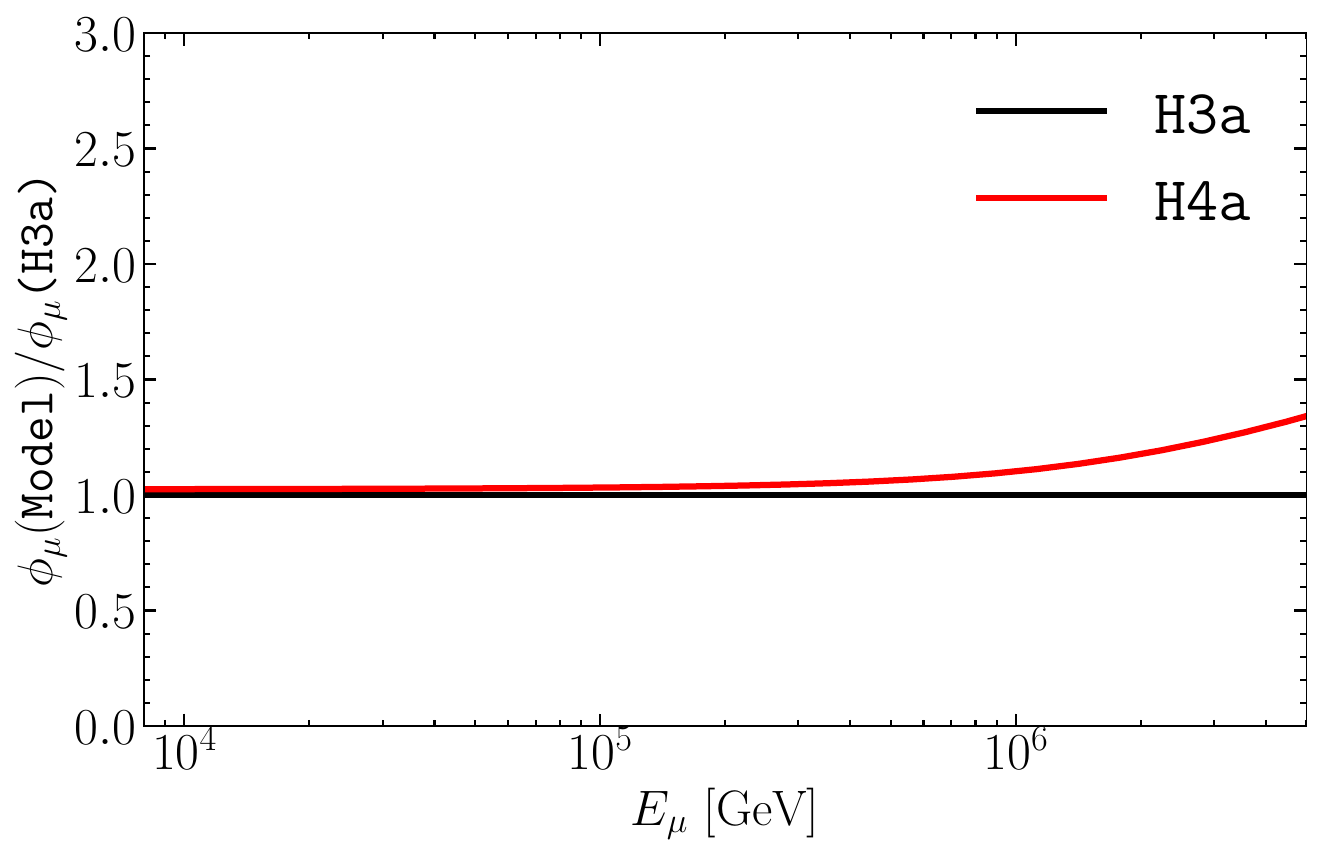}
\includegraphics[width=0.495\linewidth]{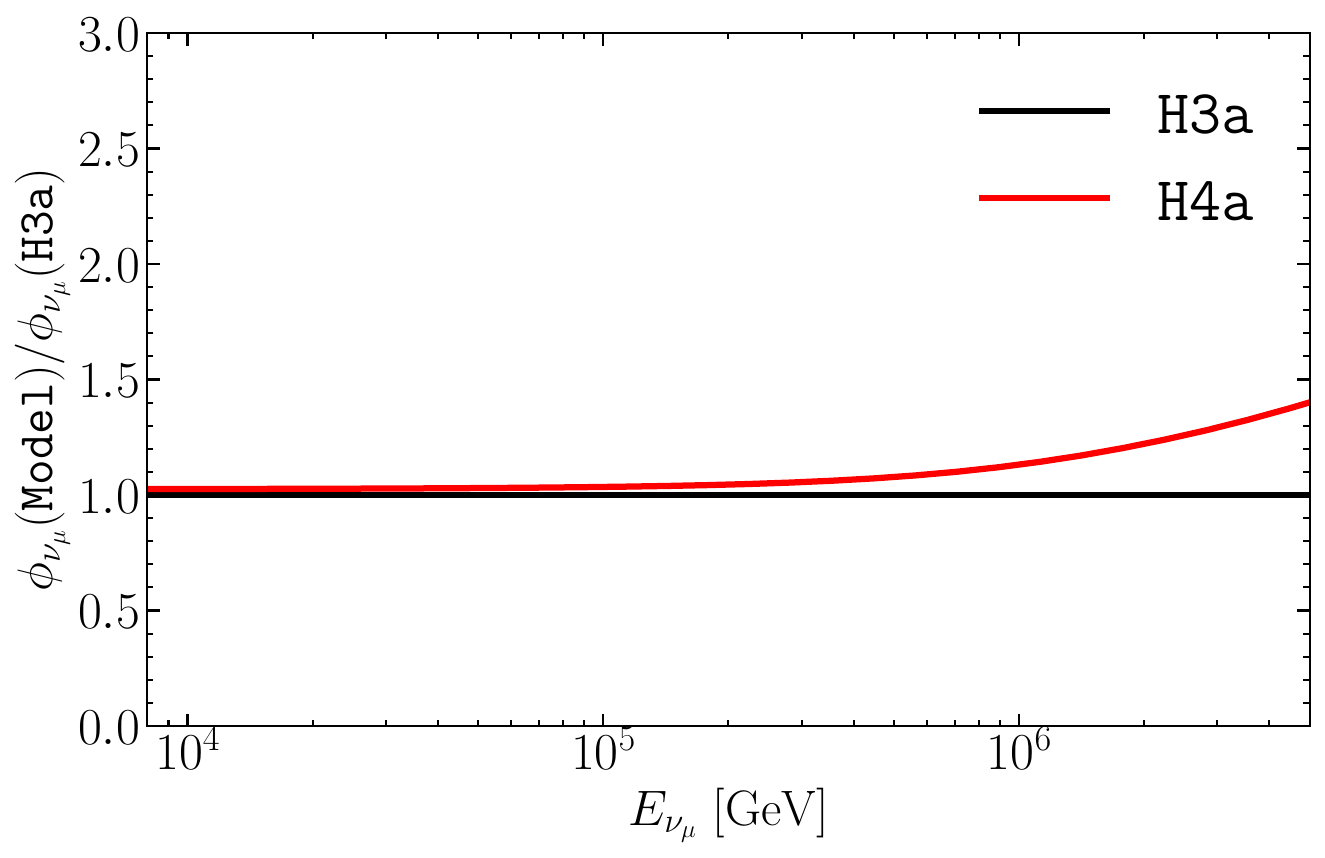}
\caption{The ratio of atmospheric lepton fluxes using as input the \texttt{H4a} all-nucleon primary cosmic ray flux model and the \texttt{H3a} all-nucleon cosmic ray flux model. Left: Angle-averaged ($\theta_{\rm zen} \leq 60^\circ$) total $\mu^+ + \mu^-$ flux ratio. Right: Angle-averaged ($\theta_{\rm zen} \leq 60^\circ$) total $\nu_\mu + \bar{\nu}_\mu$ flux ratio.}
\label{fig:mu_uncertainity_crflux_models}
\end{figure}

In~\cref{fig:mu_uncertainity_crflux_models}, we show the ratio of the fluxes evaluated with the \texttt{H4a}~\cite{Gaisser:2013bla} (sometimes called \texttt{H3p}, for instance in ref.~\cite{IceCube:2016umi}) cosmic ray model relative to the \texttt{H3a} model for muons (left) and muon neutrinos (right). Both \texttt{H3a} and \texttt{H4a} are implemented in \texttt{MCEq} as 3-Peters cycle~\cite{Peters:1961mxb} compositions with five mass groups. However, \texttt{H4a} assumes a composition that becomes proton dominated above the ankle, whereas \texttt{H3a} retains a mixed composition. The \texttt{H4a} model predicts a larger high-energy lepton flux, leading to an enhancement of approximately 35\% for muons and 40\% for muon neutrinos at $E = 5 \times 10^6$ GeV, however, this results in only modest changes to the fitted intrinsic charm and unflavored scaling parameters. For the Regge1 intrinsic charm model, the best fit values shift to $w_c^{\rm intr}=0.94 \times 10^{-4}, \ \alpha_{\rm unfl}=1$ and $w_c^{\rm intr}=0, \ \alpha_{\rm unfl}=3.91$ when using \texttt{H4a} instead of \texttt{H3a}, compared to the values reported in~\cref{unfl_table}. 

By exploiting their new measurements of the cosmic ray energy spectrum and mean logarithmic mass \cite{Lv:2024wrs}, together with other high-energy cosmic-ray data, the LHAASO  collaboration models the cosmic ray spectrum and composition with three galactic and two extragalactic components.  Their model has a harder spectrum and lighter composition in the PeV cosmic ray energy range. This new cosmic ray flux model can yield energy dependent variations in the prompt atmospheric lepton fluxes from charm by $-20\% \ -\  +50\%$ for the interval $E=10^4-10^6$ GeV. In any case, the increase around $10^6$ GeV is not sufficient to fully account for the measured muon flux. Further investigations of the impact of the spectrum and composition of galactic cosmic rays are ongoing \cite{future}.  
    
There are uncertainties in modeling hadronic interactions of cosmic rays with air targets. In ref. \cite{Yanez:2023lsy}, Ya\~nez and Fedynitch tune a data-driven model (\texttt{daemonflux}) that includes hadronic interaction and cosmic ray flux uncertainties. The tune is calibrated to atmospheric muon flux data, primarily below $E_\mu=10^4$ GeV. They compare the resulting neutrino flux from \texttt{daemonflux} with the atmospheric neutrino fluxes from  \texttt{Sibyll-2.3d}, which yields nearly identical inclusive atmospheric lepton fluxes as \texttt{Sibyll-2.3c} \cite{Riehn:2019jet}. The \texttt{Sibyll-2.3d} muon neutrino flux is within approximately $\pm 5\%$ of the \texttt{daemonflux} muon neutrino flux for $E_{\nu_\mu}=10^4-10^6$ GeV. There is a little more variation in the electron neutrino flux, with the \texttt{Sibyll-2.3d} electron neutrino flux approximately $15\%$ higher for $E_{\nu_e}=10^4$ GeV, and closer to 5\%\ higher for $E_{\nu_e}=10^5$ GeV. Our conclusion is that using a data-driven model such as \texttt{daemonflux} would not change our estimates about uncertainties in the atmospheric muon flux predictions, so there remains an apparent deficit in atmospheric muon flux predictions at high energy.

Different intrinsic charm models can, in principle, yield widely varying atmospheric flux predictions. We have assumed a functional form for the differential cross section for charm hadron production via the intrinsic charm mechanism in nucleon-air interactions according to \cref{eq:dsdx-ic}. The resulting contributions from intrinsic-charm-produced $D^-,\ \bar{D}^0$, and $\Lambda_c$ to the muon and muon neutrino flux have essentially the same shape. The primary effect of changing the fragmentation model is  a shift in the overall normalization of the intrinsic charm contribution, which is reflected in the fitted value of $w_c^{\rm intr}$, as shown in table \ref{tab:w_intr_table} 
and table \ref{unfl_table}.
The reason the fitted $w_c^{\rm intr}$ values differ among the models is that the Regge models produce larger fractions of $\Lambda_c$ baryons at higher energies compared to $D$ mesons, whereas the HLM model yields almost equal distributions of energetic $D$ and $\Lambda_c$ hadrons.
Because of kinematics,
the leptons from $D$ meson decays can emerge with higher energy fractions than from $\Lambda_c$ decays (see figs.~\ref{fig:dN_dxdy} and~\ref{fig:dN_dxdy_lambda} in  Appendix \ref{appendix:a}).
The branching ratios for $D$ semileptonic decays are also larger than for $\Lambda_c$ semileptonic decays. The HLM model gives a relatively larger intrinsic charm contribution to the muon and muon neutrino fluxes, so the best fit normalization $w_c^{\rm intr}$ is smaller for HLM fragmentation compared to the Regge models. In other words, the physical differences between the models manifest primarily as a change in the fitted intrinsic charm  normalization, rather than in the shape of the resulting lepton spectra. (See also Appendix \ref{appendix:z}.) Overall, whether $w_c^{\rm intr}$ is fitted to the muon flux or the muon neutrino flux, all three valence-like fragmentation models for intrinsic charm give values for $w_c^{\rm intr}$ that are roughly comparable among each other.

\section{Discussion}\label{sec:discussion}
\begin{figure}[]
    \centering
    \includegraphics[width=0.495\linewidth]{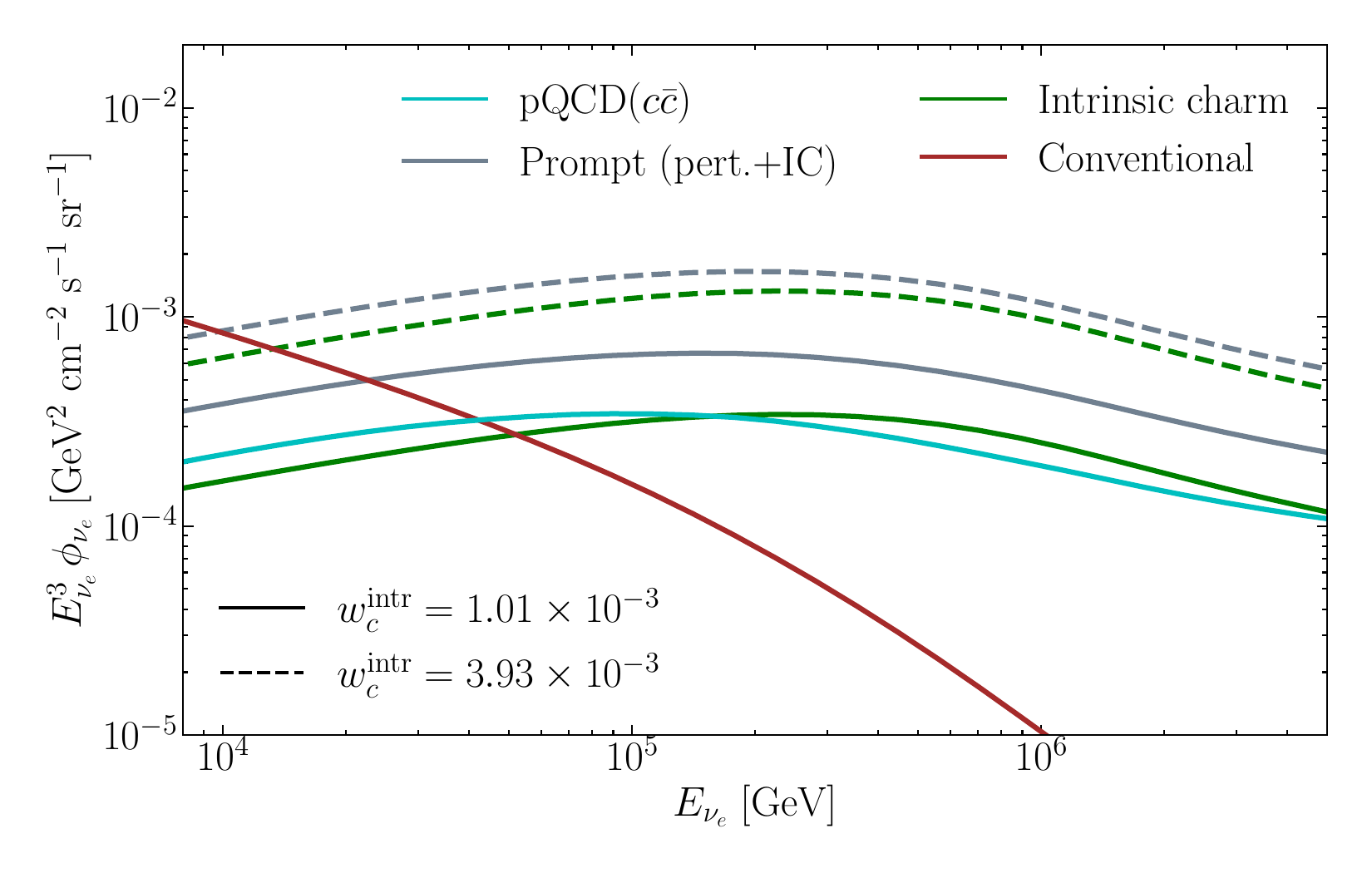}
 \includegraphics[width=0.495\linewidth]{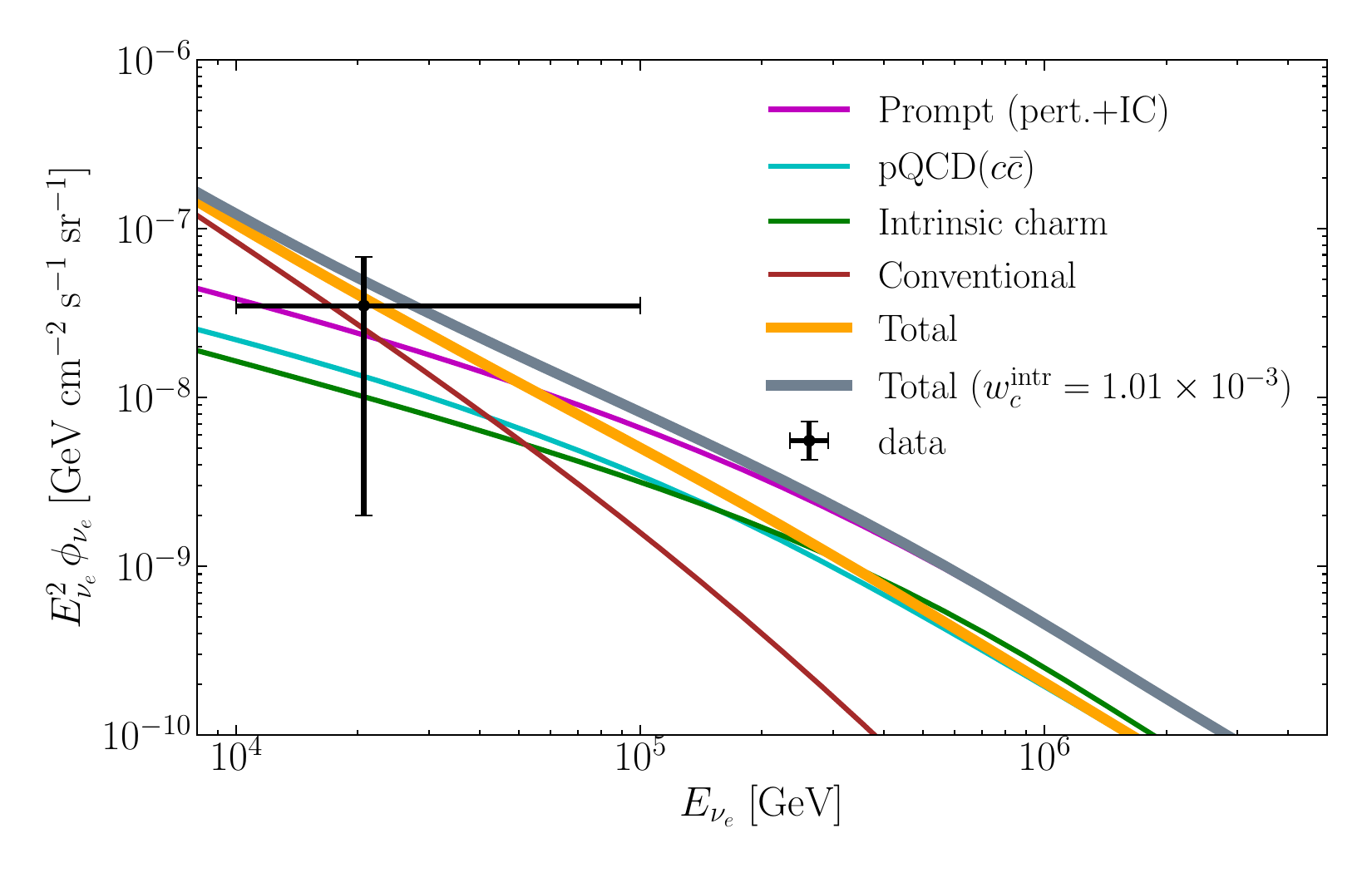}    
    \caption{Left: Angle-averaged (up-to $\theta_{\rm zen} \leq 60^\circ$) electron neutrino flux ($\nu_e+\bar\nu_e$) scaled by $E_{\nu_e}^3$ with the inclusion of intrinsic charm with $w_c^{\rm intr}$ values of $3.93\times 10^{-3}$ (dashed lines) and $1.01\times10^{-3}$ (solid lines). Right: Angle-averaged for full zenith angle range, the $\nu_e+\bar\nu_e$ flux contributions scaled by $E_{\nu_e}^2$, along with a data point from IceCube \cite{IceCube:2015mgt}.}
    \label{fig:E_neutrinos}
\end{figure}

Measurements of the high-energy atmospheric $\nu_e+\bar\nu_e$ flux may shed light on the prompt flux contribution.
While the prompt atmospheric flux of $\nu_e+\bar\nu_e$ is very nearly equal to the prompt atmospheric $\nu_\mu+\bar\nu_\mu$ flux, the conventional fluxes have larger differences. The conventional atmospheric $\nu_e+\bar\nu_e$ flux is about a factor of 20 lower than the conventional $\nu_\mu+\bar\nu_\mu$ flux because nearly 100\% decays of $\pi^+$ lead to $\nu_\mu + \mu^+$, and at the energies under consideration here, the muons do not decay. Kaon production and decay dominate as the source of conventional atmospheric $\nu_e+\bar\nu_e$ flux. Consequently, the prompt $\nu_e+\bar\nu_e$ flux becomes larger than the corresponding conventional flux at energies lower than in the case of $\nu_\mu+\bar\nu_\mu$. This is illustrated in fig. \ref{fig:E_neutrinos}, where we show both pQCD and two versions of the prompt flux, both with and without including intrinsic charm. 
In the left panel of \cref{fig:E_neutrinos}, the  $\nu_e+\bar\nu_e$ flux is scaled by $E_{\nu_e}^3$ (as with the prior flux figures).
As already noted, the dashed lines that include $w_c^{\rm intr}=3.93\times 10^{-3}$ violate the upper limit for the prompt $\nu_\mu+\bar\nu_\mu$ flux. 

The angle-averaged atmospheric electron neutrino flux was measured by DeepCore/IceCube  for $E_\nu=80$ GeV$~-~6$ TeV \cite{IceCube:2012jwm} and up to a highest energy bin at $\log_{10}(E_\nu/{\rm GeV})=4.0-5.0$ by IceCube \cite{IceCube:2015mgt}. The right panel of fig. \ref{fig:E_neutrinos} shows the fluxes in the left panel, now scaled by $E_{\nu_e}^2$, and the data point of the highest energy bin of the IceCube atmospheric electron neutrino flux measurement \cite{IceCube:2015mgt}. 
The measurements of the atmospheric 
$\nu_e+\bar\nu_e$ flux is difficult at high energy, in part because the diffuse astrophysical neutrino flux is large. For reference, the
IceCube's  high-energy starting event measurements of the per-flavor astrophysical flux in the energy region of $E_\nu=100$ TeV, assuming a single power law, is
$E_\nu^2\phi_{\nu+\bar\nu}^{\rm per\ flavor}\sim 2\times 10^{-8}(E_\nu/100\ {\rm TeV})^{-2.5}$ \cite{IceCube:2020acn,IceCube:2024fxo,IceCube:2025dlr}. To the extent that the diffuse astrophysical neutrino flux is inferred from $\nu_\mu+\bar{\nu}_\mu$ measurements, it may be possible to understand the relative contributions of the atmospheric and prompt
$\nu_e+\bar\nu_e$ fluxes in the measurements. 

The value of $w_c^{\rm intr}=1.01\times 10^{-3}$ that saturates the IceCube upper limit on the prompt neutrino flux translates to a maximum contribution of intrinsic charm to the $p$-air cross section of $0.1\%$ in our framework for intrinsic charm, given the results for pQCD($c\bar{c}$). It is difficult to directly connect this result with the parton distribution function picture of intrinsic/fitted charm, given that they do not correspond to exactly the same mechanism (in our approach, IC leads to the production of two charmed hadrons, one including the charm and another the anticharm quark, whereas in the PDF case, this does not necessarily happen). Recent studies of intrinsic charm in parton distribution functions suggest that the fraction of the proton momentum carried by intrinsic charm is comparably small; it may be of order $\sim 0.5$\% \cite{Guzzi:2022rca,Ball:2022qks}, although there is not a consensus that a fitted/intrinsic charm PDF is an established component.

\begin{figure}[]
    \centering
\includegraphics[width=0.6\linewidth]{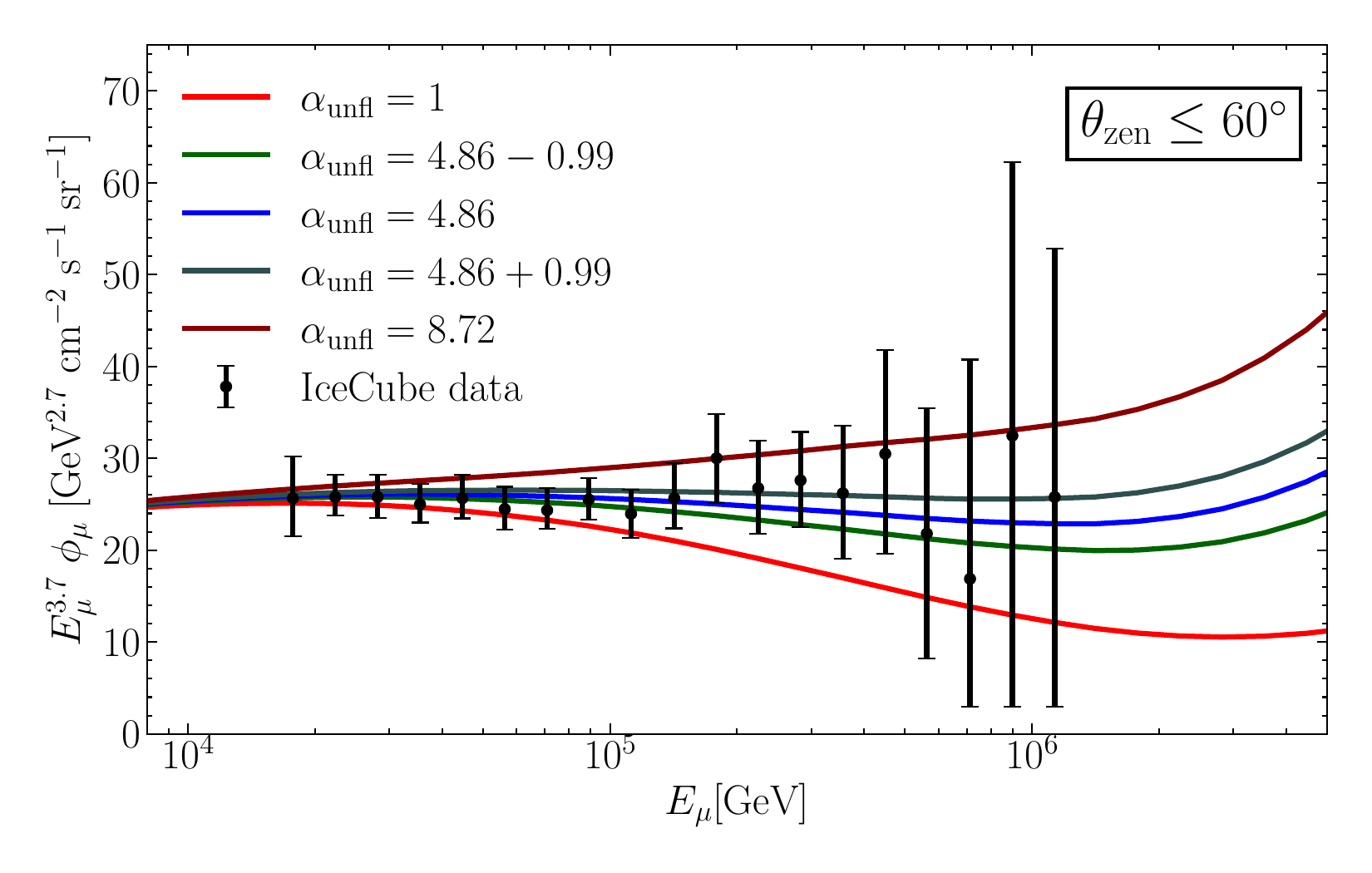}
      \caption{Angle-averaged ($\theta_{\rm zen}\leq 60^\circ$) atmospheric $\phi_{\mu^++\mu^-}$ scaled by $E_\mu^{3.7}$ with $w_c^{\rm intr}=0$ and $\alpha = 1,\ 3.87,\ 4.86,\ 5.85$ and $8.72$. The IceCube data from ref. \cite{IceCube:2015wro} are also shown.}
    \label{fig:scaled-muons-alpha}
\end{figure} 

The large experimental error bars at high energies make conclusive statements about the role of prompt contributions to the atmospheric $\mu^++\mu^-$ flux difficult. The spectral shapes of the prompt contributions from charm and from unflavored mesons to the $\mu^++\mu^-$ flux are nearly identical in the energy range of interest, so we can trade enhancements of the charm contributions for enhancements of the light unflavored contributions.

We again show the large experimental errors in the measured $\mu^++\mu^-$ flux scaled by $E_\mu^{3.7}$ in fig. \ref{fig:scaled-muons-alpha}, where the experimental data are compared to theory predictions for different values of $\alpha_{\rm unfl }$ and no intrinsic charm, $w^c_{\rm intr}=0$.  
In particular, 
the figure shows curves for $\alpha_{\rm unfl}=1$, and $\alpha_{\rm unfl}=4.86^{+0.99}_{-0.99}$,
corresponding to the $\chi^2$ best-fit value with $1\sigma$ errors.
We also show a curve for $\alpha=8.72$, to show a more extreme scaling. It is still consistent within $2\sigma$ uncertainties. 
All of the scaling factors multiply the unflavored meson contributions to the prompt flux according to interactions modeled by \texttt{Sibyll-2.3c}.
As the lower panel of fig. \ref{fig:mu_uncertainity_int_models} shows, \texttt{DPMJETIII306} would lead to a larger high-energy contribution. However, its energy dependence cannot account for the apparent excess of atmospheric muons in the hundreds of TeV energy range.

\section{Conclusions}\label{sec:conclusion}
With the usual models of the cosmic ray spectrum and composition, IceCube's measurements of the $\mu^++\mu^-$
atmospheric flux  at high energies suggest a missing prompt component. We have focused on fluxes at zenith angles $\theta\leq 60^\circ$ to avoid issues with potential mis-reconstruction of muon events in the experimental data \cite{IceCube:2015wro}. Since the prompt 
$\nu_\mu+\bar\nu_\mu$ and $\mu^++\mu^-$
fluxes from charm are nearly equal, an increase of the prompt $\mu^++\mu^-$
flux from charm to improve agreement with the high-energy muon data (which would correspond to taking 
$w_c^{\rm intr}=3.93\times 10^{-3}$) can violate the upper bounds on the prompt $\nu_\mu + \bar{\nu}_\mu$ flux. Indeed, in addition, such a high value of $w_c^{\rm intr}$ would lead to an excess of very forward production of $\Lambda_c^+$ in low-energy collisions since we have assumed that $w_c^{\rm intr}$ is energy independent. The value of $w_c^{\rm intr}=1.01\times 10^{-3}$ ($w_c^{\rm intr}=5.34\times 10^{-4}$) determined from the 2025 (2016) IceCube prompt neutrino limit is sufficiently small that the intrinsic charm cross section is less than $\sim 7\%$ (4\%)\ of the perturbative cross section for 
fixed-target $pp$ production of $c\bar{c}$ at $E_p=10^5$ GeV and less than $\sim 3\%$ (2\%) for $E_p\gtrsim 10^6$ GeV, well within the next-to-leading order perturbative \cite{Nason:1987xz,Mangano:1991jk,
Bhattacharya:2015jpa,Bhattacharya:2016jce,Garzelli:2015psa} and measurement 
\cite{PHENIX:2006tli,HERA-B:2007rfd,ALICE:2012inj,STAR:2012nbd,ATLAS:2015igt,LHCb:2013xam}  uncertainties for open charm production.  However, this smaller value of $w_c^{\rm intr}$ still underpredicts the atmospheric $\mu^+ + \mu^-$ flux at high energy, as shown in fig.~\ref{fig:Neutrino_flux_withIC}.

We have used IceCube's upper bound on the prompt atmospheric $\nu_\mu+\bar\nu_\mu$ flux from measurements of through-going muons \cite{Abbasi:2025rmj} together with \texttt{Sibyll-2.3c} in \texttt{MCEq} and various hadron-level models for intrinsic charm, to constrain the intrinsic charm contribution. This IceCube's upper bound relies on a \texttt{Sibyll-2.3c}-based \texttt{MCEq} evaluation of charm production and 
an input \texttt{H3a} cosmic ray flux.  
With either cascades or northern tracks and a single power law astrophysical flux, the 90\%\ CL upper bound on the prompt neutrino flux normalization is 1.34 and 3.02 times the \texttt{Sibyll-2.3c} prompt neutrino flux, respectively \cite{Bottcher:2025hcz}. Upper limits vary with the assumed astrophysical neutrino flux, with normalization factors ranging between 1.92-2.99 \cite{Abbasi:2025rmj}, and whether or not combined fits are made. 

We have relied on flux predictions from \texttt{Sibyll-2.3c} to represent the contribution of NLO perturbative charm pair production and decay of leptons, giving rise to the prompt lepton fluxes. As shown in
Appendix \ref{appendix:b}, there are large uncertainties in the perturbative QCD evaluation of the prompt atmospheric flux. 
If the \texttt{Sibyll-2.3c} flux is enhanced/diminished by a factor to coincide with the upper/lower limit of the uncertainty band of the NLO perturbative QCD evaluation of charm pair production, there is less/more room for intrinsic charm. Thus, the exact extent to which intrinsic charm is constrained here is model dependent, with a dependence that will indeed decrease when more accurate QCD calculations will become available.
Our point is that, in any case the IceCube limits on the prompt atmospheric 
$\nu_\mu+\bar\nu_\mu$ flux put significant constraints on the possibility that intrinsic charm can account for IceCube's high energy $\mu^++\mu^-$ measurements.

The measurements suggest a prompt component that could otherwise come from prompt decays of light unflavored mesons, as we discussed  in \cref{sec:results} with an enhancement factor of $\alpha_{\rm unfl}=4.86$ larger than the \texttt{MCEq} default contribution.
The IceCube Collaboration reports that 
the combined heavy-flavor prompt plus unflavored prompt contributions to the atmospheric muon flux correspond to 4.75 times the ERS flux with a 90\% CL range of 2.33--9.34 times the ERS flux,  assuming the \texttt{H3a} all-nucleon cosmic ray energy spectrum \cite{IceCube:2015wro}. Their conclusion is consistent with their angular analysis as well. IceCube's results are consistent with the enhancements of the prompt contributions to the muon flux discussed here.

The extraction of the prompt atmospheric muon and neutrino fluxes from neutrino telescope data remains a topic of tremendous interest. In combination, SHiP \cite{Alekhin:2015byh,Bai:2018xum} and the forward physics program \cite{
Anchordoqui:2021ghd, Feng:2022inv,FPF:2025bor} 
with neutrinos (see Appendix \ref{appendix:b} for an expanded discussion) and studies of unflavored meson production at accelerator experiments, together with a better modeling of the unflavored component in Monte Carlo event generators for (astro)-particle physics and more precise tunings of the relevant parameters of the latter~\cite{Albrecht:2025kbb},
will be important to better interpret existing and future data, potentially resolving the tension between current theoretical predictions and measurements of the high-energy atmospheric $\mu^++\mu^-$ flux. Furthermore, refined results on the cosmic ray spectrum and composition, e.g., as recently reported by the LHAASO collaboration \cite{Lv:2024wrs}, may have an impact on predictions for prompt atmospheric leptons.

\paragraph{Note Added}
This version of the manuscript differs from the version first appearing in arXiv in December 2025 because it includes the new IceCube limit on prompt neutrino fluxes~\cite{Abbasi:2025rmj} that appeared in December 2025, almost simultaneously to our work. Although the new experimental results have some impact on part of the numerical values extracted and discussed in this work, the conclusions and main message remain essentially unchanged.

\acknowledgments
We are grateful to Anatoli Fedynitch for help and tips on the use of the MCEq framework, to Felix Riehn for discussions on the physics implemented in the \texttt{Sibyll} event generator and comparisons with our QCD predictions and to Sergey Ostapchenko and Johannes Bl\"umlein for sharing with us their considerations on intrinsic charm. The model options for intrinsic charm used in this work were suggested by Sergey Ostapchenko.
This work was supported in part by the US Department of Energy under grant DE-SC-0010113, 
by the Deutsche Forschung Gemeinschaft within the FOR2926 Research Unit (Project ID 40824754) and under Germany’s Excellence Strategy – EXC 2121 “Quantum Universe” – 390833306. For facilitating portions of this research, DG wishes to acknowledge the Center for Theoretical Underground Physics and Related Areas (CETUP*), The Institute for Underground Science at Sanford Underground Research Facility (SURF), and the South Dakota Science and Technology Authority for hospitality and financial support, as well as for providing a stimulating environment.

\appendix
\section{Three-body decays of charm mesons and baryons}
\label{appendix:a}

The energy distributions of leptons from charm meson semi-leptonic decays such as $D^0\to K^- \mu^+\nu_\mu$
are often evaluated using the matrix element squared for the charm quark decay.
For the spin-summed and spin-averaged matrix element squared for the decay $c\to s \mu^+ \nu_\mu$ one obtains
\begin{equation}\label{matrix_element}
    \overline{|M|^2}= 64\, G_F^2|U_{cs}|^2(p_s \cdot p_{\nu_\mu})\, (p_{\bar{\mu}}\cdot p_c)=64\, G_F^2|U_{cs}|^2m_cE_\mu p_\nu(m_c-E_\mu+p_\mu\cos\theta_{\mu\nu})\,,
\end{equation}
where the last expression is obtained by evaluating the 4-vector scalar products and employing energy-momentum conservation in the rest frame
of the charm quark, and $U_{cs}$ is the quark dependent part of the matrix element. Here, $m_c$ is the charm quark mass, $E_\mu$ and $p_\mu$ are the $\mu^+$ energy 
and total momentum, respectively, $p_\nu$ is the magnitude of the
neutrino 3-momentum and $\theta_{\mu\nu}$ is the angle between the anti-muon and neutrino momenta.

Integrating out the strange quark momentum, integrating over $\cos\theta_{\mu\nu}$ which eliminates the delta function associated to energy conservation,
and integrating over the remaining azimuthal and solid angles of ${\bf p}_\mu$ and ${\bf p}_\nu$ we obtain
\begin{equation}
    d\Gamma= 16\,\pi^2 G_F^2|U_{cs}|^2\int dE_\mu \int dp_\nu\, E_\mu(m^2-m_cE_\mu)\,,
\end{equation}
with $m^2\equiv(m_c^2+m_\mu^2-m_s^2)/2$. We define $\mu_s\equiv m_s^2/m_c^2$ and $\mu_\mu\equiv m_\mu^2/m_c^2$.

We illustrate the impact of final state masses on the $\mu^+$ energy distribution in the charm rest frame and in the frame where the charm particle has energy $E_c$.
In the charm rest frame,
since we are interested in the $\mu^+$ energy distribution, one has to perform a final
integration over the neutrino momentum within its kinematic limits. Further defining $x\equiv 2E_\mu/m_c$, this yields 
\begin{eqnarray}
    \label{eq:dGamma_dx}
\frac{d\Gamma}{dx}&=& \pi^2 G_F^2m_c^5\frac{\left[{x^2}- 4\mu_\mu\right]^{1/2}x
     \left[1+\mu_\mu - \mu_s - x \right]^2}{1+\mu_\mu-x}\,, \end{eqnarray}
where $x$ is constrained to the interval
\begin{equation}\label{x_range}
   x_{\rm min}\equiv\frac{2m_\mu}{m_c}\leq x\leq 1+ \mu_\mu-\mu_s\equiv x_{\rm max}\,.
\end{equation}
In the limit $m_\mu=0$ this leads to
\begin{equation}\label{dGamma_dx0}
    \frac{d\Gamma}{dx}= \,\pi^2 G_F^2m_c^5\frac{x^2\left[1- \mu_s 
    -x\right]^2}{1-x}\,,
\end{equation}
for $0\leq x\leq1-\mu_s$, consistent with the literature. We note that in charm decay, the $\mu^+$ is the leptonic antiparticle, while in muon decay $\mu\to \nu_\mu e\bar\nu_e$, the corresponding antiparticle is the $\bar\nu_e$.

Finally, one has to perform a Lorentz boost with a Lorentz factor $\gamma$, corresponding
to velocity $\beta$, to the laboratory frame,
\begin{equation}
    E_\mu=\gamma E_\mu^0\left(1+\beta\frac{p_\mu^0}{E_\mu^0}\cos\theta\right)\,,
\end{equation}
where the quantities with superscript 0 now refer to the charm quark rest frame and thus to the $\mu^+$ energy and  momentum appearing in the previous equations. Defining $y\equiv E_\mu/E_c$, where $E_c=\Gamma m_c$ is now the charm quark energy in the laboratory frame,
one obtains
\begin{equation}
    y=\frac{x}{2}+\frac{\beta}{2}\left[{x^2}-4\mu_\mu \right]^{1/2}\cos\theta\,,
\end{equation}
and
\begin{equation}\label{eq:dGamma_dy}
\frac{d\Gamma}{dy}=\int_{x_{\rm min}(y)}^{x_{\rm max}(y)}\frac{dx}{2}[{x^2}-4\mu_\mu ]^{-1/2}\frac{d\Gamma}{dx}\,,    
\end{equation}
where $d\Gamma/dx$ is given by eq.~(\ref{eq:dGamma_dx}).
Here, the kinematic range of $y$ is given by
\begin{equation}
    \frac{x_{\rm max}}{2}-\frac{\beta}{2}\left[{x^2}-4\mu_\mu \right]^{1/2}\leq y\leq
     \frac{x_{\rm max}}{2}+\frac{\beta}{2}\left[{x^2}-4\mu_\mu\right]^{1/2}\,,
\end{equation}
and the integration boundaries in eq. (\ref{eq:dGamma_dy}) are given by
\begin{equation}\label{dGamma_dy}
    x_{\rm min}(y)=\max(x_{\rm min},x_-)\,,\quad x_{\rm max}(y)=\min(x_{\rm max},x_+)\,,
\end{equation}
where $x_{\rm min}$ and $x_{\rm max}$ are given by eq. (\ref{x_range}) and
\begin{equation}\label{dGamma_dy2}
x_\pm\equiv2\gamma^2\left[y\pm\left(y^2-\frac{y^2+\beta^2\mu_\mu}{\gamma^2}\right)^{1/2}\right]\,.
\end{equation}
As long as the matrix element has the structure of eq. (\ref{matrix_element}),
the above equations can be applied to any other 3-body decays by substituting $m_c$ and $m_s$ with the relevant particle
masses.

\begin{figure}
    \centering
\includegraphics[width=0.495\linewidth]{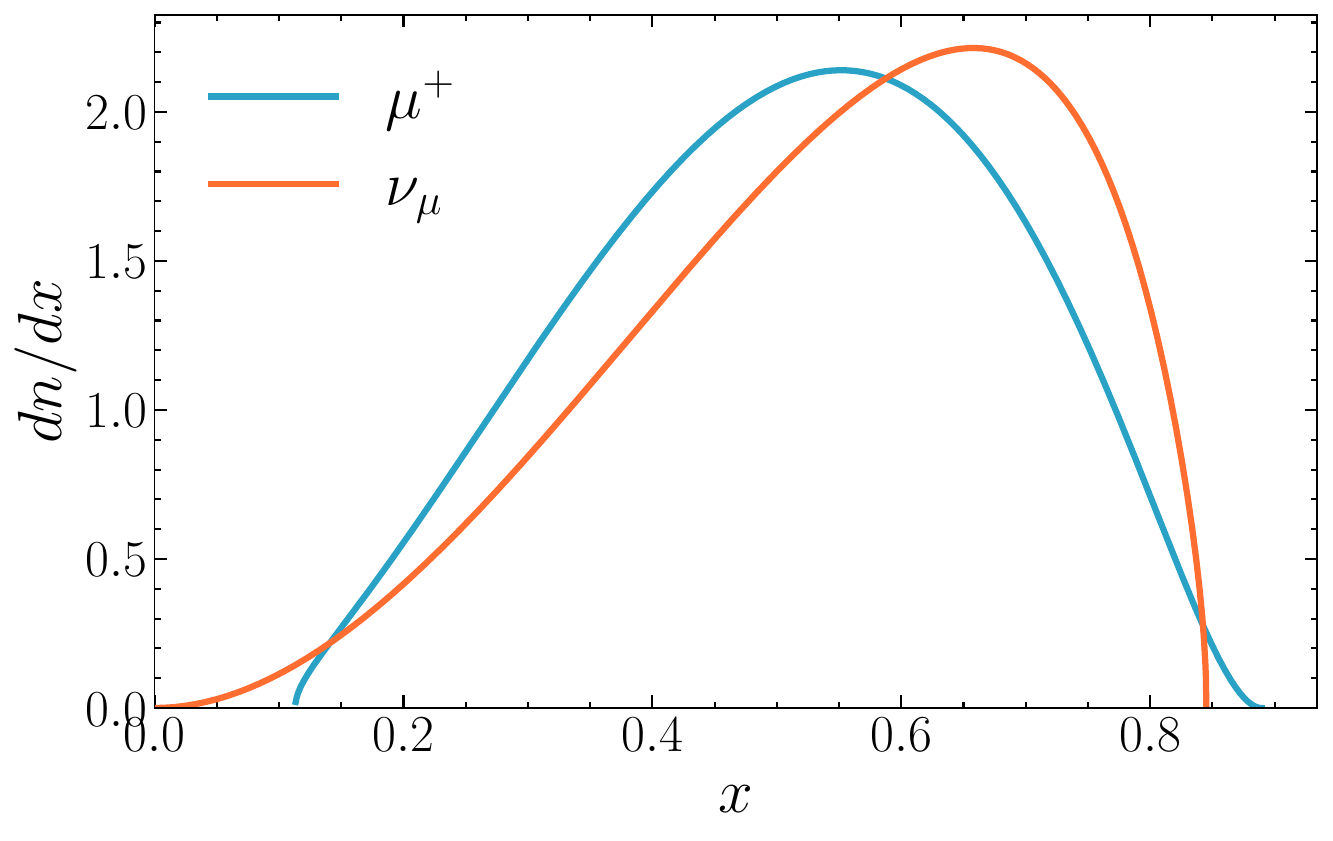}
\includegraphics[width=0.495\linewidth]{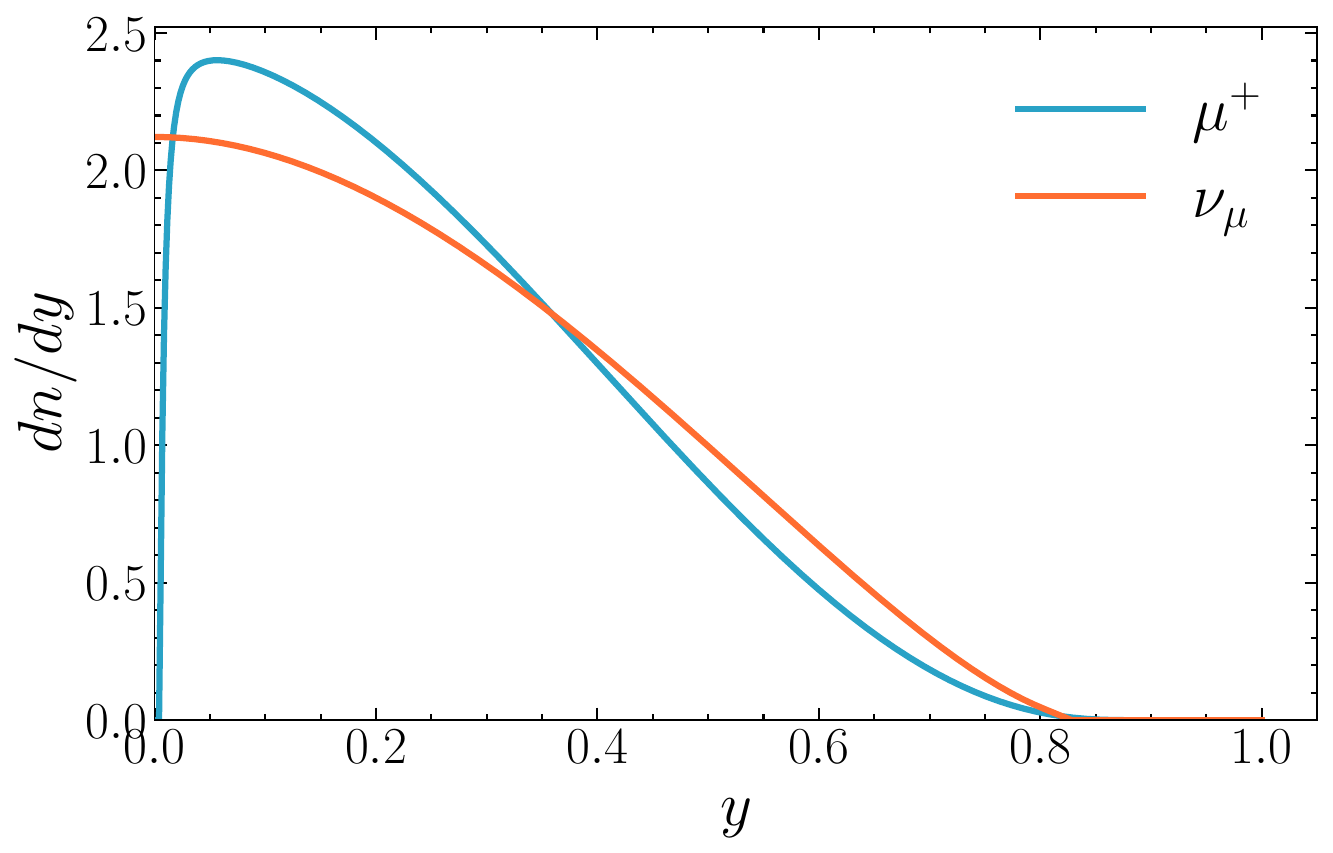}
    \caption{The center-of-mass frame $x=2E_{\mu}^0/m_{D}$ (left panel) and the boosted $y=E_{\mu}/E_{D}$ (right panel) distributions of $\mu^+$ and $\nu_\mu$ from $D^-$ decay, in the case of massive muon.}
    \label{fig:dN_dxdy}
\end{figure}

Figure \ref{fig:dN_dxdy} shows the impact of the muon mass on the $\mu^+$ normalized energy distribution in the charm meson rest frame where $x=2E_{\mu}^0/m_{D}$ (left) and in the lab frame where the charm meson has energy $E_c$ and $y=E_{\mu}/E_{D}$ (right).
A similar procedure can be applied to the $\nu_\mu$ energy distribution. In Fig. \ref{fig:dN_dxdy}, we show the $\mu^+$ and $\nu_\mu$ distributions in $x$ and $y$ from $D^0\to K^-\mu^+\nu_\mu$. 

\begin{figure}[]
    \centering
\includegraphics[width=0.495\linewidth]{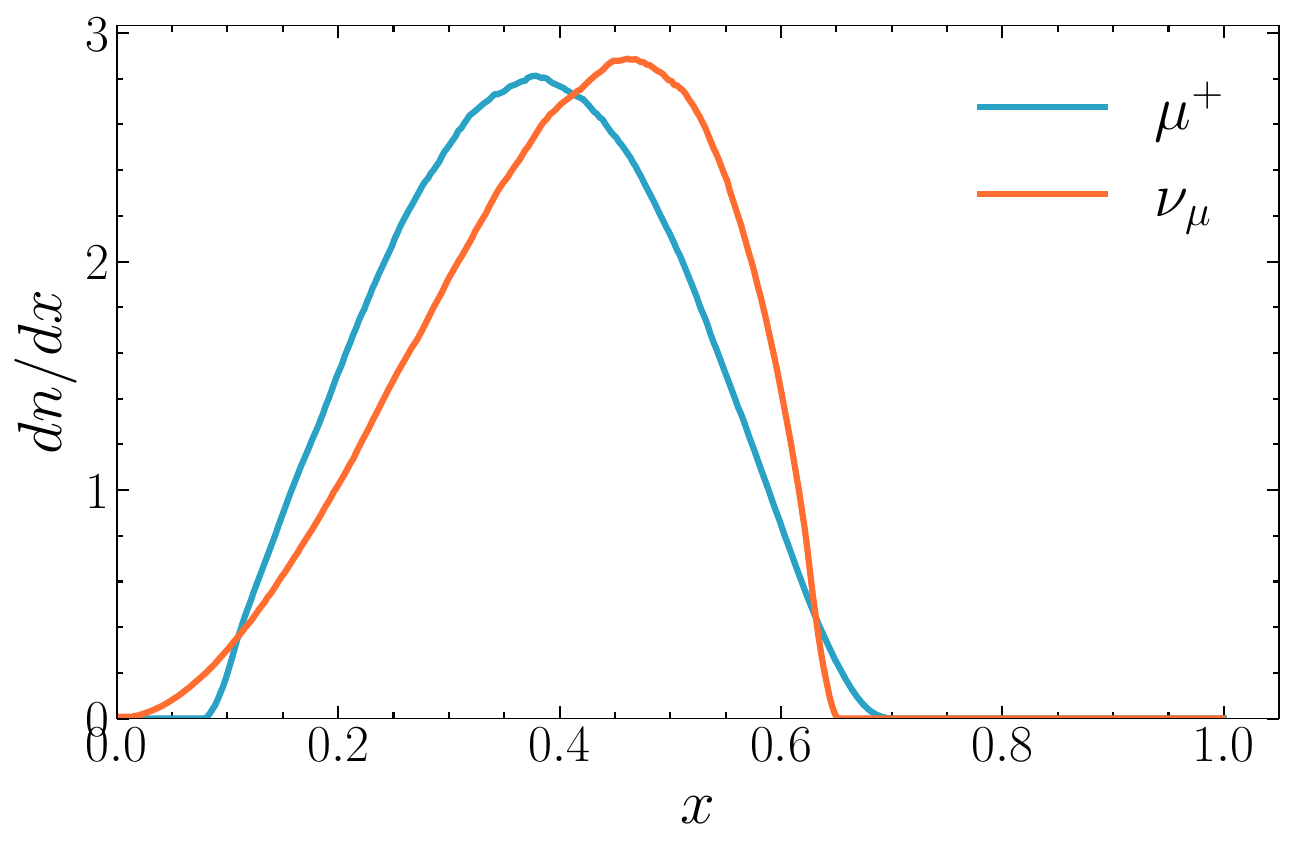}
\includegraphics[width=0.495\linewidth]{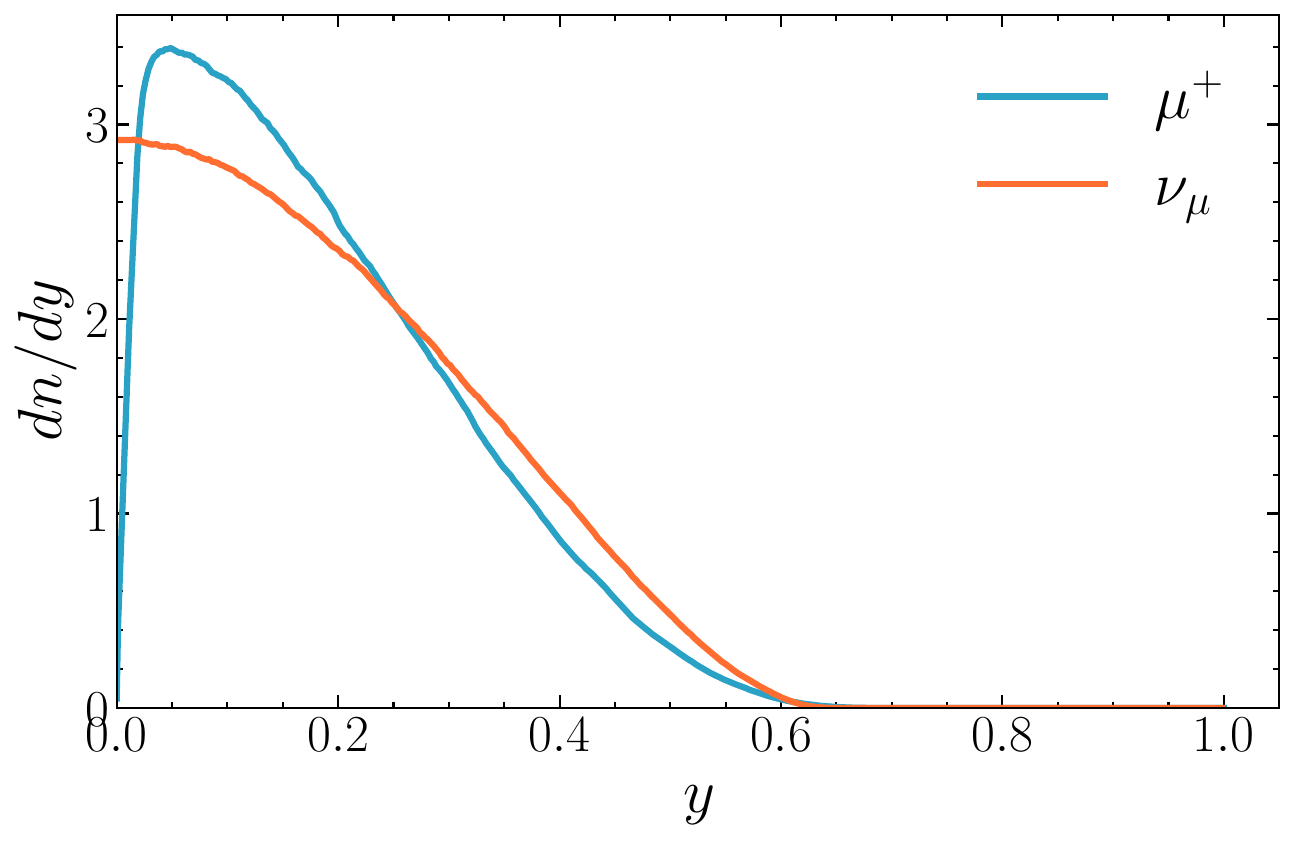}
    \caption{Rest frame $x= 2E_\ell^0/(M_{\Lambda_c}) $ (left panel) and boosted frame $y = E_\ell/E_{\Lambda_c} $ (right panel) distributions of $\mu^+$ and $\nu_\mu$ from $\Lambda_c$ decays.
    }
    \label{fig:dN_dxdy_lambda}
\end{figure}

\begin{table}[]
\centering

\vskip 0.1in

\begin{tabular}{|l|c|c|}

\hline
  $h$
  & Spectral Index $\gamma$ & $Z_{h\nu}^{\rm scaling}/Z_{h\mu}^{\rm scaling}$  \\ 

\hline
\hline
$D^\pm$ & 3 & 1.15 \\
 & 2.7 & 1.13 \\

\hline
$\Lambda_c^+$ & 3 & 1.16 \\
 & 2.7 & 1.14 \\
\hline 

\end{tabular}
\caption{The comparison of $Z_{h\mu}^{\rm scaling}$ and $Z_{h\nu_\mu}^{\rm scaling}$ for the decay of $h \equiv D^\pm,\,  \Lambda_c^+$ to muons and muon neutrinos in the scaling approximation where the cosmic ray flux energy spectrum is $\sim E^{-\gamma}$.}
\label{tab:zmom-leptons}
\end{table}

Traditionally, the $\mu^+$ energy distribution from $\Lambda_c$ decays
is evaluated using three decay form factors \cite{Buras:1976dg,Pietschmann:1984en}. (See also the appendix of the preprint version of ref. \cite{Bugaev:1998bi}.) We evaluated the energy distributions for $\mu^+$ and $\nu_\mu$ in an effective three-body semi-leptonic decay including the muon mass and an effective final-state strange hadron mass $m_{s}^{\rm eff}=1.27$ GeV \cite{Bugaev:1998bi}.
The $\mu^+$ and $\nu_\mu$ $x$-distribution
in the rest frame and $y$-distribution in the boosted frame are shown in the left and right panel of fig. \ref{fig:dN_dxdy_lambda}. 

It was noted already in Ref.~\cite{Illana:2010gh,Volkova:2011zza} that, due to the $V-A$ couplings of leptons in the weak charged current, neutrinos carry about 15\% more of the parent charm hadrons' momentum than the muon. We confirm this with our distributions in the $Z$-moment approximation. In this approximation, the lepton flux is proportional to $Z_{h\ell}$, for hadron $h$ and lepton $\ell=\mu,\nu$, where
\begin{equation}
\label{eq:zdecay}
    Z_{h\ell}(E) = \int_E^{\infty} dE_h \Biggl( \frac{\phi_h(E_h)}{\phi_h(E)}\Biggr) \frac{d_h(E)}{d_h(E_h)}\frac{1}{\Gamma (h)}\frac{d\Gamma(h\to \ell)}{dE}\,,
\end{equation}
in terms of the lepton energy $E$, hadron energy $E_h$, the flux of hadrons $\phi_h$ that depends on energy and the distance traveled in a mean $h$ decay time $d_h(E)\propto E$. For an incident cosmic ray flux that has power law scaling $\phi_{\rm CR}\propto E^{-\gamma}$, charm hadrons in the prompt regime scale as $\phi_{h}\propto E^{-\gamma+1}$ \cite{Lipari:1993hd,Gondolo:1995fq}. The decay $Z$-moment is energy independent in this approximation and can be written as 
\begin{equation}
    Z_{h\ell}^{\rm scaling} = \int_0^1 dy\, y^{\gamma-1}\frac{1}{\Gamma (h)}\frac{d\Gamma(h\to \ell)}{dy}\,.
\end{equation}
Table \ref{tab:zmom-leptons} shows the ratio $Z_{h\nu}^{\rm scaling}/Z_{h\mu}^{\rm scaling}$ for $D^\pm$ and $\Lambda_c^+$ semileptonic decays, for $\gamma=2.7$ and $\gamma=3$ in a scaling cosmic ray energy spectrum $\sim E^{-\gamma}$. In the case of $D^0$ and $\bar D^0$ decays the ratio remains the same as for $D^\pm$, since both use eq.~(\ref{eq:dGamma_dy}).

\section{$Z-$ moments and the prompt atmospheric lepton flux}
\label{appendix:z}

The results in this paper are based on  \texttt{MCEq} evaluations of the atmospheric lepton fluxes, adapted to include intrinsic charm. One feature that appears for the prompt muon fluxes is the similar shapes of the flux predictions from light unflavored mesons, pQCD($c\bar{c}$) and intrinsic charm, given an input cosmic-ray spectrum and composition. This is a consequence of the fact that the cascade equations  interaction cross sections do not change rapidly with energy in the relevant energy range and that the flux of muons at a given energy come from spectrum-weighted integrals. 
Additionally for the prompt flux, there is little dependence on the differences in interaction lengths  of the promptly decaying hadrons  since they decay before they interact.

The definition of $Z-$moment in eq. (\ref{eq:source_term1}) emphasizes the importance of the cosmic-ray nucleon spectrum weighting of the integral. Overall, the average of the outgoing energy of charm and the normalization of the charm cross section are most important. The spectrum-weighted integral of the differential charm energy distribution reduces the dependence of the prompt atmospheric lepton fluxes on the charm energy distribution near its maximum energy, where uncertainties can be quite large (due to the combination of very large$-x$ and very small$-x$ gluon PDFs for perturbative QCD), compared to the dependence on the same uncertainties of the lepton fluxes from charm hadrons close to their maximum energies at fixed collision energy, e.g., at a forward physics facility experiment at the LHC \cite{Bai:2021ira}.

As discussed in detail in refs. \cite{Lipari:1993hd} and \cite{Gondolo:1995fq}, the evaluation of the prompt muon flux at sea level using the $Z-$moment approach 
can be carried out through the
approximation
\begin{eqnarray}
\label{eq:muon-approx}
    \phi_\mu (E)&\simeq& \sum_h \frac{Z_{Nh}Z_{h\mu}}{1-Z_{NN}}\phi_{\rm CR}(E)
    \,.
    \label{eq:muon-lowe}
\end{eqnarray}
The production $Z-$moments $Z_{Nh}$ for $h=\eta,\rho^0,J/\psi,\phi,\eta',\omega$ and charm hadrons (pQCD and intrinsic) are defined in eq. (\ref{eq:source_term1}), where $N$ is the incident cosmic-ray nucleon and $\phi_N(E)=\phi_{\rm CR}(E)$ is the cosmic-ray spectrum per nucleon at the top of the atmosphere. Equation \ref{eq:muon-lowe} is applicable below charm hadron energies of $\sim 4\times 10^7$ GeV--$2\times 10^8$ GeV, depending on the hadron (see, e.g., ref. \cite{Gondolo:1995fq}) and to higher energies for the very short-lived unflavored mesons.

For a power-law cosmic ray spectrum, since hadronic cross sections are weakly dependent on energy, $Z_{Nh}$ is nearly energy independent. As discussed at the end of Appendix \ref{appendix:a}, the decay moment $Z_{h\mu}$ in eq. (\ref{eq:zdecay}) is also nearly energy independent, as is the cosmic-ray nucleon re-interaction $Z-$moment, $Z_{NN}$. Deviations from a strict power-law cosmic ray spectrum introduces a mild energy dependence. Each hadron that contributes to the prompt muon flux therefore involves a weakly energy dependent coefficient that multiplies the cosmic-ray nucleon flux. This accounts for the nearly identical shapes of the pQCD($c\bar{c})$, intrinsic charm and light unflavored hadron contributions to the prompt atmospheric muon flux at sea level in the energy range considered here, despite different energy distributions such as the outgoing $D$-meson distributions from pQCD and from intrinsic charm in cosmic ray-air collisions.

\section{Prompt atmospheric neutrino fluxes from perturbative QCD evaluations}
\label{appendix:b}
Despite the relatively small fraction of charm production in the total proton-air cross section in the atmosphere, it has long been understood that the prompt decays of charm hadrons into neutrinos and muons are important contributions to high energy atmospheric lepton fluxes \cite{Volkova:1980sw,Volkova:1987gh}. In this Appendix, we compare the prompt atmospheric $\nu_\mu+\bar\nu_\mu$ flux  in the energy range of $2\times 10^3-10^5$ GeV from \texttt{Sibyll-2.3c} with some representative predictions based on perturbative QCD. The comparison is limited to this neutrino energy range since this is where the IceCube upper bounds on the prompt atmospheric $\nu_\mu+\bar\nu_\mu$ flux apply \cite{IceCube:2016umi,Abbasi:2025rmj}. 

In hadron-hadron collisions, gluon-gluon fusion to produce a $c\bar{c}$ pair dominates the charm production cross section in perturbative QCD in collinear factorization \cite{Nason:1987xz,Mangano:1991jk}. The early prompt 
atmospheric neutrino flux evaluations have been succeeded by evaluations from a large number of groups, e.g., in refs. \cite{Gondolo:1995fq,Pasquali:1998xf,Gelmini:1999ve,Enberg:2008te,Bhattacharya:2015jpa,Bhattacharya:2016jce,Gauld:2015yia,Gauld:2015kvh,Garzelli:2015psa,Garzelli:2016xmx,Ostapchenko:2022thy,Bai:2022xad,Benzke:2017yjn,Zenaiev:2019ktw,PROSA:2015yid}.
Most of the next-to-leading order (NLO) perturbative QCD evaluations of charm pair production are performed in the 3-flavor number scheme in the collinearly-factorized parton model.  Non-perturbative contributions to the differential cross sections for charm-hadron production include the fragmentation of charm quarks to hadrons. 
Different evaluations of the prompt atmospheric lepton fluxes from charm have similar predictions (i.e. predictions whose uncertainty bands overlap). The predictions vary because of  parton distribution function choices, the value of the charm quark mass and renormalization/factorization scale dependence.

The prompt atmospheric lepton fluxes at very high energies depend on small parton-$x$ distributions of the gluon parton distribution function (PDF) which have the largest variations between PDF sets, since for $E_\nu,\ E_\mu>10^6$ GeV, the gluon PDF is probed below $x\simeq 10^{-5}$, at the border of the region where already existing measurements have probed the gluon PDF (see, e.g., ref. \cite{Zenaiev:2019ktw}). At energies below $E_\nu=10^5$ GeV, PDF uncertainties within a given PDF set choice are small, although there are some differences associated with choice of PDF set, as seen in, e.g., refs. \cite{Bhattacharya:2016jce,Gauld:2015kvh,Zenaiev:2019ktw,PROSA:2015yid}. The charm pole mass is assumed to be $m_c=1.4-1.5$ GeV, with variations of order $\pm 0.15$ GeV to $\pm 0.20$ GeV that introduce additional uncertainties. Finally, the choice of renormalization and factorization scales and the convention  
of considering variations of the renormalization/factorization scale ratio  between factors of 1/2 and 2 provides the largest uncertainties. On the one hand, the scale uncertainties are proxies for missing higher-order uncertainties. 
On the other hand, one might question up to which extent the scale variation procedure considered above is a measure of the actual uncertainties, also considering the fact that LHC experimental data have much lower uncertainties.

\begin{figure}
    \centering
    \includegraphics[width=0.5\linewidth]{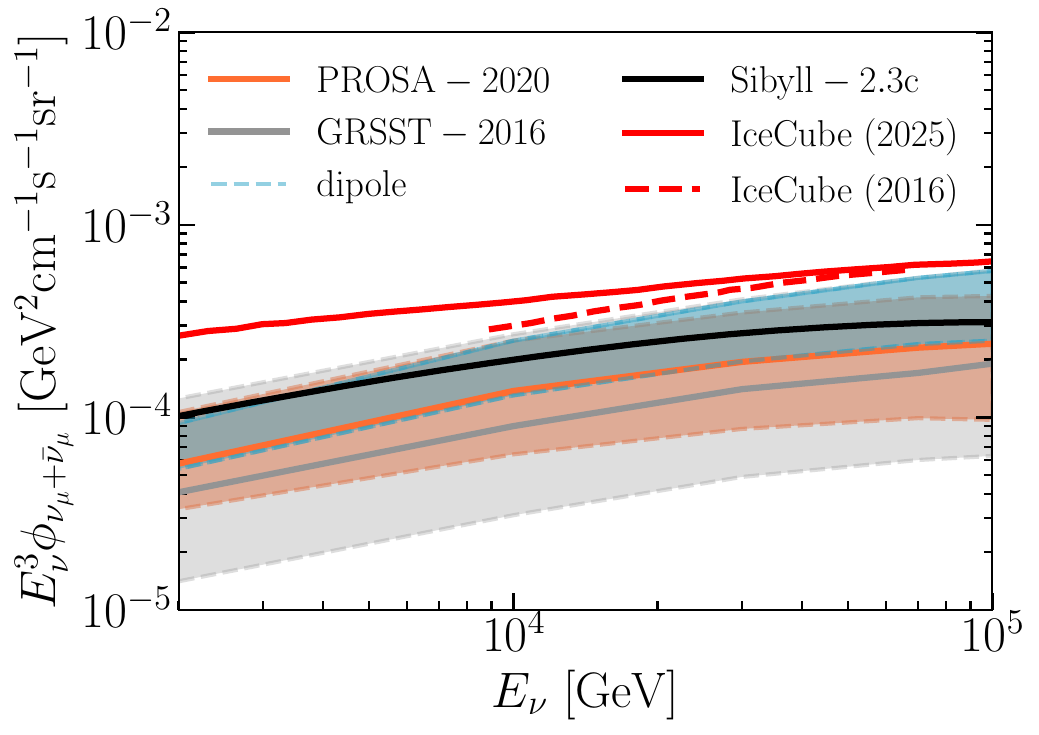}
    \caption{A comparison of the central NLO QCD predictions (solid curves) from Gauld et al. (GRRST) \cite{Gauld:2015kvh} and  the PROSA collaboration \cite{Zenaiev:2019ktw}, together with their respective error bands that include PDF uncertainties, charm mass dependence and renormalization/factorization scale dependence. Also shown a band for the range of dipole model predictions from \cite{Bhattacharya:2016jce}. The solid black curve shows the baseline \texttt{Sibyll-2.3c} results from \texttt{MCEq}. The red solid (dashed) curve shows the 2025 (2016) IceCube upper bounds \cite{Abbasi:2025rmj,IceCube:2016umi} based on the \texttt{Sibyll-2.3c} (dipole model \cite{Enberg:2008te}) evaluation. 
    \label{fig:flux-compare}}
\end{figure}

Figure \ref{fig:flux-compare} shows a comparison of two NLO perturbative QCD evaluations of the prompt atmospheric $\nu_\mu+\bar\nu_{\mu}$ flux in the energy range $E_\nu=2\times 10^3-10^5$ GeV. The orange curve and associated uncertainty band is from the PROSA collaboration \cite{Zenaiev:2019ktw}. The grey curve and error band show the results of Gauld et al. in 
ref. \cite{Gauld:2015kvh}. In the same plot, we show a blue band which shows the range of dipole model predictions from ref. \cite{Bhattacharya:2016jce}. Finally, the black curve is the \texttt{Sibyll-2.3c} result. All of the curves were evaluated using the \texttt{H3a} cosmic-ray flux model. 
The figure confirms that the \texttt{Sibyll-2.3c} evaluation of charm production and decay in cosmic ray-air interactions is consistent with the NLO perturbative QCD evaluation. 
For reference, we also show the IceCube upper bounds from 2025 \cite{Abbasi:2025rmj} (solid red curve) and from 2016 \cite{IceCube:2016umi} (dashed red curve) that are based on the shape of the prompt flux from \texttt{Sibyll-2.3c} and from the dipole model \cite{Enberg:2008te,Bhattacharya:2016jce}, respectively.

Given its consistency with NLO QCD-based approaches to charm production and the fact that the recent IceCube upper bound on the prompt atmospheric neutrino flux relies on the shape of the \texttt{Sibyll-2.3c} flux, we adopt the \texttt{Sibyll-2.3c} prompt atmospheric lepton fluxes as our baselines. So as not to confuse the \texttt{Sibyll-2.3c} charm results with intrinsic charm as introduced in Section \ref{sec:intrinsiccharm}, we denote the \texttt{Sibyll-2.3c} charm contributions to the atmospheric muon and muon neutrino fluxes as ``pQCD,'' with the clarification in the main text that \texttt{Sibyll-2.3c} is a phenomenological model based on soft and hard minijets.

For higher neutrino energies, above $\sim 10^6$ GeV, where the prompt atmospheric muon neutrino flux dominates the conventional flux contributions, the uncertainty bands in the perturbative QCD evaluations expand. As has been emphasized in,  e.g., refs. \cite{Bai:2020ukz,Bai:2021ira,
Bai:2022jcs,Kling:2023tgr,Bhattacharya:2023zei}, 
neutrino experiments (see ref. \cite{Ariga:2025qup} for a review of experiments)
at the proposed Forward Physics Facility \cite{Anchordoqui:2021ghd,Feng:2022inv,
FPF:2025bor} will be sensitive to gluon distribution functions at large and small parton $x$ values. While the kinematic overlap of charm contributions to the Forward Physics Facility neutrino fluxes with charm contributions to the  prompt atmospheric neutrino fluxes with $E_\nu\sim 10^6$ GeV is partial \cite{Bai:2022xad}, neutrino flux measurements at the Forward Physics Facility would be sensitive to small-$x$ gluon distribution function and to intrinsic charm. We emphasize that the possibility of using these data in actual PDF fits or to measure intrinsic charm effects would still require disentangling neutrino fluxes from neutrino cross sections, as well as a very good control of the neutrino fluxes, identifying precisely the components due to charm production and decay.  
Already, a neutrino flux analysis from the first data from the FASER experiment \cite{FASER:2024ref} has been published \cite{John:2025qlm}, in which future  sensitivities to intrinsic charm is considered, albeit with a hybrid approach using fixed-flavor number scheme with 3 flavors in collinear factorization and intrinsic charm incorporated separately with another scheme \cite{Maciula:2022lzk}. Future measurements will drive refinements to treatments of intrinsic charm.

\bibliographystyle{JHEP}
\bibliography{sample}

\end{document}